\RequirePackage{fix-cm}
\documentclass[smallextended]{svjour3}       % onecolumn (second format)
\smartqed  % flush right qed marks, e.g. at end of proof
\usepackage{graphicx}
\usepackage{color}
\begin{document}

\title{Interaction phenomena between lump and solitary wave of a generalized (3 + 1)-dimensional variable-coefficient
nonlinear-wave equation in liquid with gas bubbles\thanks{Project
supported by National Natural Science Foundation of China (Grant No
81960715) and Science and Technology Project of Education Department
of Jiangxi Province(GJJ151079).}}

%\titlerunning{Jimbo-Miwa equationn}        % if too long for running head

\author{ Jian-Guo Liu$^{*}$ \and Wen-Hui Zhu$^{*}$ \and Yan He$^{*}$ \and Ya-Kui Wu$^{*}$%etc.
}


\institute{Jian-Guo Liu(*Corresponding author),  Yan
He(*Corresponding author)\at
              College of Computer, Jiangxi University of
Traditional Chinese Medicine, Jiangxi 330004, China\\
              Tel.: +8613970042436\\
              \email{20101059@jxutcm.edu.cn(J.G. Liu)}\\
              \email{274667818@qq.com(Y. He)}          %  \\
 \and
            Wen-Hui Zhu(*Corresponding author) \at Institute of artificial intelligence, Nanchang Institute of Science and Technology,
             Jiangxi 330108, China\\
             \email{415422402@qq.com}\and
            Ya-Kui Wu(*Corresponding author) \at School of science, Jiujiang University, Jiangxi
             332005, China\\
             \email{2180501008@cnu.edu.cn}
              }

\date{Received: date / Accepted: date}

\maketitle

\begin{abstract}
In this paper,  a generalized (3 + 1)-dimensional
variable-coefficient nonlinear-wave equation is studied in liquid
with gas bubbles. Based on  the Hirota's bilinear form and symbolic
computation, lump and interaction solutions between lump
 and solitary wave are obtained. Their interaction phenomena is
 shown in some 3d graphs and contour plots, which include a periodic-shape
 lump
 solution, a parabolic-shape lump solution, a cubic-shape lump solution, interaction solutions between lump
 and one solitary wave, and between lump and two solitary waves.
  The spatial structures called the bright lump wave and the bright-dark lump wave are discussed.
   Interaction behaviors of two bright-dark lump waves and a periodic-shape bright lump wave  are also presented. \\


\end{abstract}

\section{Introduction}
\label{intro} \quad In some branches of science and engineering such
as fluid mechanics, quantum mechanics, particle physics, mass
transfer,plasma physics, nanoliquids and biological mathematics
[1-5], Nonlinear partial differential equations (NPDES) are used to
describe many nonlinear phenomena and wave propagation
characteristics. As the lump solutions of the NPDES are the special,
powerful destructive ocean wave in the real world, it is important
to search for the lump solutions of the NPDES, especially the
constant coefficient NPDES have attracted the attention of many
scholars [6-14].

\quad Recently, a generalized (3 + 1)-dimensional nonlinear-wave
equation has been presented as [15]
\begin{eqnarray}
[4 u_t+4 u u_x+u_{xxx}-4 u_x]_x+3
   (u_{yy}+u_{zz})=0,
\end{eqnarray}which describes a liquid with gas bubbles in the three dimensional case.

\quad However, the variable-coefficient NPDES  provide us more real
phenomena in the inhomogeneities of media and non-uniformities of
boundaries than corresponding constant-coefficient counterparts in
some physical cases. In this paper, a generalized (3 +
1)-dimensional variable-coefficient nonlinear-wave equation is
investigated [16,17]
\begin{eqnarray}
\alpha (t) u_x^2+\alpha (t) u u_{xx}+\beta (t) u_{xxxx}+\gamma (t) u_{xx}+\delta (t)
   u_{yy}+\varrho (t) u_{zz}+u_{xt}=0,
\end{eqnarray}where $u=u(x,y,z,t)$ is the wave-amplitude function. The bilinear form,
B\"{a}cklund transformation, Lax pair, infinitely-many conservation
laws, multi-soliton solutions, travelling-wave solutions and
one-periodic wave solutions are presented by virtue of the binary
Bell polynomials, the Hirota method, the polynomial expansion method
and the Hirota-Riemann method [18]. However, lump and interaction
solutions between lump
 and solitary wave of Eq. (2) have not been obtained yet, which will make the main work of our paper.

\quad This paper will be organised as follows. Section 2 obtains the
lump solutions of Eq. (2) with the aid of the Hirota's bilinear form
[19-24] and Mathematical software [25-35] and demonstrates their
physical structures by some 3d graphs and contour plots; Section 3
presents the interaction solutions between lump
 and one solitary wave;  Section 4
derives the interaction solutions between lump
 and two solitary waves; Section 5  gives the conclusion.

\section{Lump solutions of Eq. (2)} \label{sec:2}
\quad Setting  $u=12\,[ln\xi(x,y,z,t)]_{xx}$ and  $\alpha
(t)=\beta(t)$, the bilinear form of Eq. (2) can be introduced as
(see Ref. [18]) by using the multi-dimensional Bell polynomials
{\begin{eqnarray} [D_x D_t+\beta (t) D_x^4+\gamma (t) D_x^2+\delta
(t) D_y^2+\varrho (t) D_z^2] \xi\cdot \xi=0.
\end{eqnarray}
}This is equivalent to {\begin{eqnarray} &&\xi  [\beta (t)
\xi_{xxxx}+\gamma (t) \xi_{xx}+\delta (t) \xi_{yy}+\varrho (t)
\xi_{zz}+\xi_{xt}]+3 \beta (t) \xi_{xx}^2\nonumber\\&&-4 \beta (t)
\xi_x \xi_{xxx}-\gamma (t) \xi_x^2-\delta (t) \xi_y^2-\varrho (t)
\xi_z^2-\xi_t \xi_x=0.
\end{eqnarray}
}In order to seek the lump solutions of Eq. (2), we suppose
{\begin{eqnarray} \zeta&=&x \alpha _1+y \alpha _2+z
\alpha _3+\alpha _4(t),\nonumber\\
\varsigma&=&x \alpha _5+y \alpha _6+z \alpha _7+\alpha _8(t),\nonumber\\
\xi&=& \zeta^2+\varsigma^2+\alpha _9(t),
\end{eqnarray}}where $\alpha_1$, $\alpha_2$, $\alpha_3$, $\alpha_5$, $\alpha_6$ and $\alpha_7$ are
unknown constants. $\alpha_4(t)$, $\alpha_8(t)$, $\alpha_9(t)$ are
undefined real functions. Substituting Eq. (5) into Eq. (4) through
Mathematical symbolic computations, we get {\begin{eqnarray}(I):
\alpha _8(t)&=&\eta _1+\int_1^t -\frac{\left(\alpha _1^2+\alpha
_5^2\right) [\alpha _1^2 \gamma (t)+\alpha _2^2 \delta (t)+\alpha
_3^2 \varrho (t)]+\alpha _1^3 \alpha
   _4'(t)}{\alpha _1^2 \alpha _5} \, dt,\nonumber\\ \alpha_9(t)&=&\eta _2+\int_1^t [2 [\alpha _1 \left(\alpha _1 \gamma (t)+\alpha _4'(t)\right)+\alpha _2^2 \delta (t)+\alpha _3^2 \varrho (t)] [\alpha _1
   [\eta _3\nonumber\\&+&\int_1^t -\frac{\left(\alpha _1^2+\alpha _5^2\right) [\alpha _1^2 \gamma (t)+\alpha _2^2 \delta (t)+\alpha _3^2 \varrho (t)]+\alpha _1^3
   \alpha _4'(t)}{\alpha _1^2 \alpha _5} \, dt]-\alpha _5\nonumber\\&*&\alpha _4(t)]]/(\alpha _1 \alpha _5) \, dt, \alpha(t)=\beta(t)=0, \alpha_6=\frac{\alpha _2 \alpha _5}{\alpha _1},
\alpha_7=\frac{\alpha _3 \alpha _5}{\alpha _1},
\end{eqnarray}}with  $\alpha_1 \neq 0$, $\alpha_5 \neq 0$. Substituting Eq. (5) and Eq. (6) into the transformation $u=12\,[ln\xi(x,y,z,t)]_{xx}$,
we have  the following lump solution of Eq. (2) {\begin{eqnarray}
u^{(I)}&=&[12 [2 \left(\alpha _1^2+\alpha _5^2\right) [\eta
_2+\int_1^t [2 [\alpha _1 [\alpha _1 \gamma (t)+\alpha
_4'(t)]+\alpha _2^2
   \delta (t)+\alpha _3^2 \varrho (t)] [\alpha _1 [\eta _1\nonumber\\&+&\int_1^t -\frac{\left(\alpha _1^2+\alpha _5^2\right) [\alpha _1^2 \gamma (t)+\alpha
   _2^2 \delta (t)+\alpha _3^2 \varrho (t)]+\alpha _1^3 \alpha _4'(t)}{\alpha _1^2 \alpha _5} \, dt]-\alpha _5 \alpha _4(t)]]\nonumber\\&/&(\alpha _1 \alpha _5)
   \, dt+[\eta _1-\int_1^t \frac{\left(\alpha _1^2+\alpha _5^2\right) [\alpha _1^2 \gamma (t)+\alpha _2^2 \delta (t)+\alpha _3^2 \varrho
   (t)]+\alpha _1^3 \alpha _4'(t)}{\alpha _1^2 \alpha _5} \, dt\nonumber\\&+&\frac{\alpha _5 \left(\alpha _1 x+\alpha _2 y+\alpha _3 z\right)}{\alpha
   _1}]{}^2+[\alpha _4(t)+\alpha _1 x+\alpha _2 y+\alpha _3 z]{}^2]-4 [\alpha _5 [\eta _1\nonumber\\&+&\int_1^t -\frac{\left(\alpha _1^2+\alpha
   _5^2\right) [\alpha _1^2 \gamma (t)+\alpha _2^2 \delta (t)+\alpha _3^2 \varrho (t)]+\alpha _1^3 \alpha _4'(t)}{\alpha _1^2 \alpha _5} \, dt+\alpha _5
   x]\nonumber\\&+&\alpha _1 \left(\alpha _4(t)+\alpha _2 y+\alpha _3 z\right)+\alpha _1^2 x+\frac{\alpha _5^2 \left(\alpha _2 y+\alpha _3 z\right)}{\alpha
   _1}]{}^2]]/[[\eta _2\nonumber\\&+&\int_1^t [2 [\alpha _1 \left(\alpha _1 \gamma (t)+\alpha _4'(t)\right)+\alpha _2^2 \delta (t)+\alpha _3^2 \varrho
   (t)] [\alpha _1 [\eta _1\nonumber\\&-&\int_1^t \frac{\left(\alpha _1^2+\alpha _5^2\right) [\alpha _1^2 \gamma (t)+\alpha _2^2 \delta (t)+\alpha _3^2
   \varrho (t)]+\alpha _1^3 \alpha _4'(t)}{\alpha _1^2 \alpha _5} \, dt]-\alpha _5 \alpha _4(t)]]\nonumber\\&/&(\alpha _1 \alpha _5) \, dt+[\eta
   _1-\int_1^t \frac{\left(\alpha _1^2+\alpha _5^2\right) [\alpha _1^2 \gamma (t)+\alpha _2^2 \delta (t)+\alpha _3^2 \varrho (t)]+\alpha _1^3 \alpha
   _4'(t)}{\alpha _1^2 \alpha _5} \, dt\nonumber\\&+&\frac{\alpha _5 \left(\alpha _1 x+\alpha _2 y+\alpha _3 z\right)}{\alpha _1}]{}^2+\left(\alpha _4(t)+\alpha _1
   x+\alpha _2 y+\alpha _3 z\right){}^2]{}^2],
\end{eqnarray}}where $\alpha _4(t)$ is  arbitrary  function, $\eta _1$ and $\eta _2$ are integral constants.

The physical structures for  $(u^{(I)})$ are described in Fig. 1  by
the 3d graphs and contour plots. Fig. 1 shows the propagation of
solution $(u^{(I)})$ when $\gamma (t)$, $\delta(t)$, $\varrho(t)$
and $\alpha_4(t)$ select different functions. When $\gamma
(t)=\delta(t)=\varrho(t)=\cos t$ and $\alpha_4(t)=\sin t$, a
periodic-shape rational solution is listed in Fig. 1(a) and Fig.
1(d). When $\gamma (t)=\delta(t)=\varrho(t)=t$ and $\alpha_4(t)=\sin
t$, a parabolic-shape rational solution is presented in Fig. 1(b)
and Fig. 1(e). When $\gamma (t)=\cosh t, \varrho(t)=\exp t$ and
$\alpha_4(t)=\delta(t)=t$, a cubic-shape rational solution is shown
in Fig. 1(c) and Fig. 1(f).

\includegraphics[scale=0.4,bb=20 270 10 10]{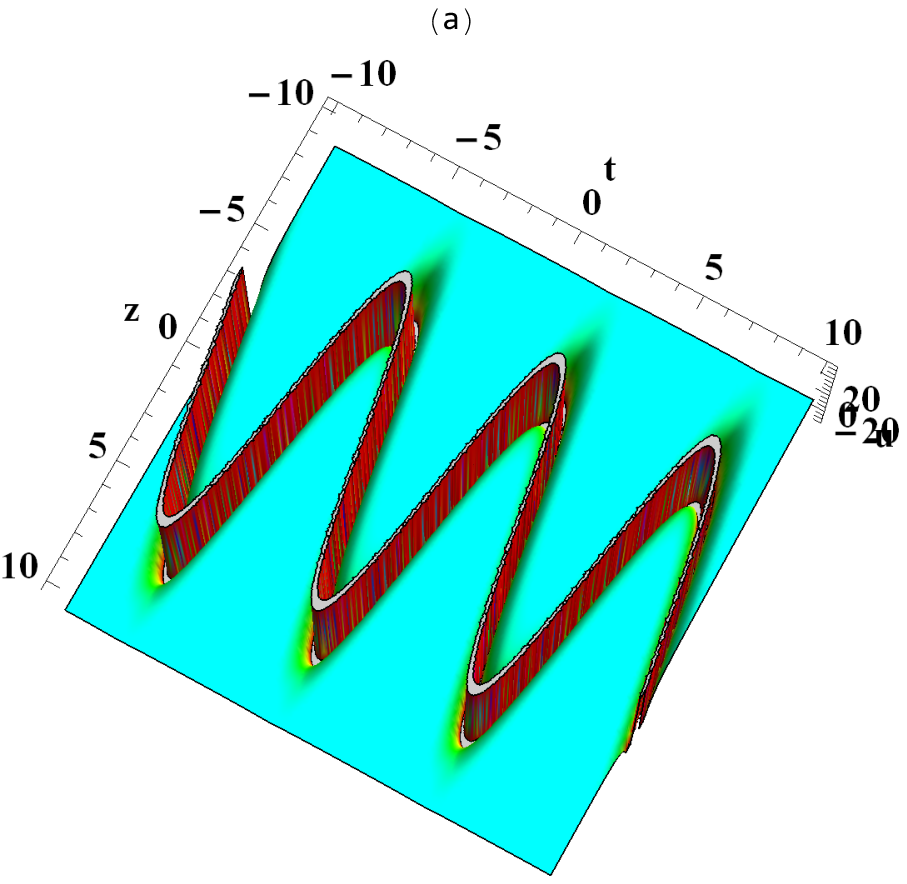}
\includegraphics[scale=0.4,bb=-255 270 10 10]{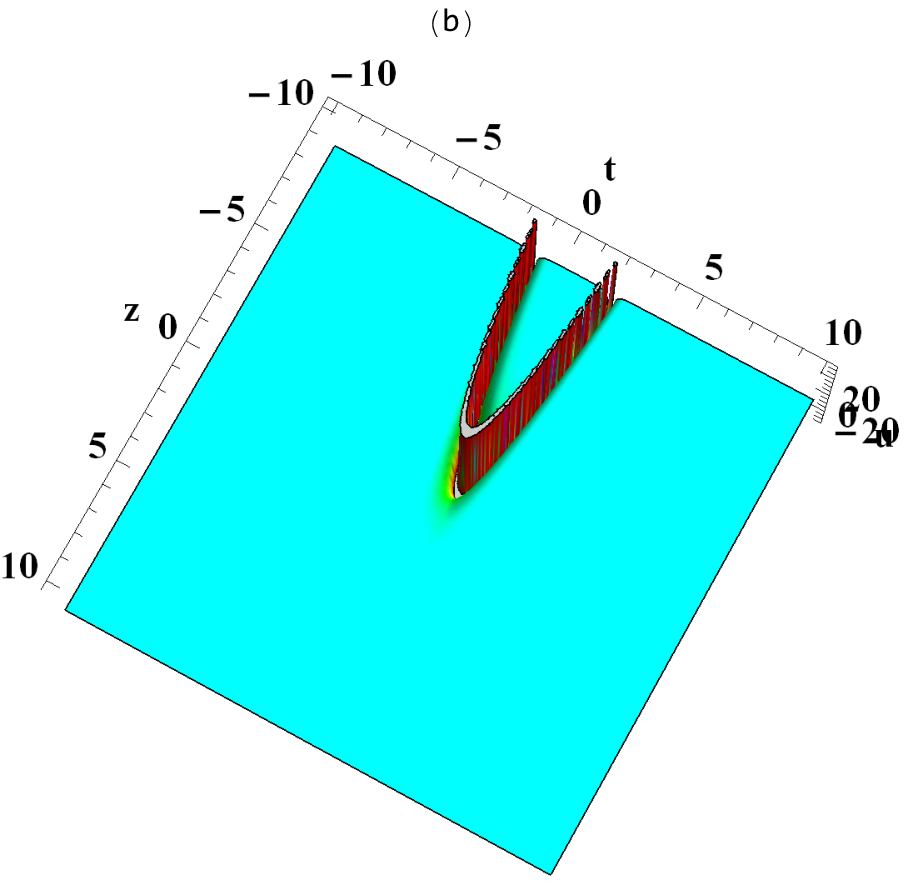}
\includegraphics[scale=0.4,bb=-260 270 10 10]{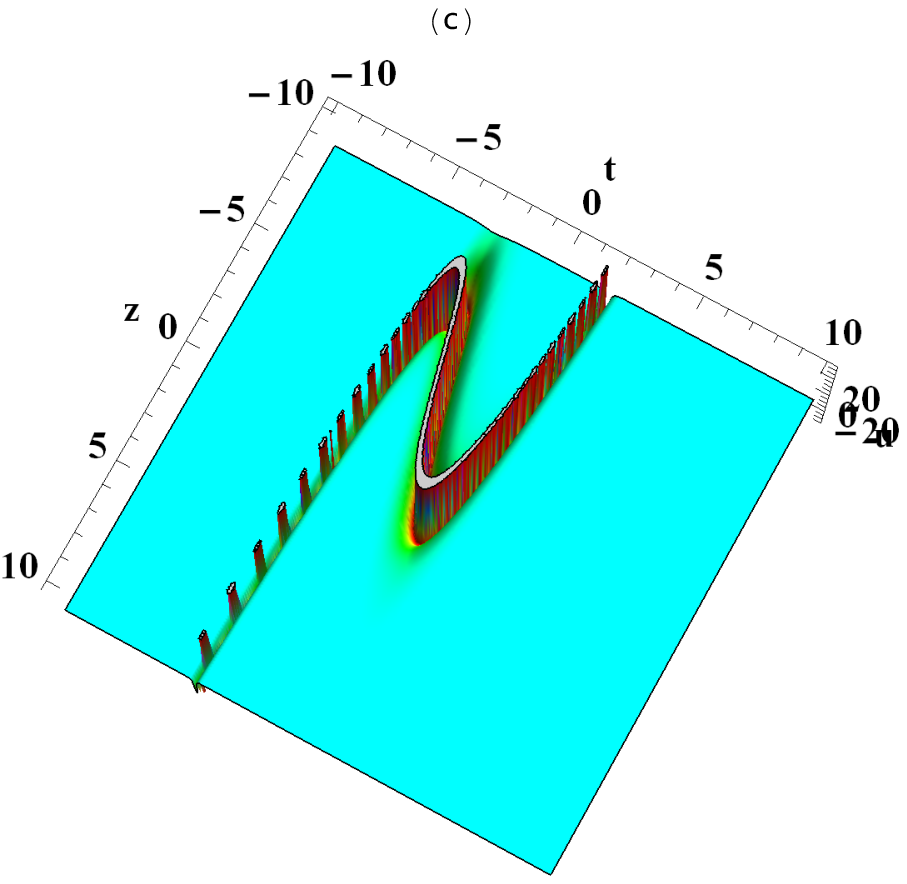}
\includegraphics[scale=0.35,bb=750 620 10 10]{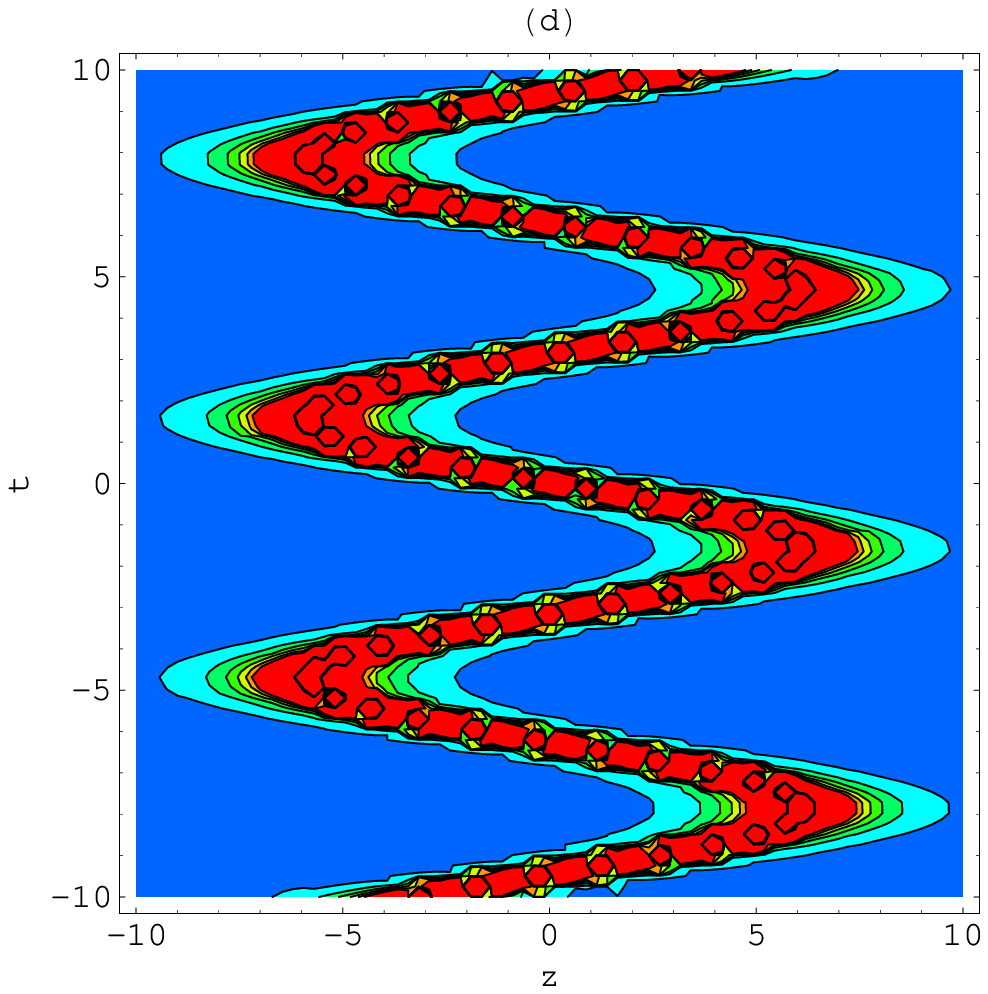}
\includegraphics[scale=0.35,bb=-290 620 10 10]{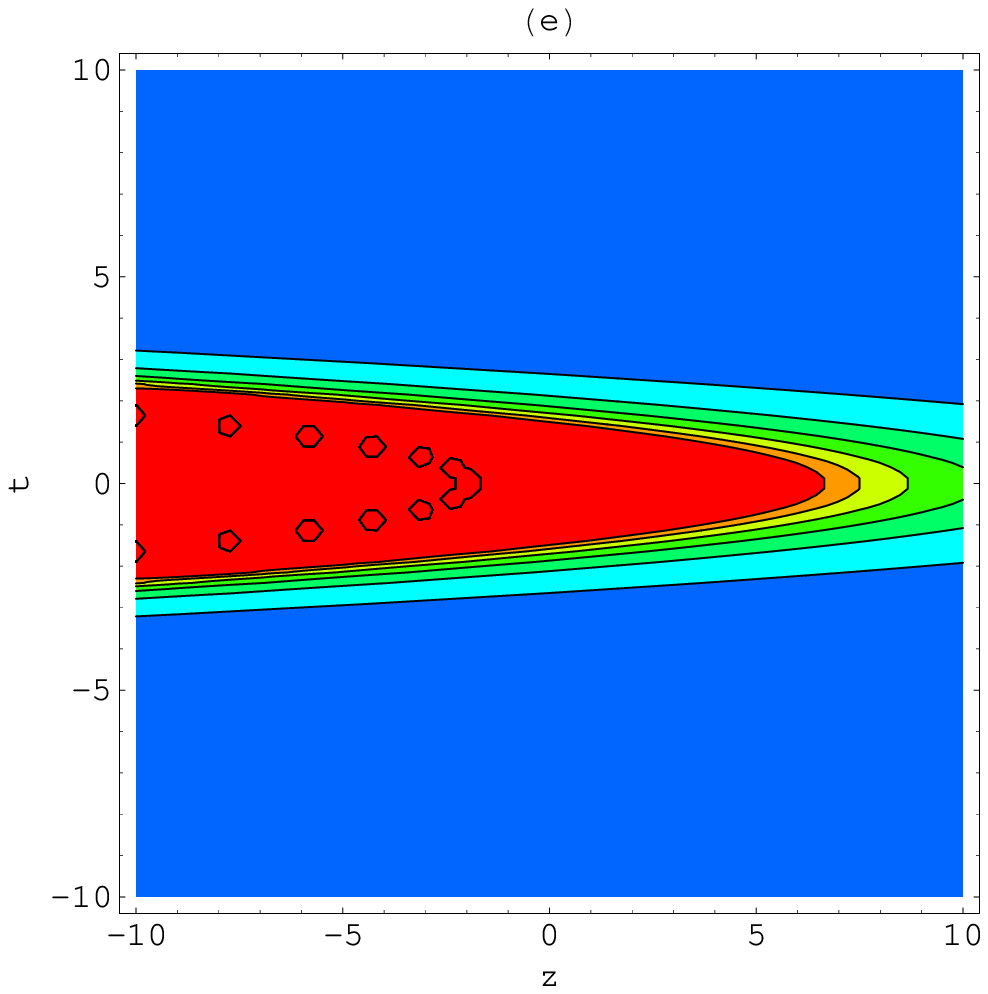}
\includegraphics[scale=0.35,bb=-290 620 10 10]{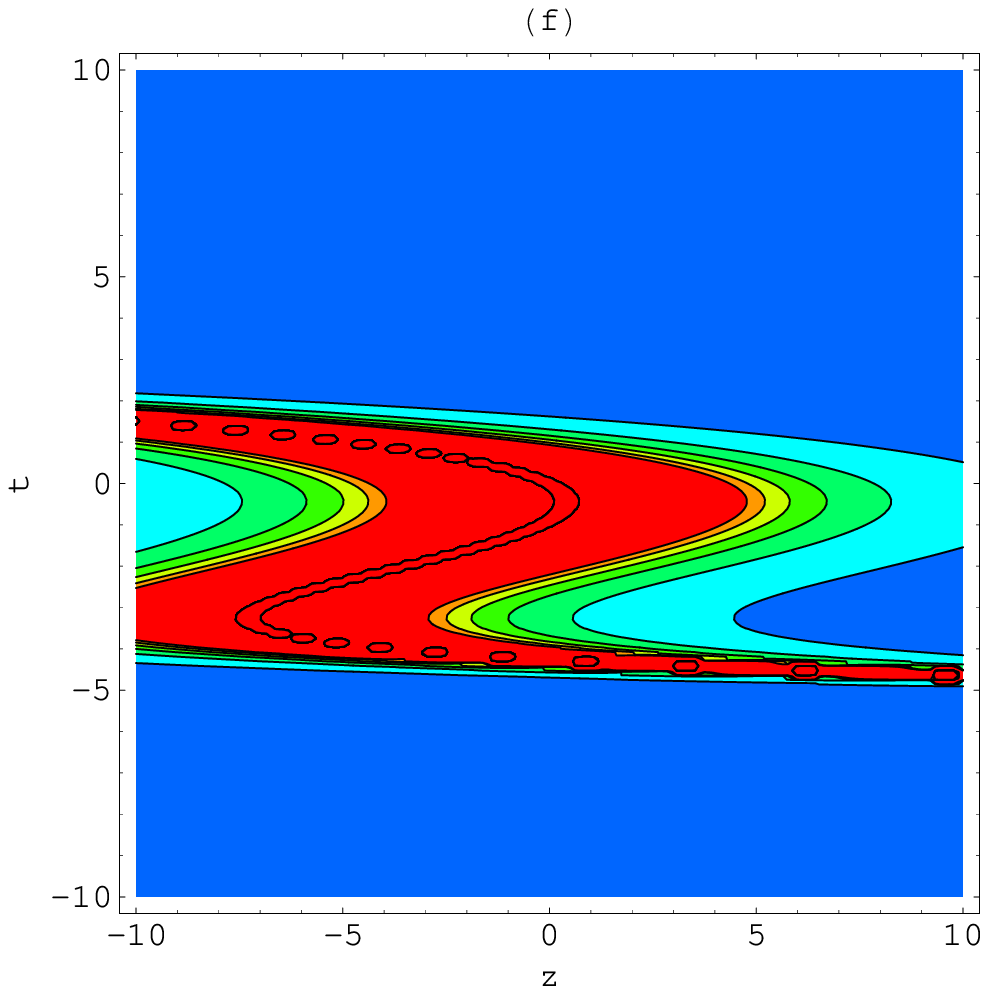}
\vspace{7.5cm}
\begin{tabbing}
\textbf{Fig. 1}. Lump solution $u^{(I)}$ with $\alpha_1= 1$,
$\alpha_2= 2$, $\alpha_3=-1$,  $\alpha _5=-3$, $x=-5$,\\
$\eta_1=\eta _2=y=0$, when $\gamma (t)=\delta(t)=\varrho(t)=\cos t,
\alpha_4(t)=\sin t$  in (a) (d),\\ $\gamma
(t)=\delta(t)=\varrho(t)=t,
\alpha_4(t)=\sin t$ in (b) (e) and $\gamma (t)=\cosh t$,\\
$\varrho(t)=\exp t$ and $\alpha_4(t)=\delta(t)=t$ in (c) (f).
\end{tabbing}

{\begin{eqnarray}(II):  \alpha _8(t)&=& \eta _3+\int_1^t \alpha _5 [-\frac{3 \left(\alpha _1^2+\alpha _5^2\right) \beta (t)}{\alpha _9}-\frac{\alpha _3^2 \varrho (t)}{\alpha _1^2}-\gamma (t)]
   \, dt,\nonumber\\ \alpha_9(t)&=&\alpha_9, \alpha_6=-\frac{\alpha _1 \alpha _2}{\alpha _5},
\alpha_7=\frac{\alpha _3 \alpha _5}{\alpha _1},  \delta(t)=-\frac{3 \alpha _5^2 \left(\alpha _1^2+\alpha _5^2\right) \beta (t)}{\alpha _2^2 \alpha _9},\nonumber\\ \alpha_4(t)&=&\eta _4+\int_1^t [-\frac{\alpha _1 [3 \left(\alpha _1^2+\alpha _5^2\right) \beta (t)+\alpha _9 \gamma (t)]}{\alpha _9}-\frac{\alpha _3^2 \varrho
   (t)}{\alpha _1}] \, dt,
\end{eqnarray}}with  $\alpha_1 \neq 0$, $\alpha_5 \neq 0$, $\alpha_9 \neq 0$ and $\alpha_2 \alpha_9 \neq 0$. Substituting Eq. (5) and Eq. (8)
into the transformation $u=12\,[ln\xi(x,y,z,t)]_{xx}$, we have  the
following lump solution of Eq. (2) {\begin{eqnarray} u^{(II)}&=&[12
[2 \left(\alpha _1^2+\alpha _5^2\right) [\alpha _9+[\int_1^t \alpha
_5 [-\frac{3 \left(\alpha _1^2+\alpha _5^2\right) \beta
   (t)}{\alpha _9}-\frac{\alpha _3^2 \varrho (t)}{\alpha _1^2}-\gamma (t)] \, dt\nonumber\\&+&\eta _3+\alpha _5 x-\frac{\alpha _1 \alpha _2 y}{\alpha _5}+\frac{\alpha _3 \alpha _5
   z}{\alpha _1}]{}^2+[\int_1^t [-\frac{\alpha _1 [3 \left(\alpha _1^2+\alpha _5^2\right) \beta (t)+\alpha _9 \gamma (t)]}{\alpha
   _9}\nonumber\\&-&\frac{\alpha _3^2 \varrho (t)}{\alpha _1}] \, dt+\eta _4+\alpha _1 x+\alpha _2 y+\alpha _3 z]{}^2]\nonumber\\&-&[4 [\alpha _1^2 [\int_1^t
   [-\frac{\alpha _1 [3 \left(\alpha _1^2+\alpha _5^2\right) \beta (t)+\alpha _9 \gamma (t)]}{\alpha _9}-\frac{\alpha _3^2 \varrho (t)}{\alpha
   _1}] \, dt]+\alpha _5 \alpha _1 [\eta _3\nonumber\\&+&\int_1^t \alpha _5 [-\frac{3 \left(\alpha _1^2+\alpha _5^2\right) \beta (t)}{\alpha
   _9}-\frac{\alpha _3^2 \varrho (t)}{\alpha _1^2}-\gamma (t)] \, dt+\alpha _5 x]+\alpha _1^3 x\nonumber\\&+&\alpha _1^2 \left(\eta _4+\alpha _3 z\right)+\alpha _3
   \alpha _5^2 z]{}^2]/(\alpha _1^2)]]/[[\alpha _9+[\eta _3+\alpha _5 x-\frac{\alpha _1 \alpha _2 y}{\alpha _5}+\frac{\alpha _3 \alpha _5
   z}{\alpha _1}\nonumber\\&+&\int_1^t \alpha _5 [-\frac{3 \left(\alpha _1^2+\alpha _5^2\right) \beta
   (t)}{\alpha _9}-\frac{\alpha _3^2 \varrho (t)}{\alpha _1^2}-\gamma (t)] \, dt]{}^2+[\eta _4+\alpha _1 x+\alpha _2 y\nonumber\\&+&\int_1^t [-\frac{\alpha _1 [3 \left(\alpha _1^2+\alpha _5^2\right) \beta (t)+\alpha _9 \gamma (t)]}{\alpha
   _9}-\frac{\alpha _3^2 \varrho (t)}{\alpha _1}] \, dt+\alpha _3 z]{}^2]{}^2],
\end{eqnarray}}where $\eta _3$ and $\eta _4$ are integral constants.

\includegraphics[scale=0.4,bb=20 270 10 10]{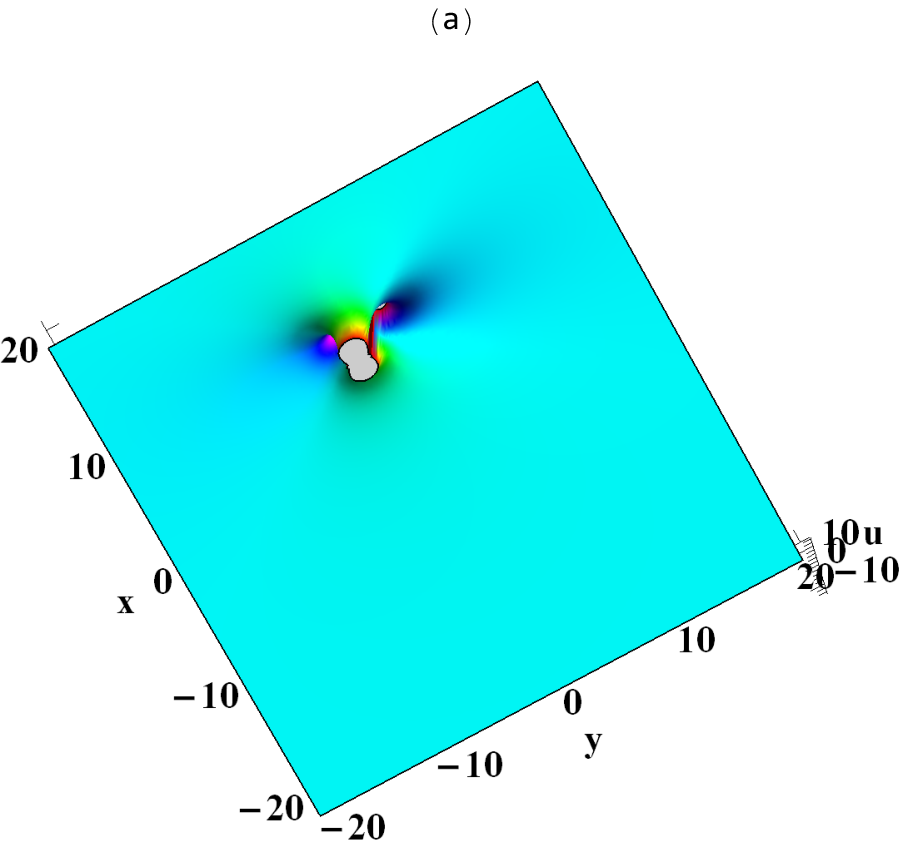}
\includegraphics[scale=0.4,bb=-255 270 10 10]{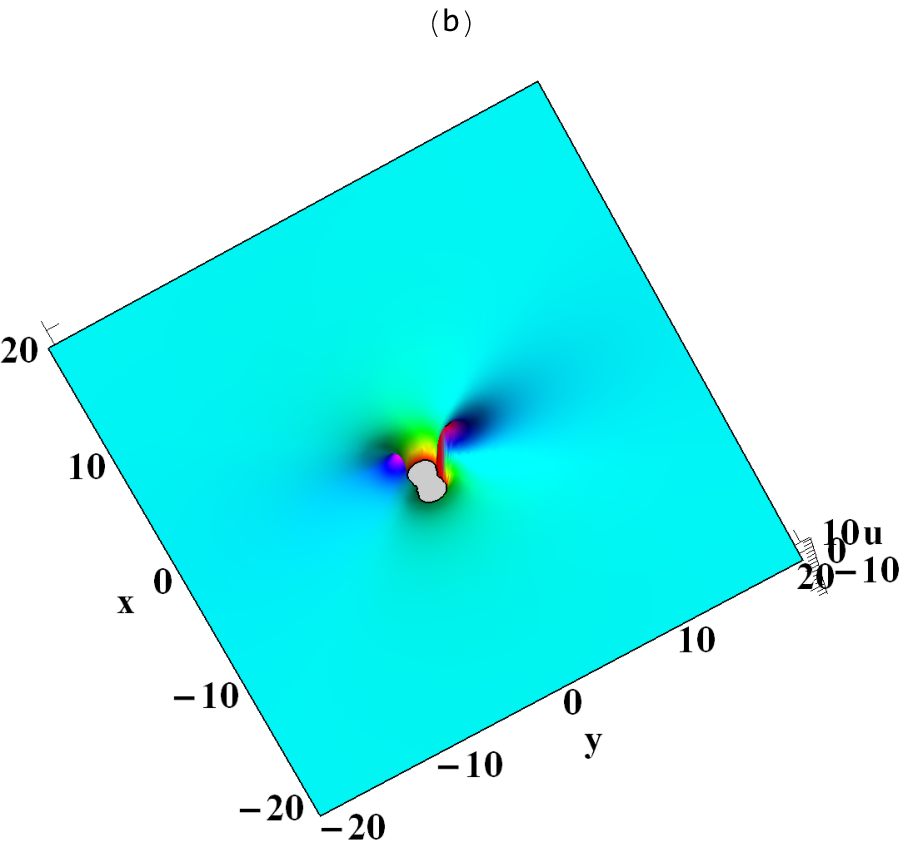}
\includegraphics[scale=0.4,bb=-260 270 10 10]{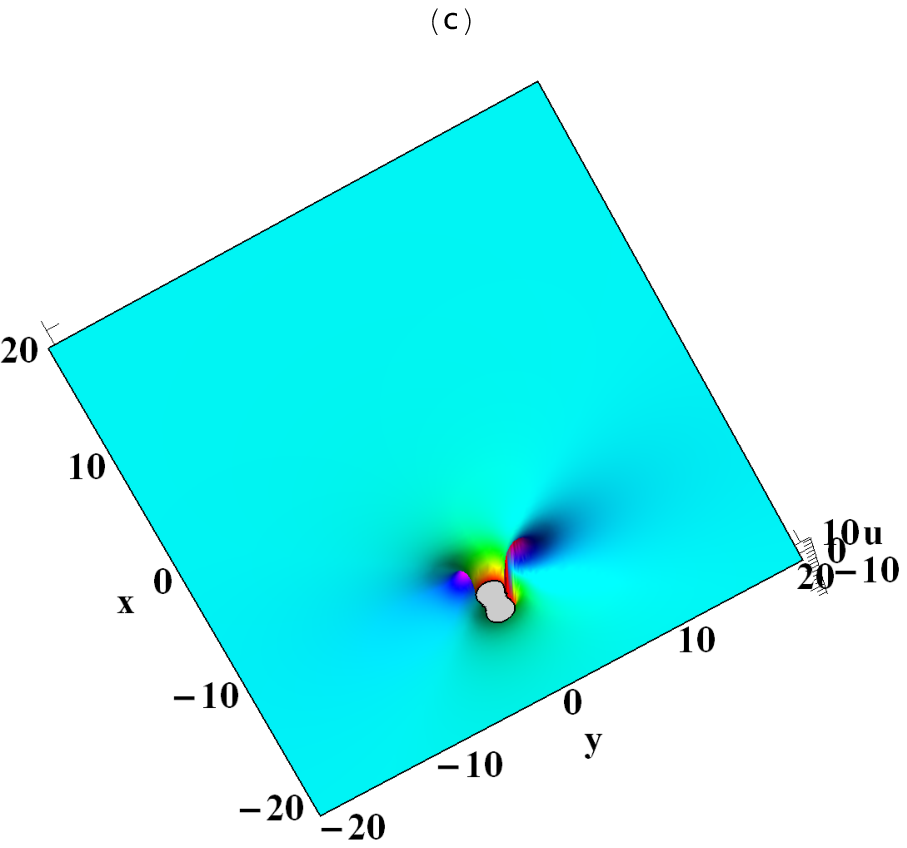}
\includegraphics[scale=0.35,bb=750 620 10 10]{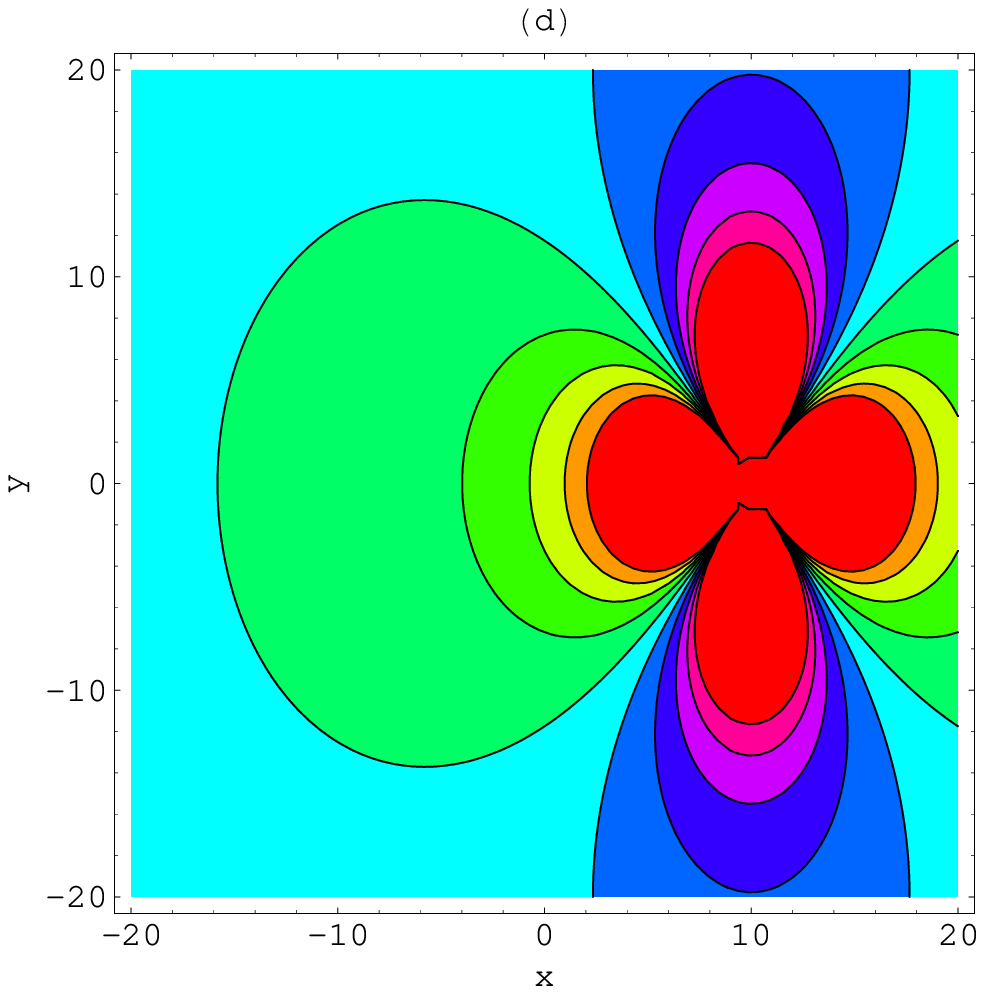}
\includegraphics[scale=0.35,bb=-290 620 10 10]{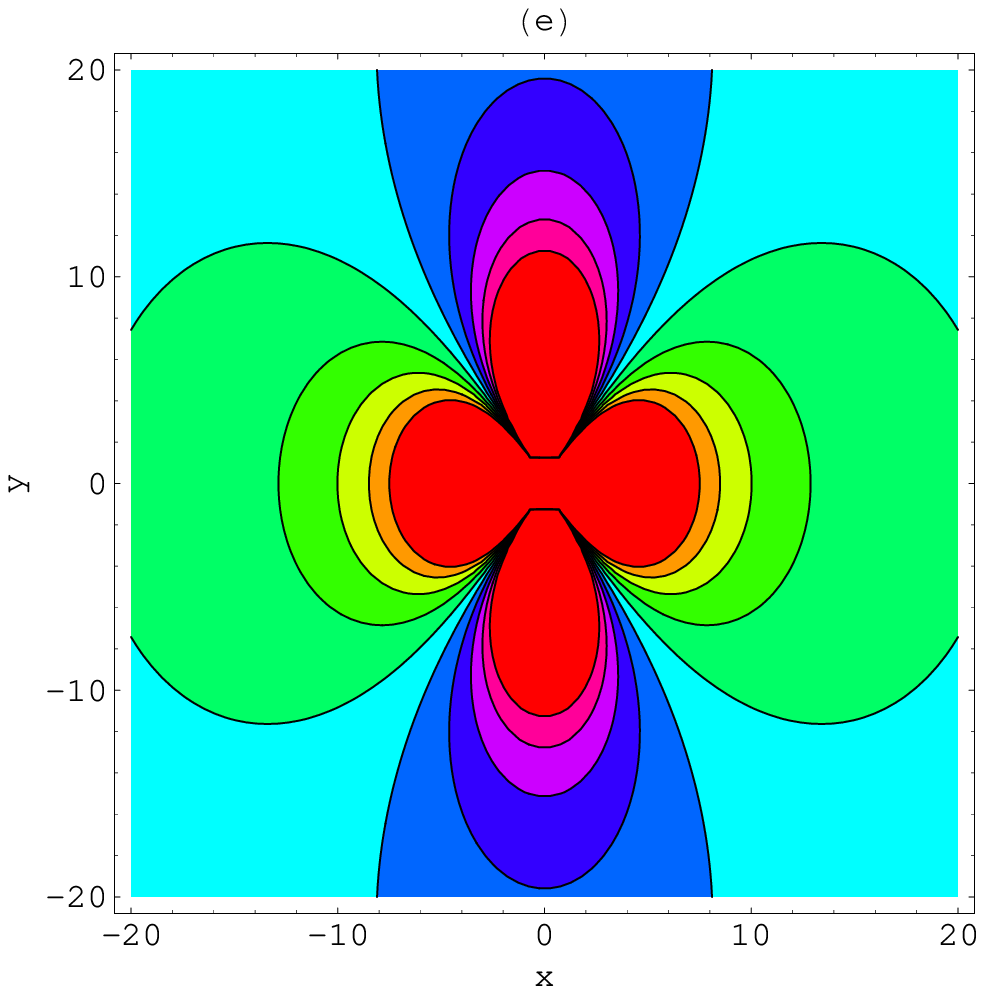}
\includegraphics[scale=0.35,bb=-290 620 10 10]{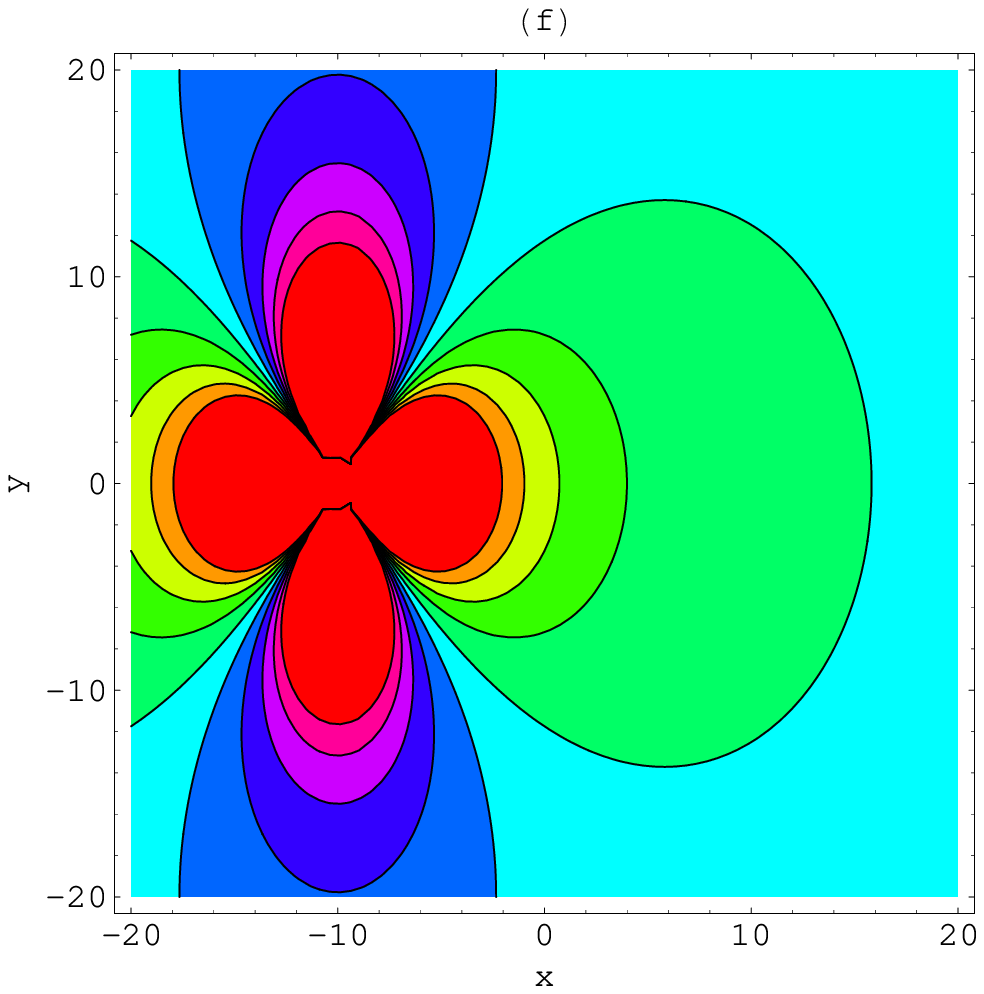}
\vspace{7.5cm}
\begin{tabbing}
\textbf{Fig. 2}. Lump solution (9) with $\alpha_1= 1$,  $\alpha_2=
2$, $\alpha_3=-1$,  $\alpha _5=\alpha_9=-3$,\\ $\eta_3=\eta _4=0$,
$z=-10$, when $t= -1$  in (a) (d), $t= 0$ in (b) (e) and\\ $t = 1$
in (c) (f).
\end{tabbing}
\newpage

\includegraphics[scale=0.4,bb=20 270 10 10]{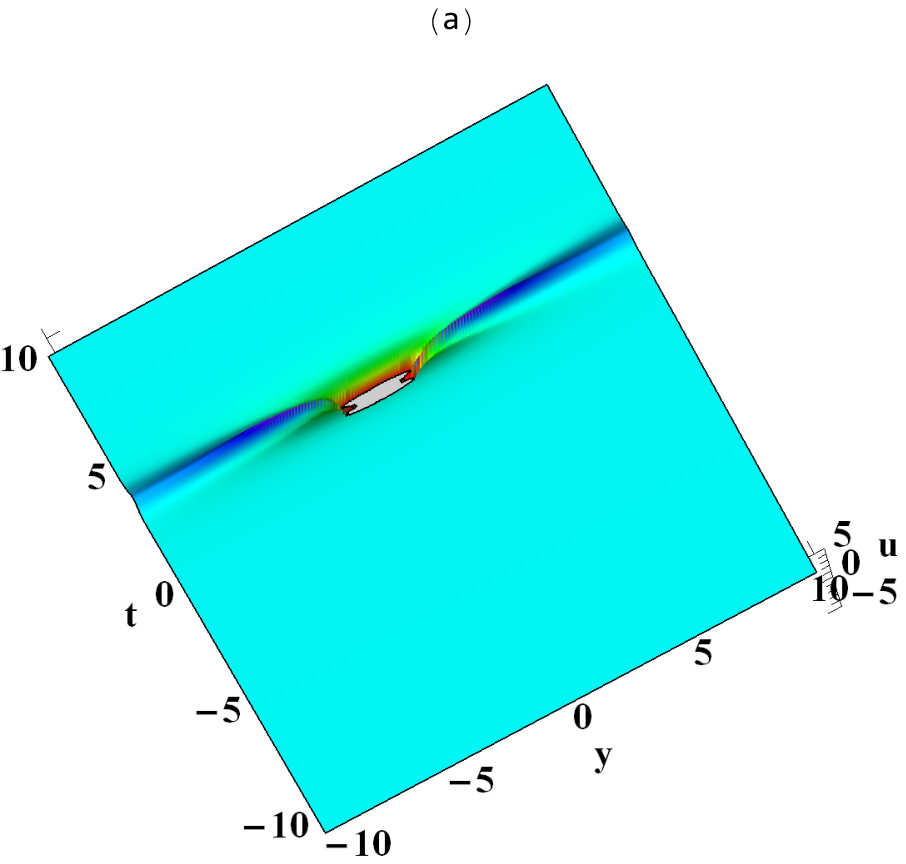}
\includegraphics[scale=0.4,bb=-255 270 10 10]{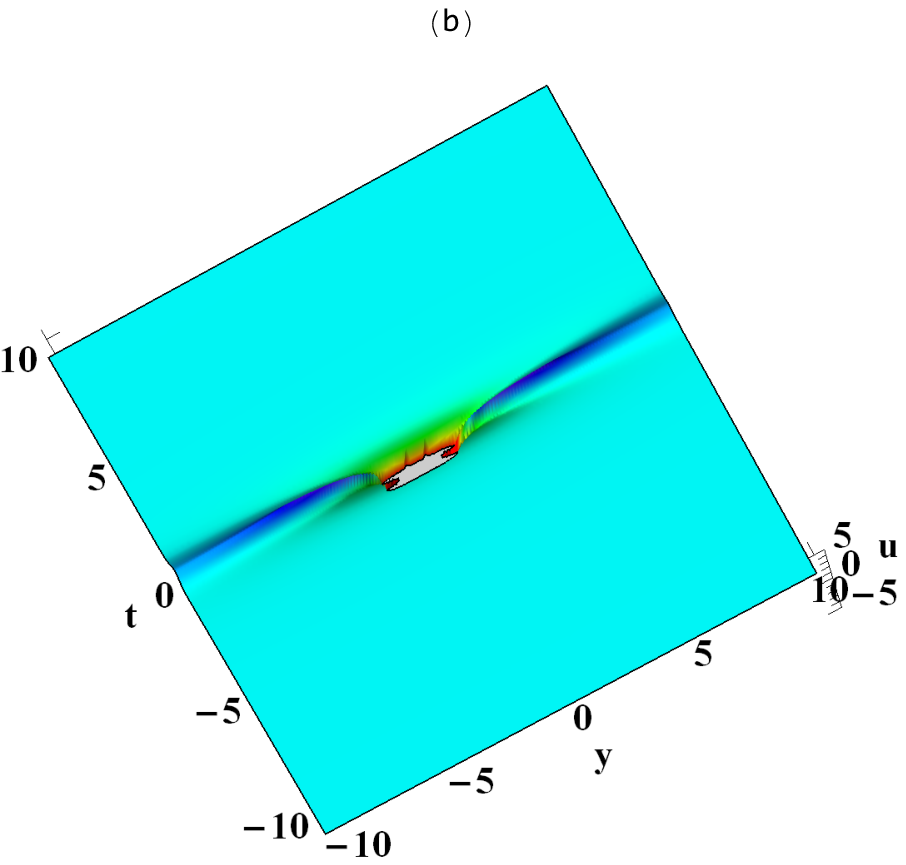}
\includegraphics[scale=0.4,bb=-260 270 10 10]{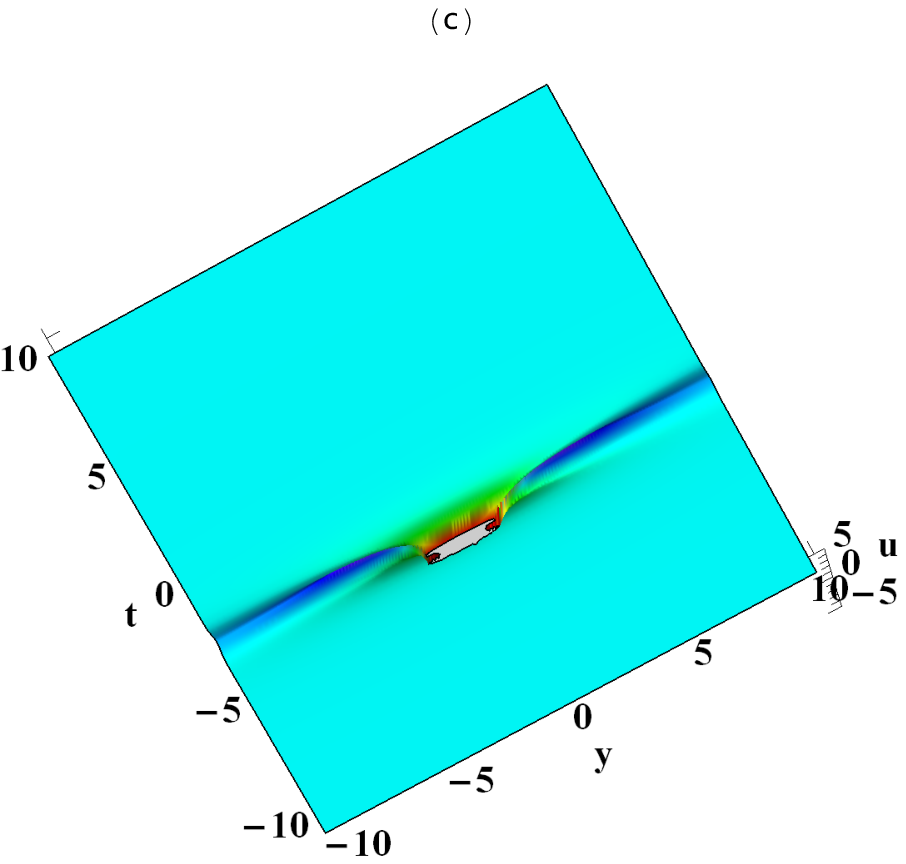}
\includegraphics[scale=0.35,bb=750 620 10 10]{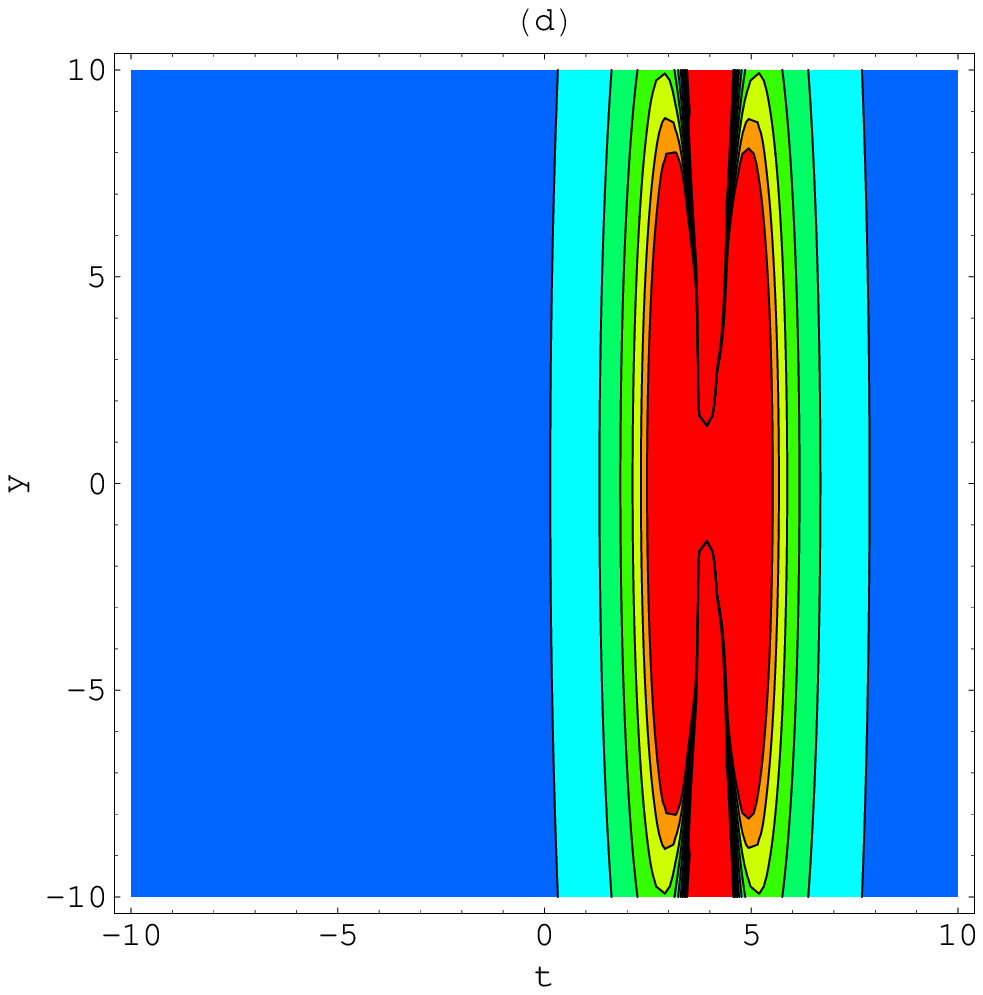}
\includegraphics[scale=0.35,bb=-290 620 10 10]{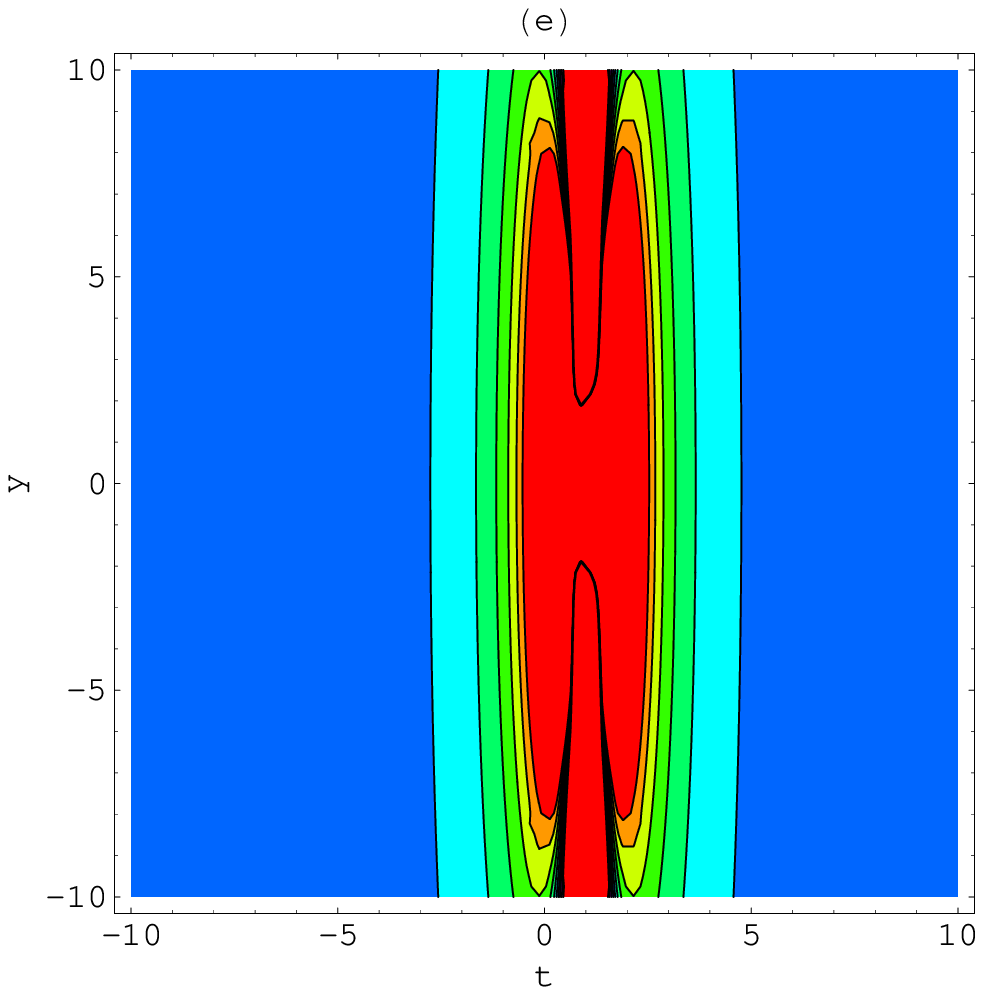}
\includegraphics[scale=0.35,bb=-290 620 10 10]{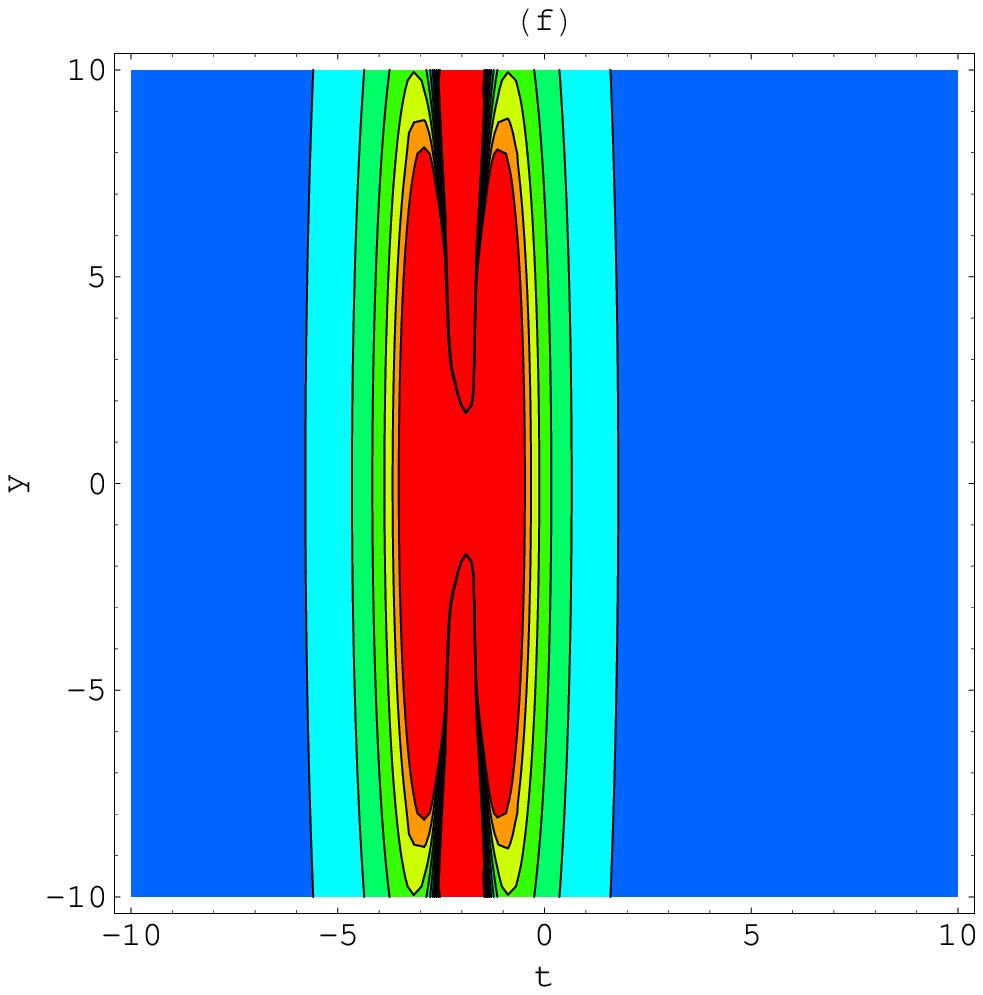}
\vspace{7.5cm}
\begin{tabbing}
\textbf{Fig. 3}. Lump solution (9) with $\alpha_1= -1$,  $\alpha_2=
2$, $\alpha_3=-1$,  $\alpha _5=3$, $\alpha_9=-3$,\\ $\eta_3=\eta
_4=z=0$, when $x= -30$ in (a) (d), $x= 0$ in (b) (e) and\\ $x = 30$
in (c) (f).
\end{tabbing}

\includegraphics[scale=0.4,bb=20 270 10 10]{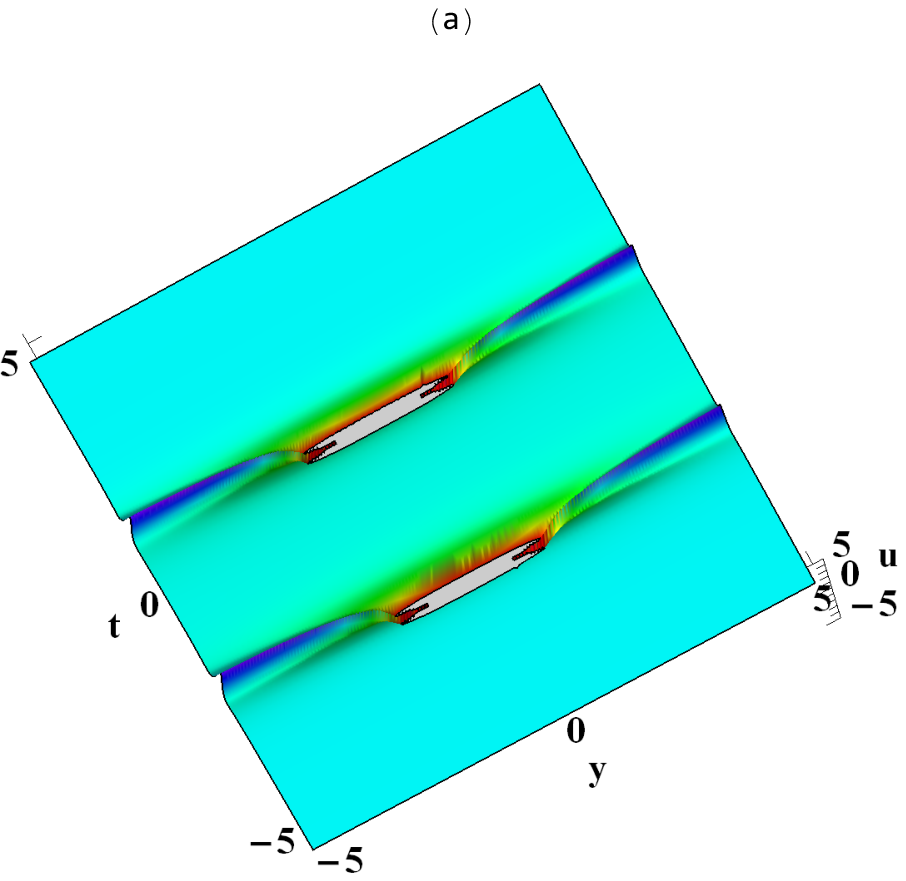}
\includegraphics[scale=0.4,bb=-255 270 10 10]{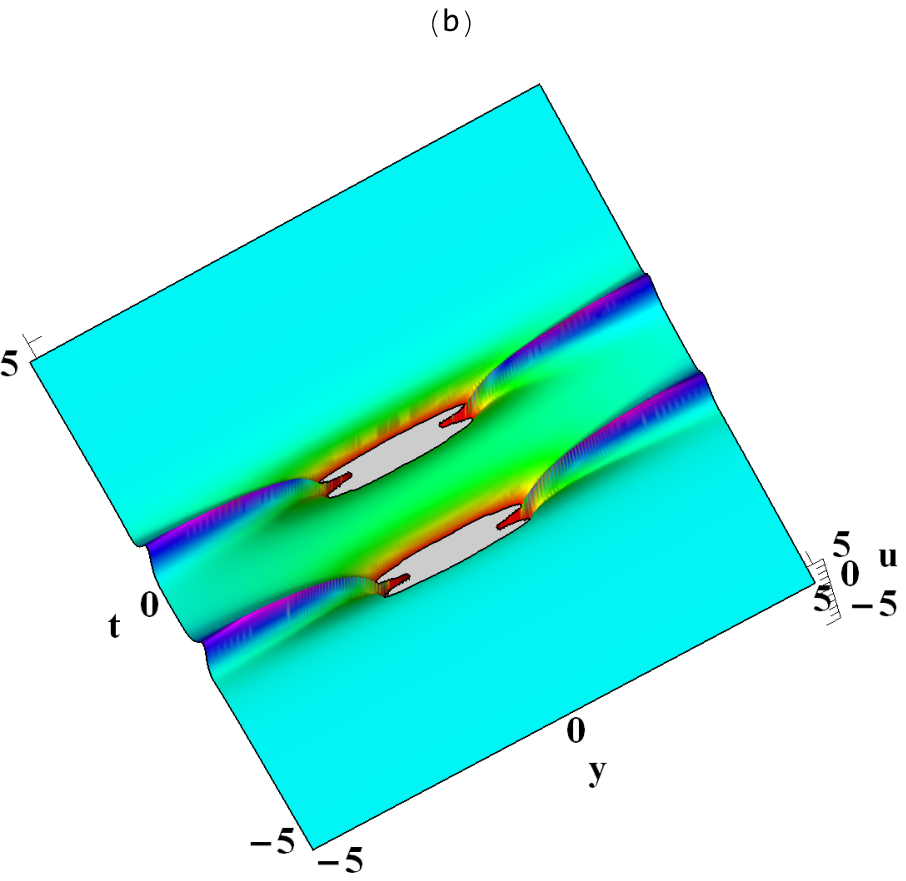}
\includegraphics[scale=0.4,bb=-260 270 10 10]{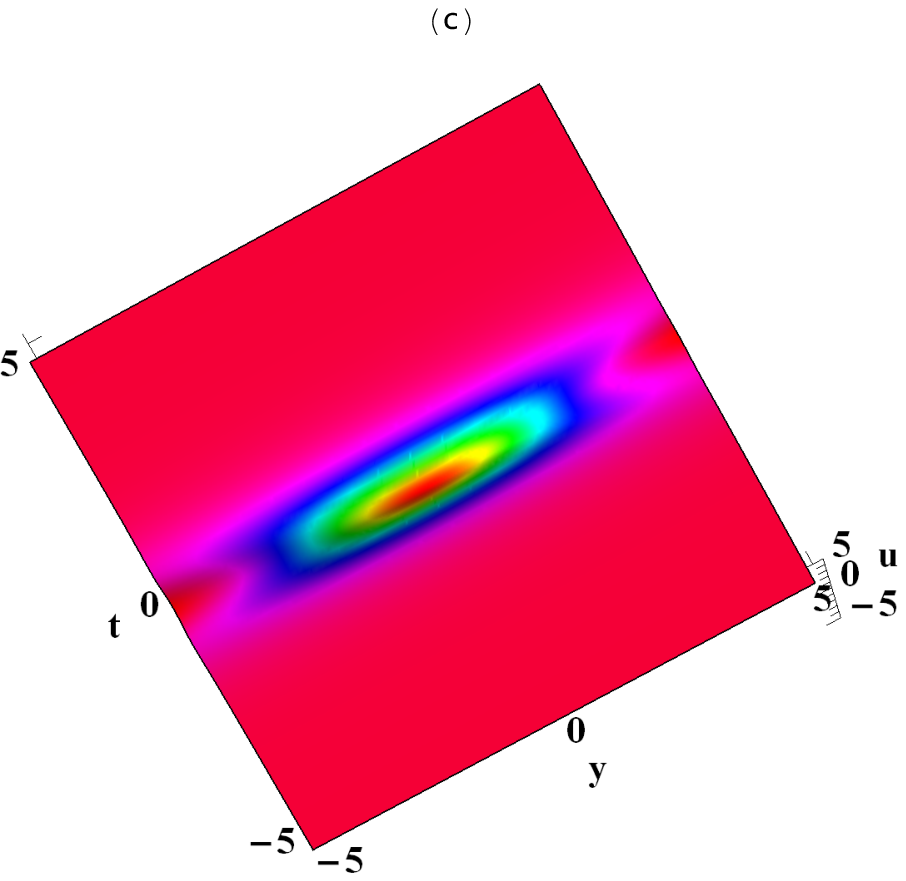}
\includegraphics[scale=0.35,bb=750 620 10 10]{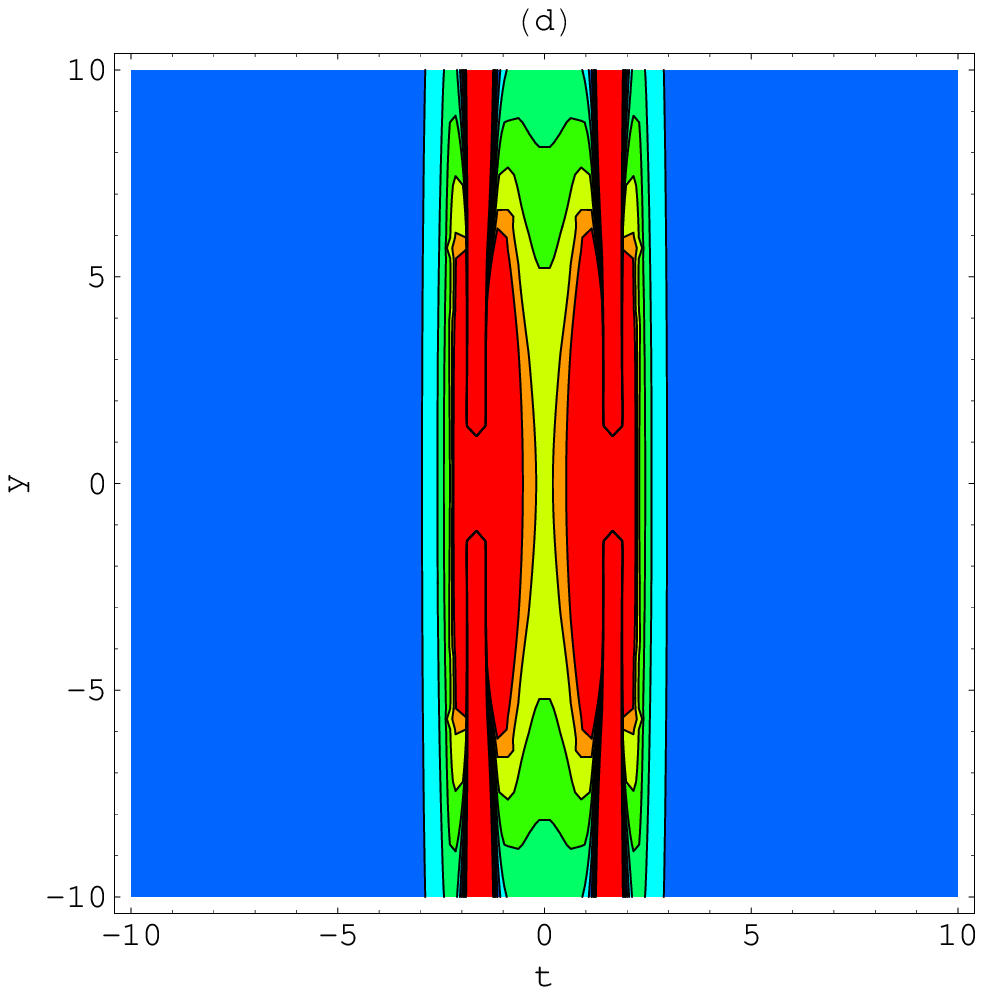}
\includegraphics[scale=0.35,bb=-290 620 10 10]{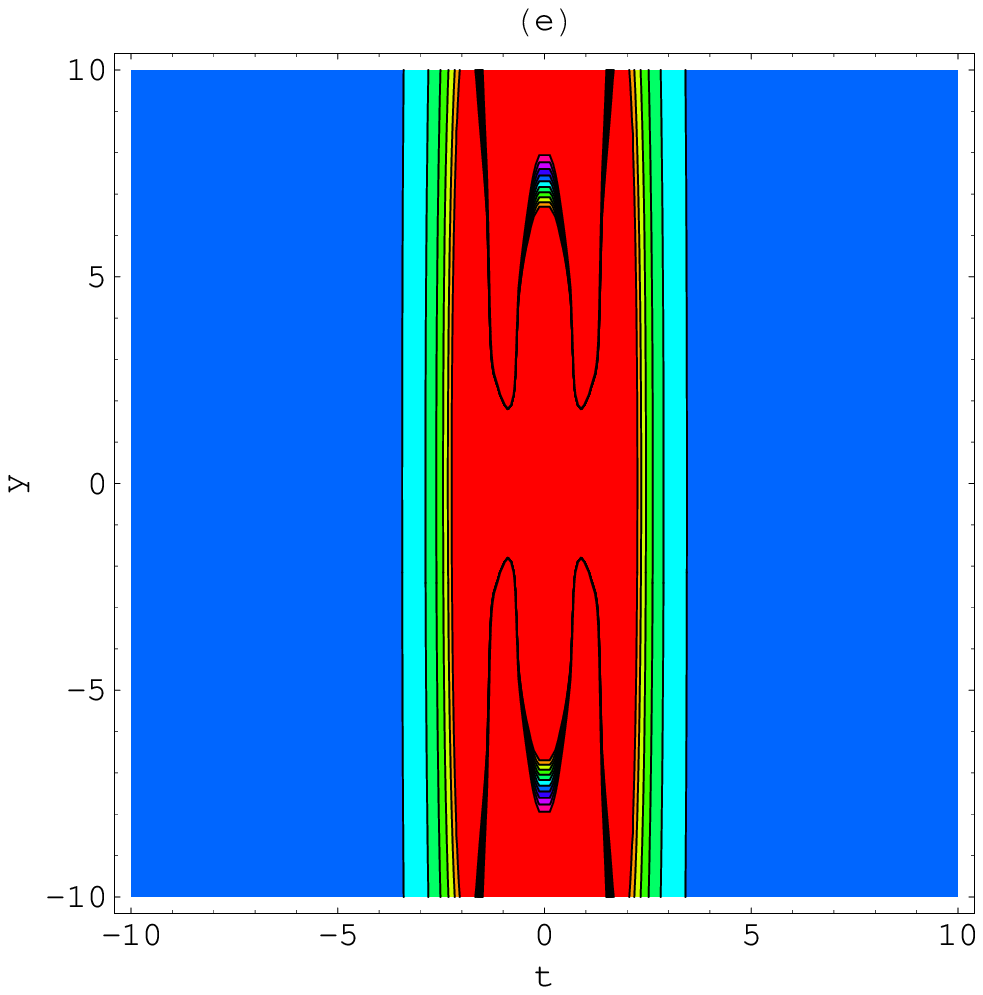}
\includegraphics[scale=0.35,bb=-290 620 10 10]{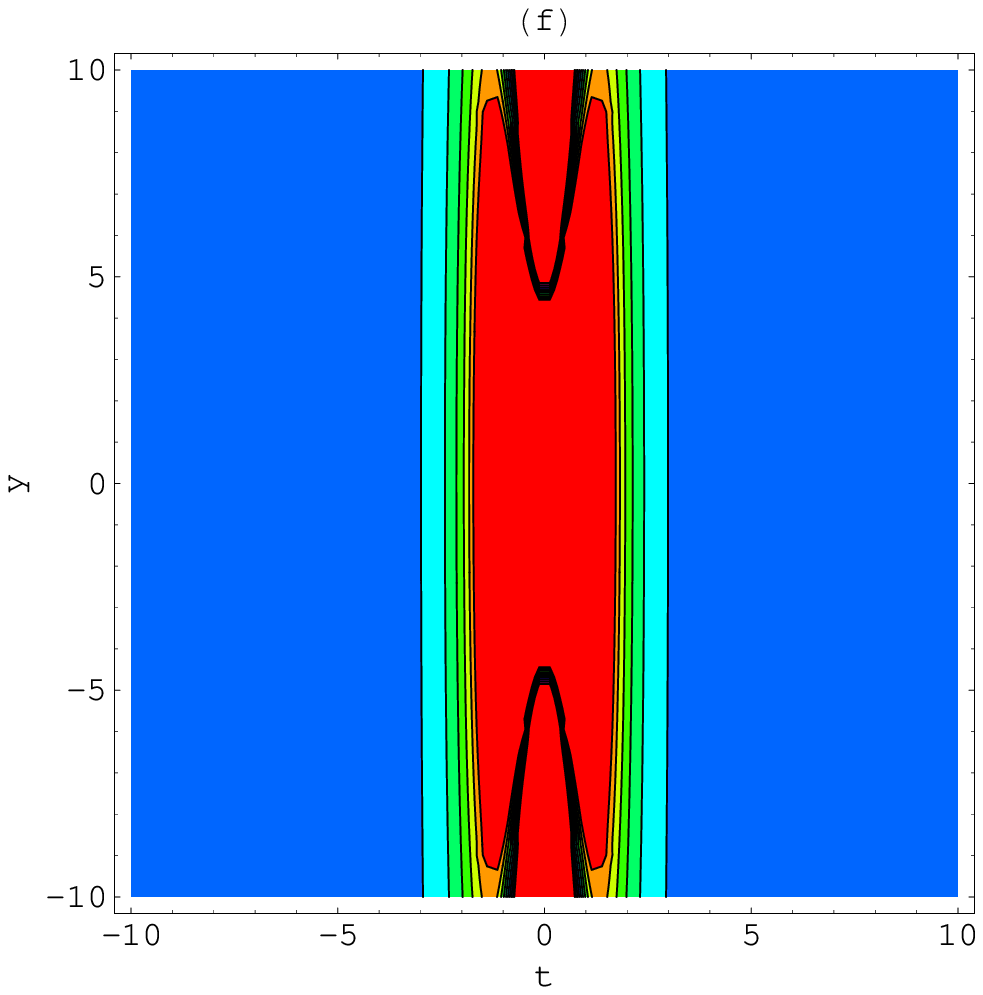}
\vspace{7.5cm}
\begin{tabbing}
\textbf{Fig. 4}. Lump solution (9) with $\alpha_1= -1$,  $\alpha_2=
2$, $\alpha_3=-1$,  $\alpha _5=3$, $\alpha_9=-3$,\\ $\eta_3=\eta
_4=z=0$, when $x= -8$  in
(a) (d), $x= 0$ in (b) (e) and\\ $x = 8$ in (c) (f).\\
\end{tabbing}

\includegraphics[scale=0.4,bb=20 270 10 10]{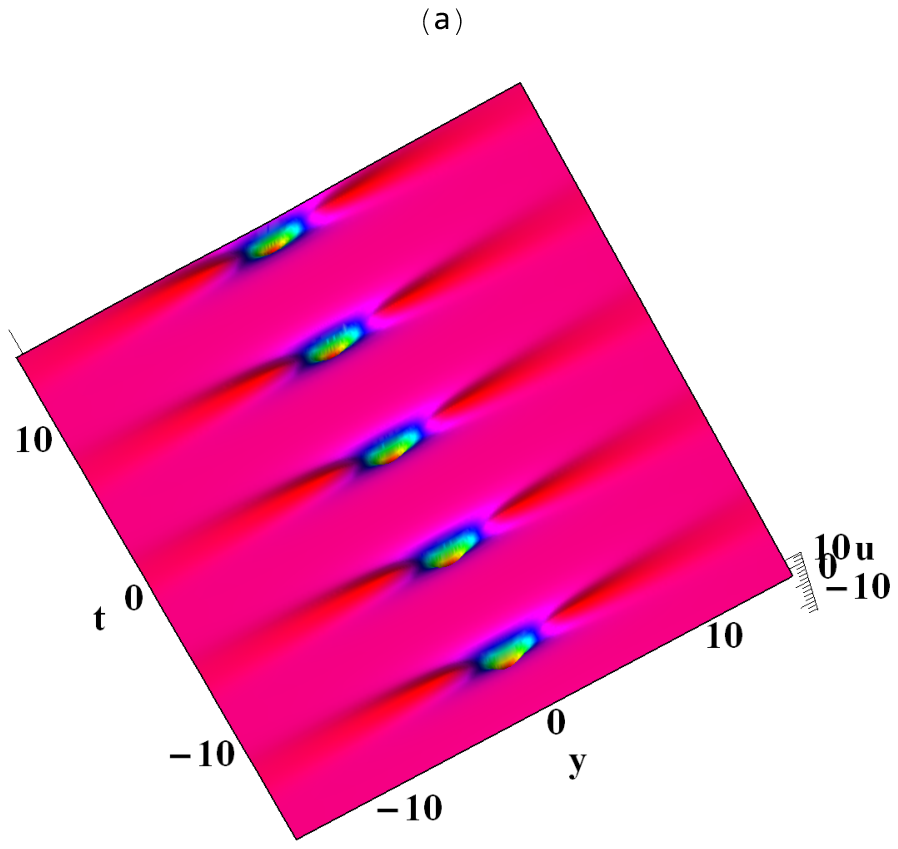}
\includegraphics[scale=0.4,bb=-255 270 10 10]{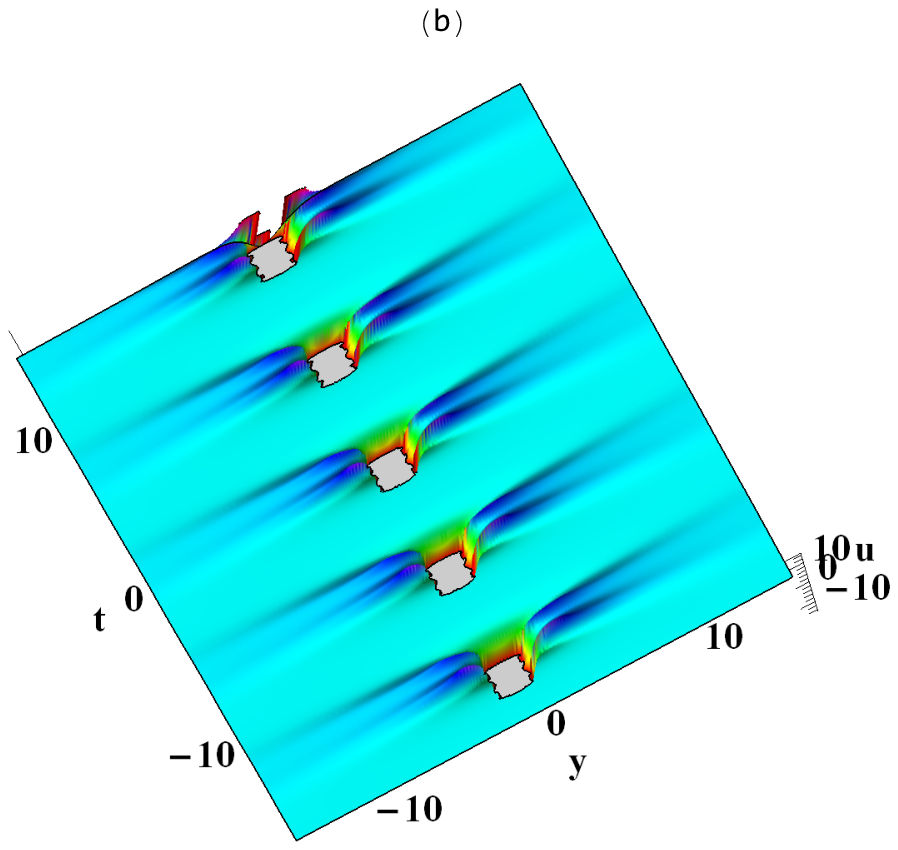}
\includegraphics[scale=0.4,bb=-260 270 10 10]{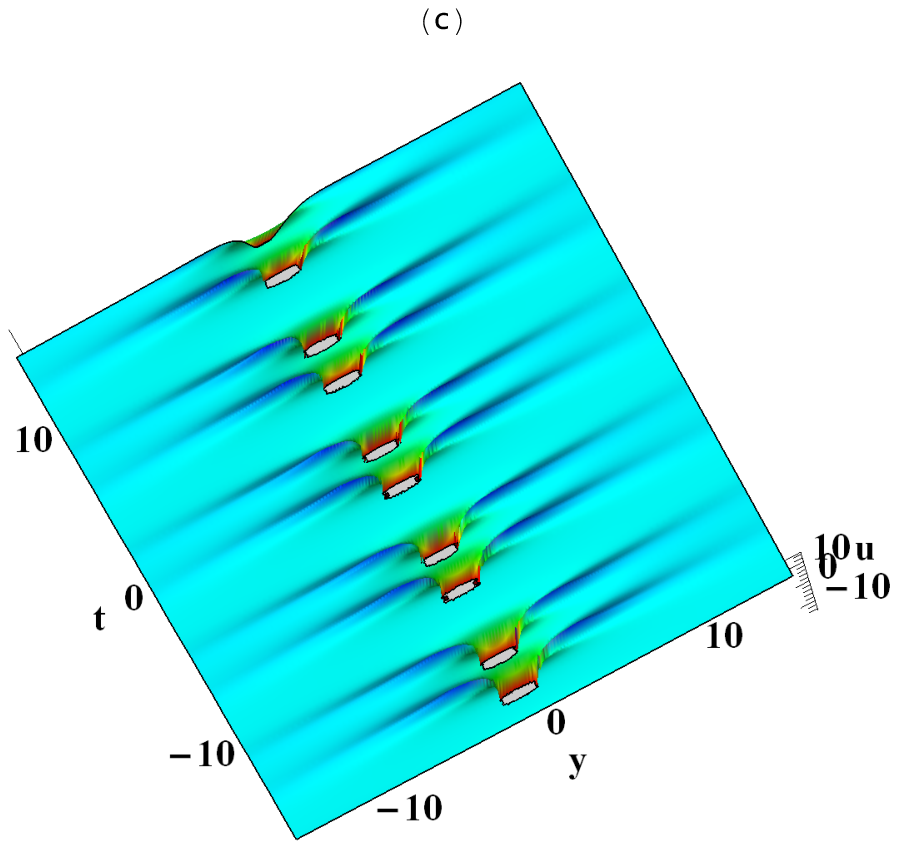}
\includegraphics[scale=0.35,bb=750 620 10 10]{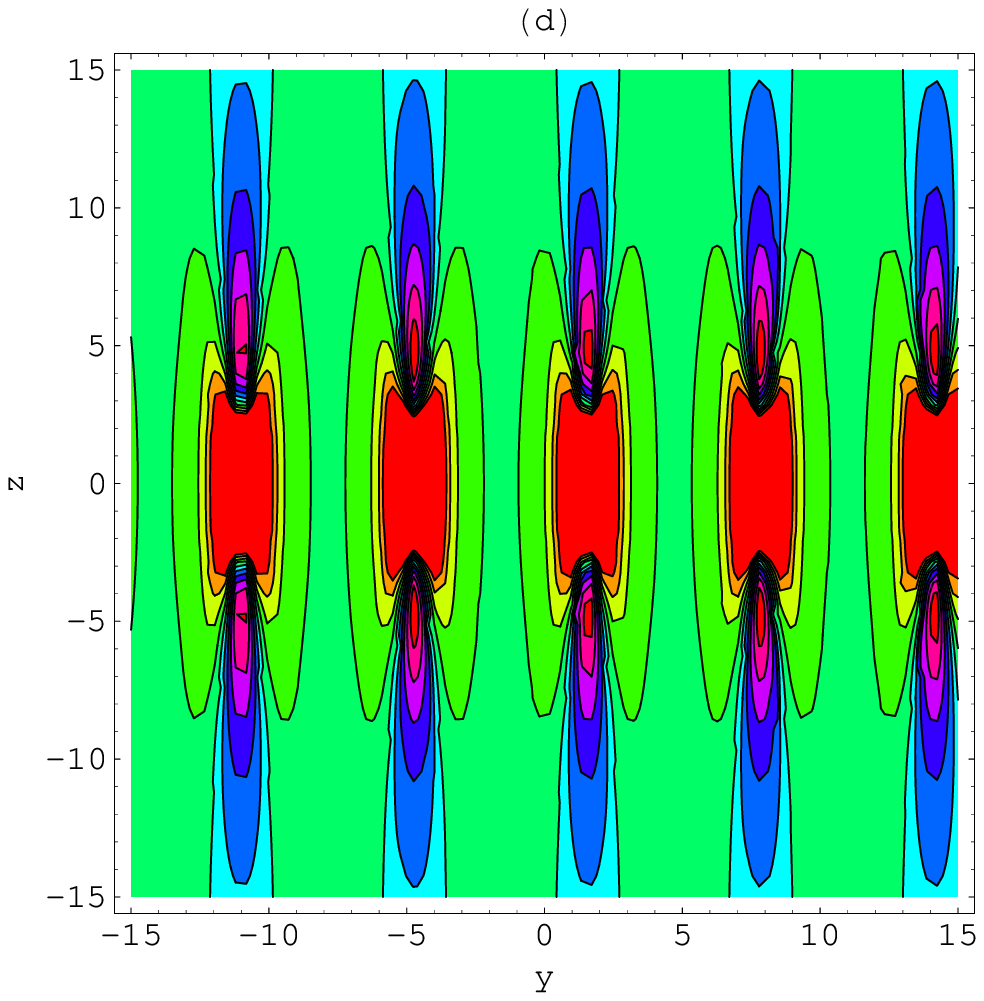}
\includegraphics[scale=0.35,bb=-290 620 10 10]{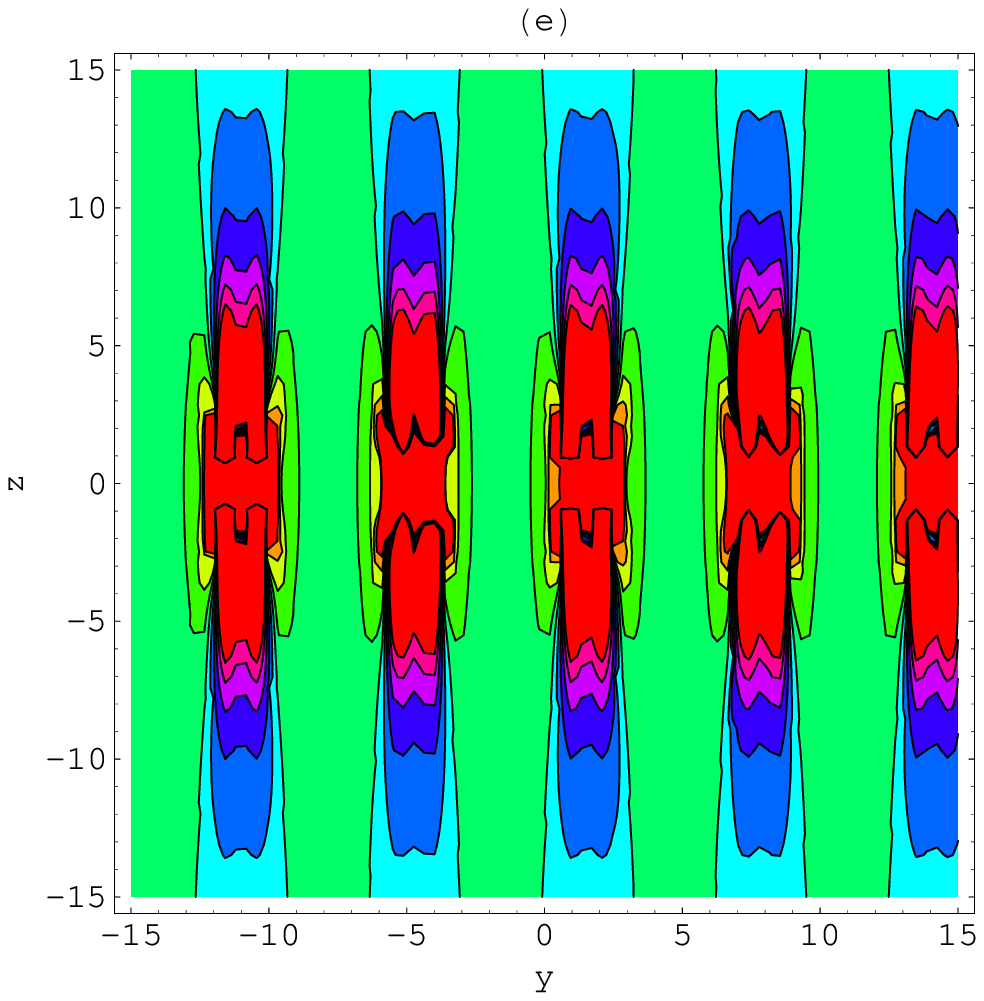}
\includegraphics[scale=0.35,bb=-290 620 10 10]{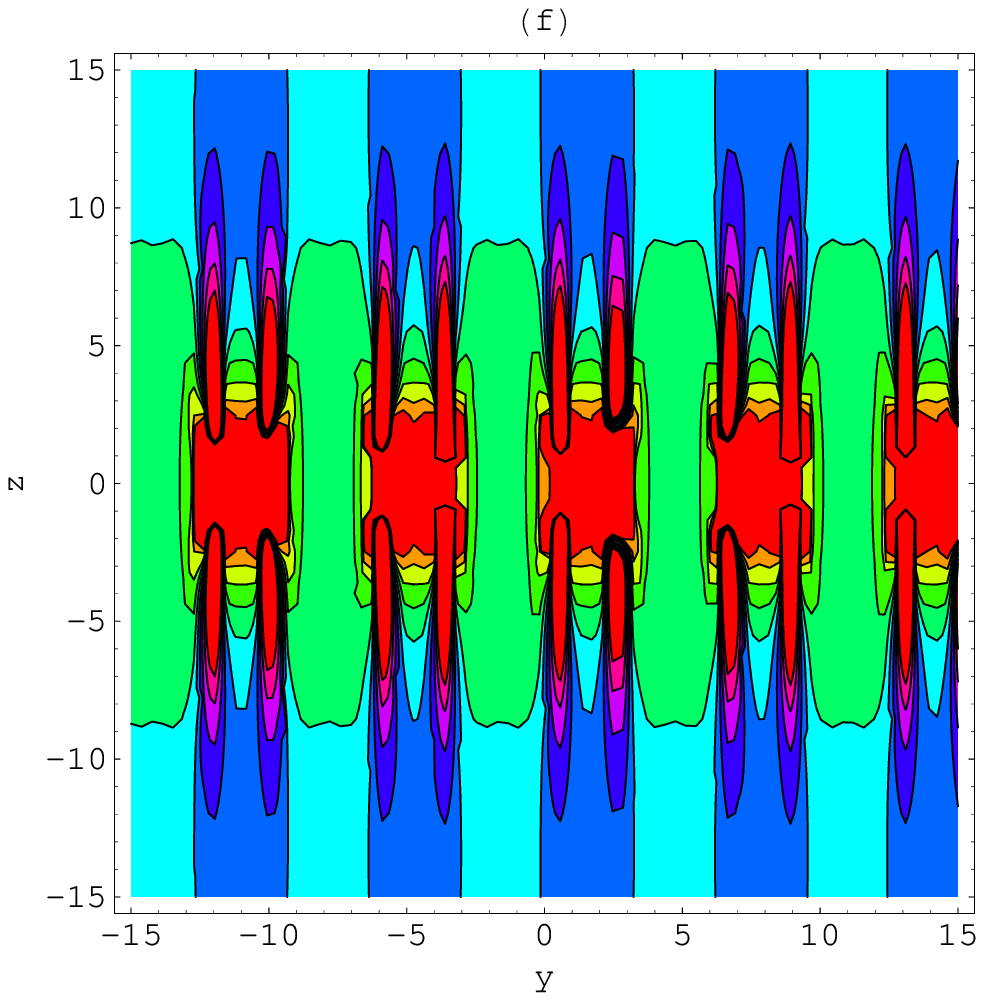}
\vspace{7.5cm}
\begin{tabbing}
\textbf{Fig. 5}. Lump solution (9) with $\alpha_1= -1$,  $\alpha_2=
2$, $\alpha_3=-1$,  $\alpha _5=3$, $\alpha_9=-3$,\\ $\eta_3=\eta
_4=z=0$, when $x= -3$  in (a) (d), $x= 0$ in (b) (e) and\\ $x = 3$
in (c) (f).
\end{tabbing}\quad Then, the physical structures for  $(u^{(II)})$ are shown in Figs.
2-5 with some 3d graphs and contour plots. When $\gamma (t)=-1,
\varrho(t)=\beta(t)=1$, the spatial structure called the bright lump
wave is seen in Fig. 2 at $t=-1; 0; 1$, the spatial structure called
the bright-dark lump wave is shown in Fig. 3 at $x=-30; 0; 30$.
 When $\gamma (t)=-t, \varrho(t)=\beta(t)=t$, interaction behaviors of two bright-dark lump waves are presented
 in Fig. 4 at $x=-8; 0; 8$. As the value of $x$ changes, the two bright-dark lump waves move towards each other, and finally merge together.
 When $\gamma (t)=\varrho(t)=\beta(t)=\cos t$, a periodic-shape bright lump wave is found in Fig. 5 at $x=-3; 0; 3$.

{\begin{eqnarray}(III):
 \varrho(t)&=&-\frac{3 \left(\alpha _1^2+\alpha _5^2\right){}^3 \beta (t)+\left(\alpha _2 \alpha _5-\alpha _1 \alpha _6\right){}^2 \alpha _9 \delta (t)}{\left(\alpha _3 \alpha
   _5-\alpha _1 \alpha _7\right){}^2 \alpha _9}, \alpha_9(t)=\alpha_9,\nonumber\\ \alpha _4(t)&=&\eta _5-\int_1^t [\alpha _1 [\alpha _5^2 \gamma (t)+\left(\alpha _2^2-\alpha _6^2\right) \delta (t)+\left(\alpha _3^2-\alpha _7^2\right) \varrho
   (t)]\nonumber\\&+&\alpha _1^3 \gamma (t)+2 \alpha _5 [\alpha _2 \alpha _6 \delta (t)+\alpha _3 \alpha _7 \varrho (t)]]/(\alpha _1^2+\alpha _5^2) \, dt, \nonumber\\
 \alpha_8(t)&=&\eta _6-\int_1^t [\alpha _5 [\alpha _5^2 \gamma (t)+\left(\alpha _6^2-\alpha _2^2\right) \delta (t)+\left(\alpha _7^2-\alpha _3^2\right) \varrho
   (t)]\nonumber\\&+&\alpha _5 \alpha _1^2 \gamma (t)+2 \alpha _1 [\alpha _2 \alpha _6 \delta (t)+\alpha _3 \alpha _7 \varrho (t)]]/(\alpha _1^2+\alpha _5^2) \,
   dt
\end{eqnarray}}with  $\left(\alpha _3 \alpha
   _5-\alpha _1 \alpha _7\right){}^2 \alpha _9\neq 0$,
   $\alpha _1^2+\alpha _5^2\neq 0$, $\eta _5$ and $\eta _6$ are integral constants.
   Substituting Eq. (5) and Eq. (10) into the transformation $u=12\,[ln\xi(x,y,z,t)]_{xx}$,
   we derive  another lump solution of Eq. (2)

\includegraphics[scale=0.4,bb=20 270 10 10]{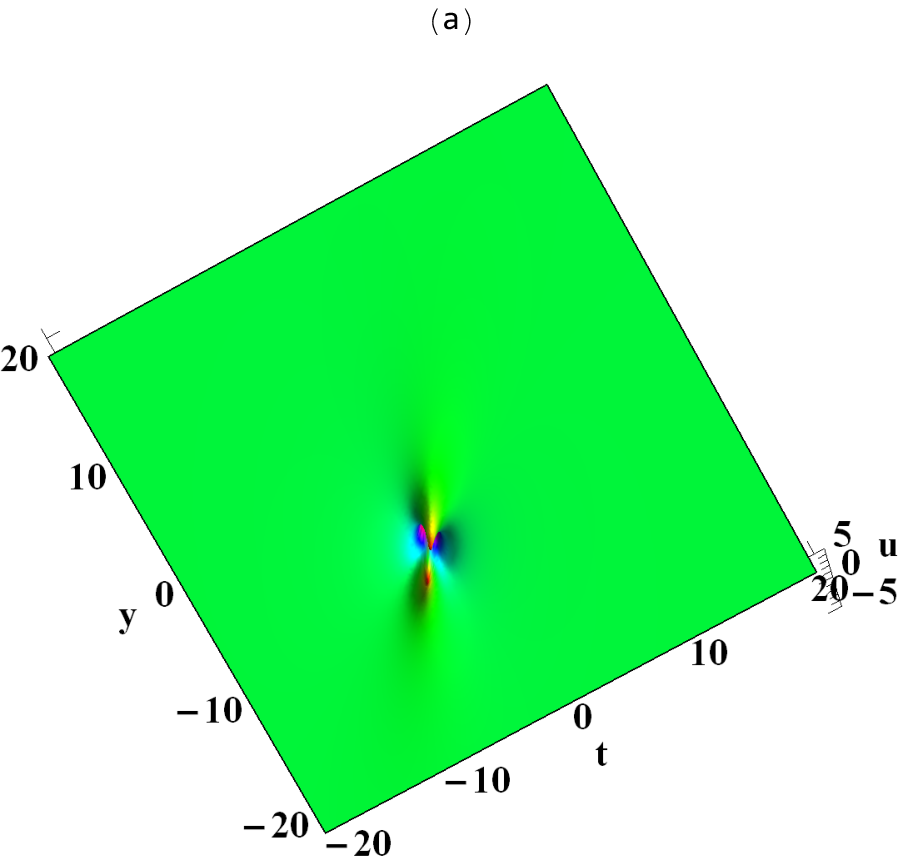}
\includegraphics[scale=0.4,bb=-255 270 10 10]{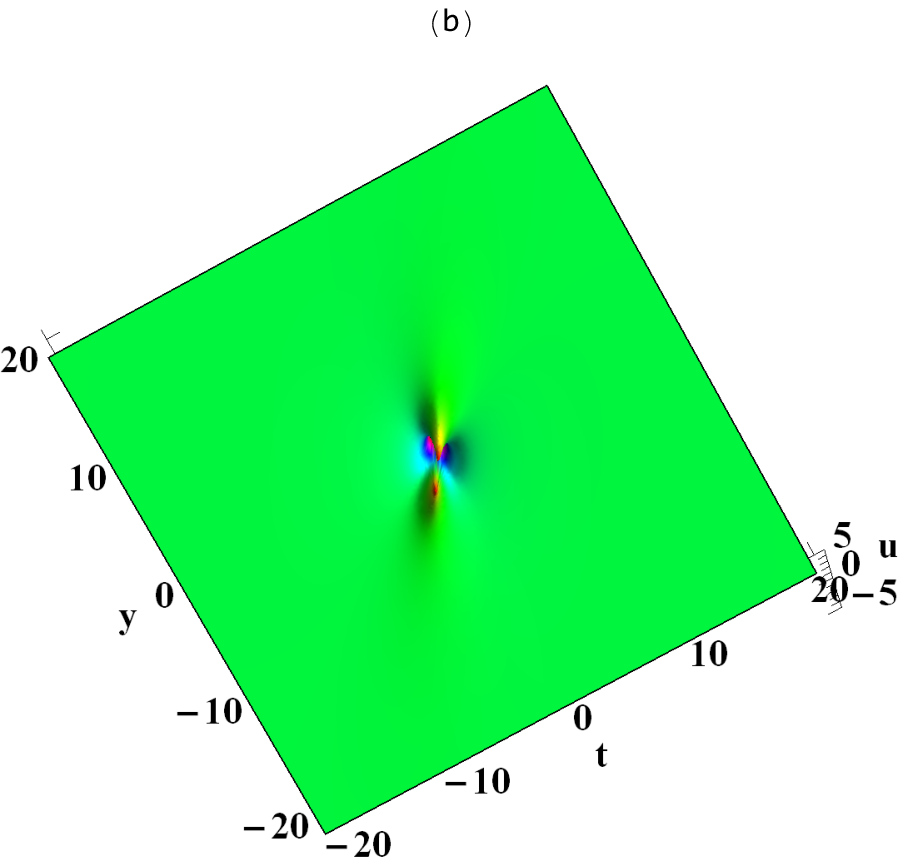}
\includegraphics[scale=0.4,bb=-260 270 10 10]{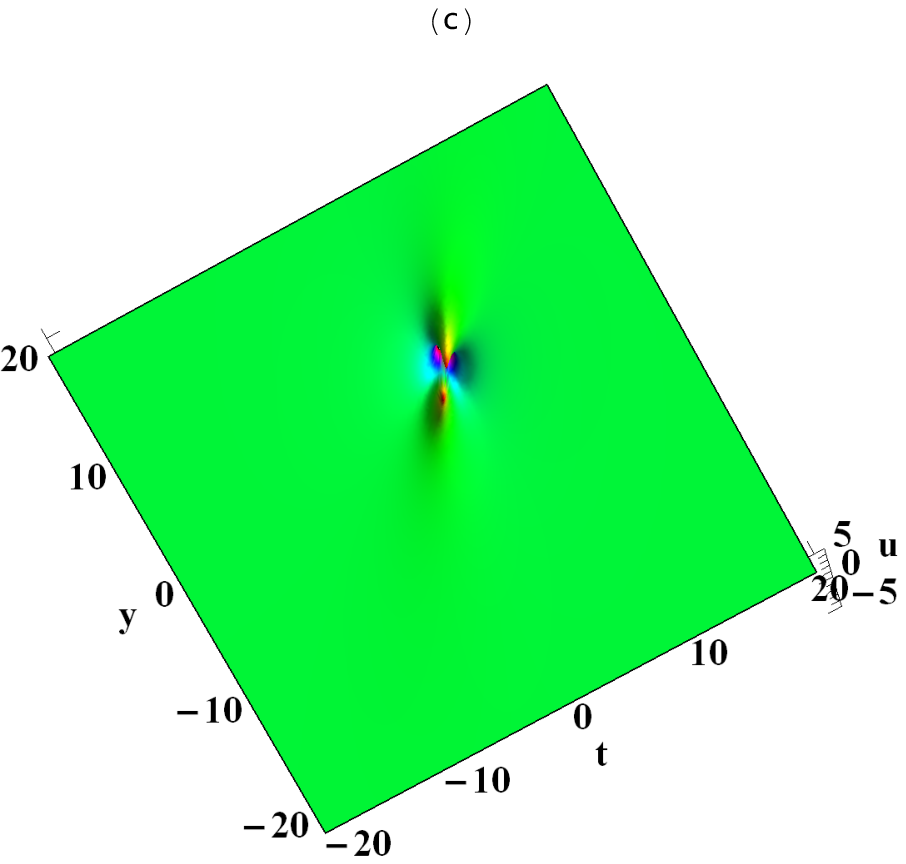}
\includegraphics[scale=0.35,bb=750 620 10 10]{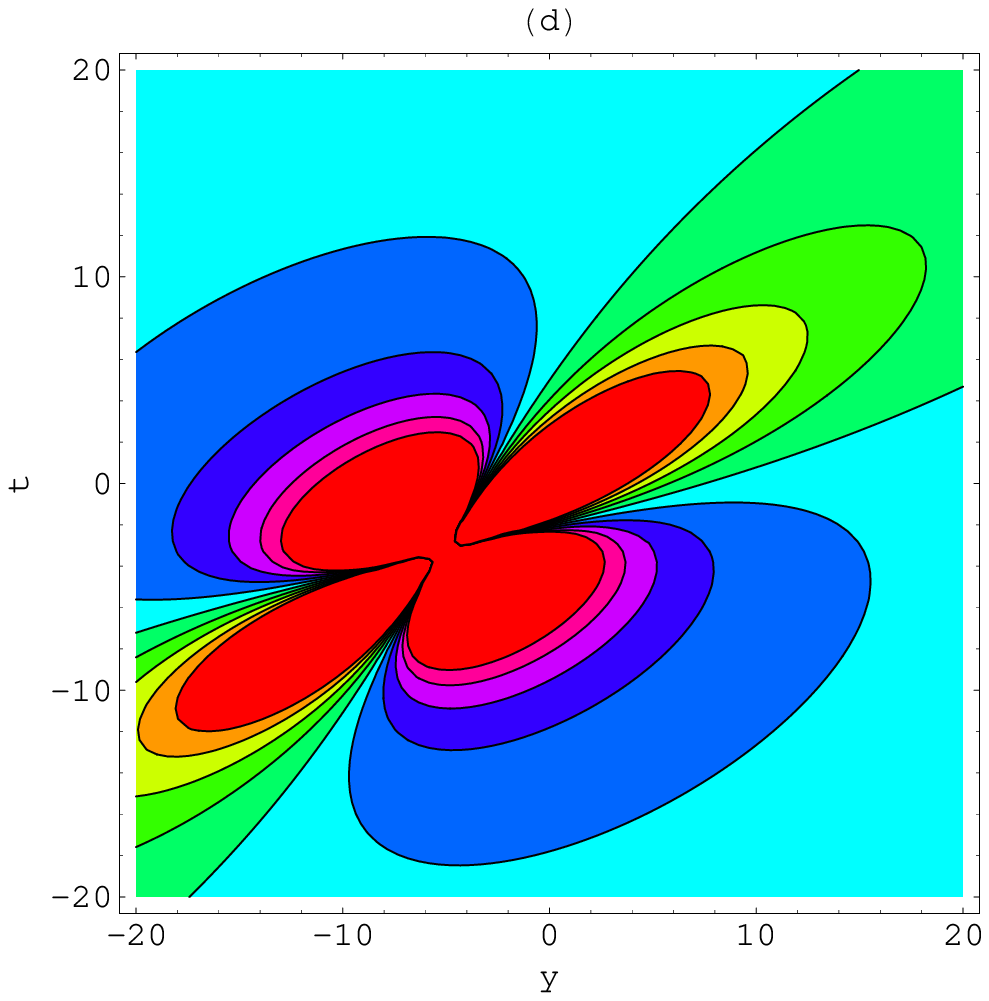}
\includegraphics[scale=0.35,bb=-290 620 10 10]{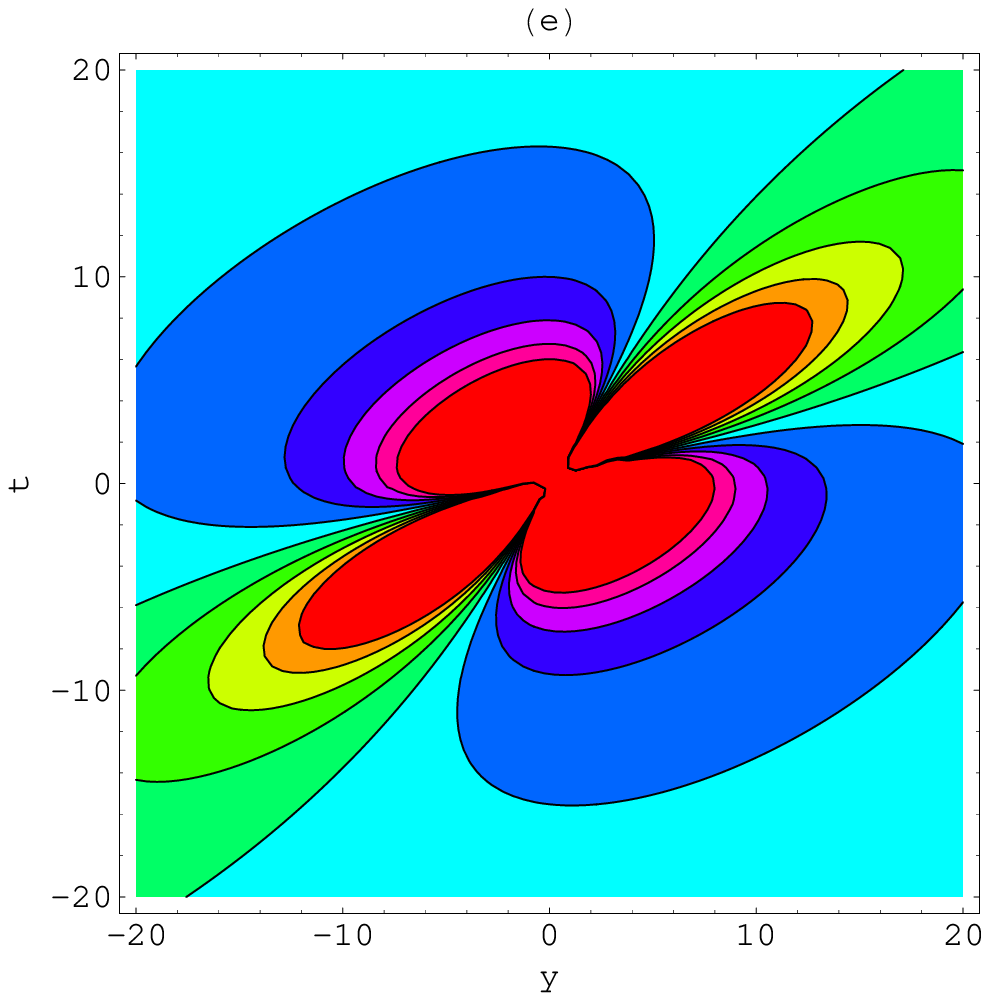}
\includegraphics[scale=0.35,bb=-290 620 10 10]{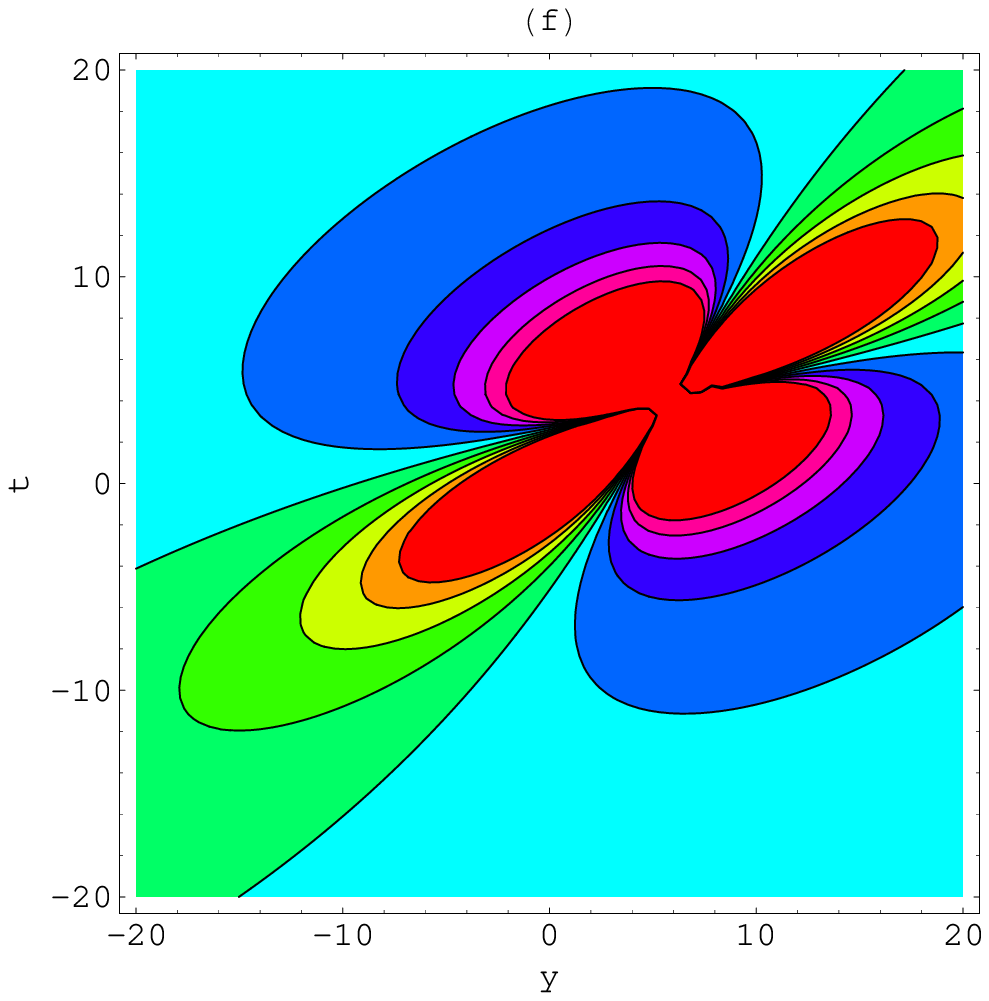}
\vspace{7.5cm}
\begin{tabbing}
\textbf{Fig. 6}. Lump solution (11) with $\alpha_1=
\alpha_2=\alpha_3=\alpha _5=-1$, $\alpha_6=\alpha_7= \eta_5=3$,\\
$\eta _6=-2$, $\alpha_9=2$, $z=0$, when $x= -10$  in (a) (d), $x= 0$
in (b) (e) and\\ $x = 10$ in (c) (f).
\end{tabbing}

\includegraphics[scale=0.4,bb=20 270 10 10]{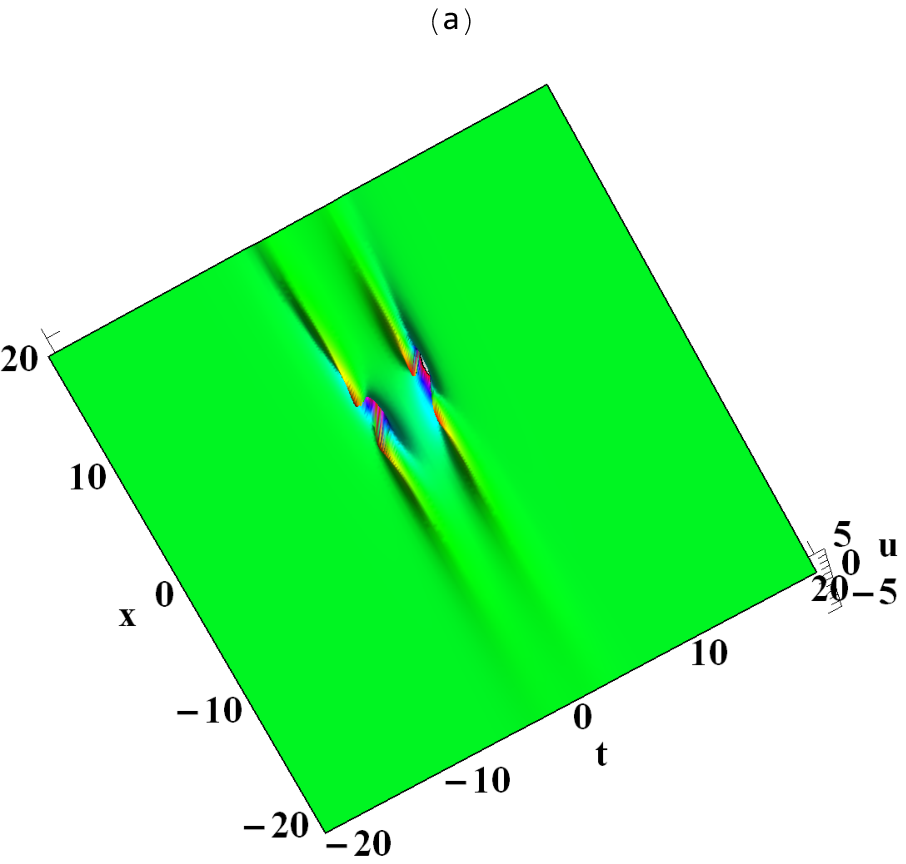}
\includegraphics[scale=0.4,bb=-255 270 10 10]{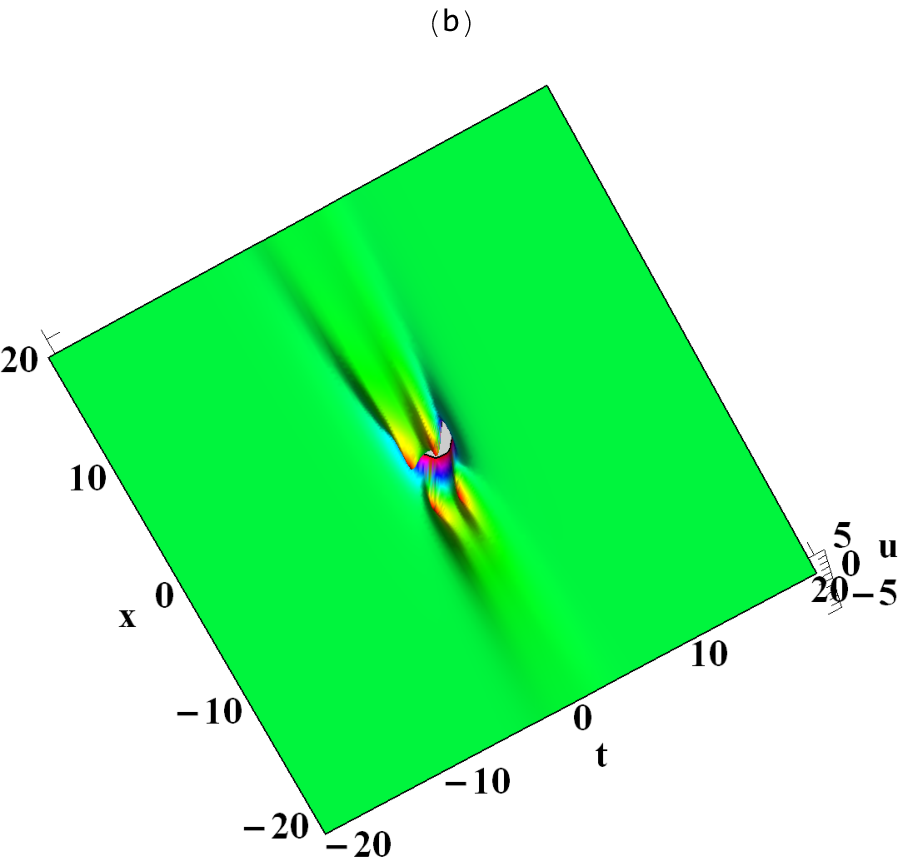}
\includegraphics[scale=0.4,bb=-260 270 10 10]{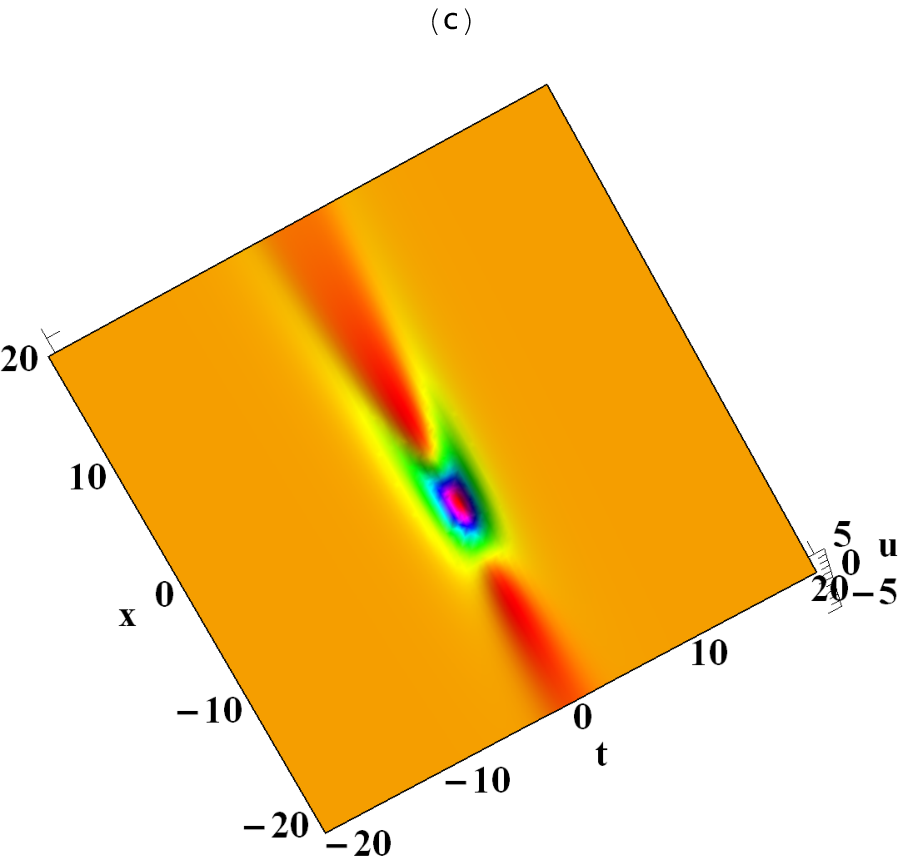}
\includegraphics[scale=0.35,bb=750 620 10 10]{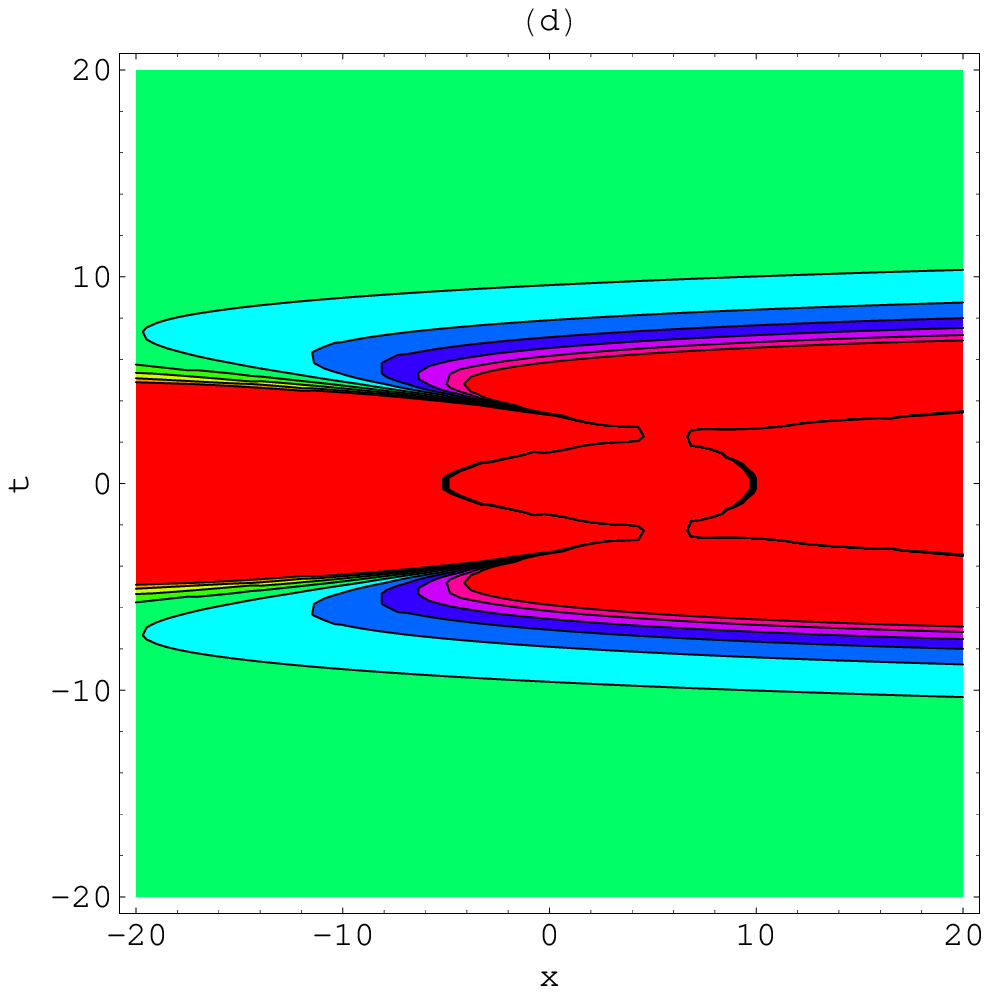}
\includegraphics[scale=0.35,bb=-290 620 10 10]{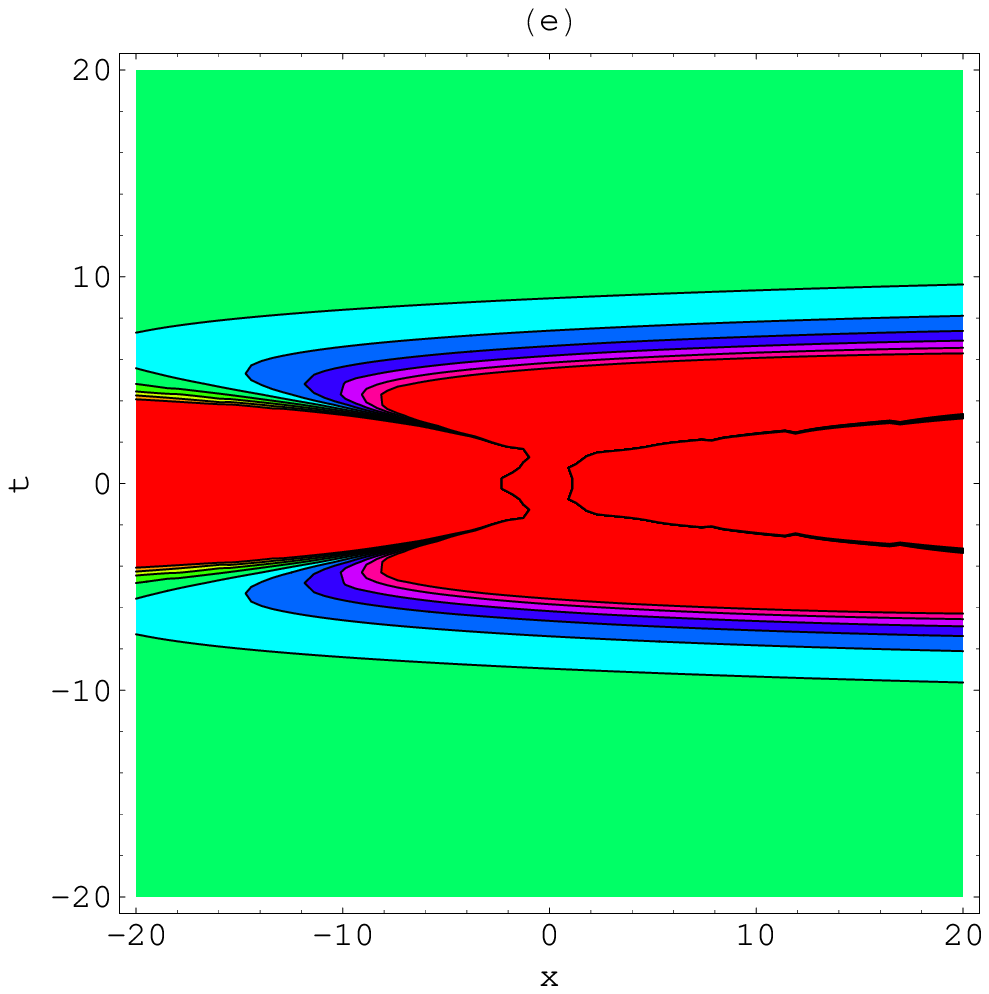}
\includegraphics[scale=0.35,bb=-290 620 10 10]{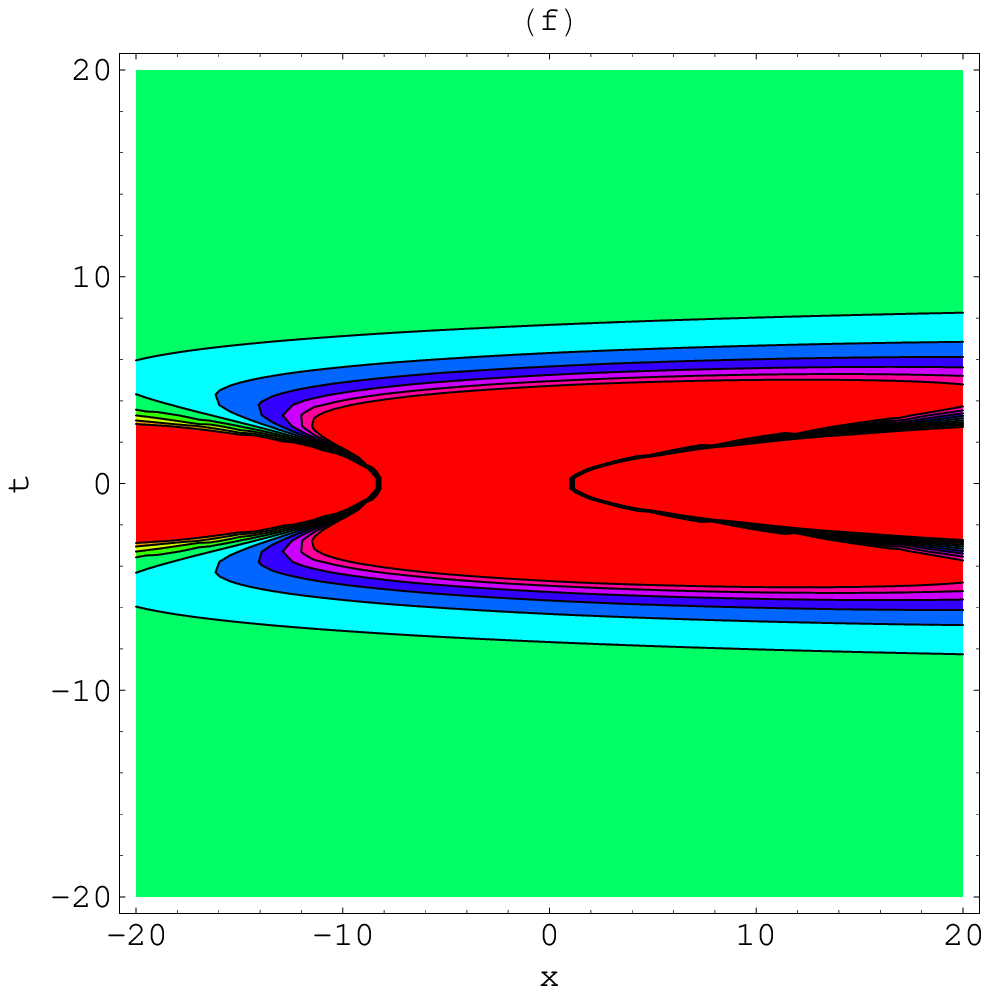}
\vspace{7.5cm}
\begin{tabbing}
\textbf{Fig. 7}. Lump solution (11) with $\alpha_1= \alpha_3=\alpha
_5=-1$, $\alpha_2=1$, $\alpha_6=-3$,\\ $\alpha_7= 3$, $\eta
_6=\eta_5=z=0$, $\alpha_9=2$, $z=0$, when $y= -3$  in (a) (d), $y=
0$ in\\ (b) (e) and $y = 3$ in (c) (f).
\end{tabbing}

{\begin{eqnarray} u^{(III)}&=&[12 [2 \left(\alpha _1^2+\alpha
_5^2\right) [\alpha _9+[\alpha _4(t)+\alpha _1 x+\alpha _2 y+\alpha
_3 z]{}^2+[\alpha _8(t)+\alpha _5
   x\nonumber\\&+&\alpha _6 y+\alpha _7 z]{}^2]-[2 \alpha _1 [\alpha _4(t)+\alpha _1 x+\alpha _2 y+\alpha _3 z]+2 \alpha _5 [\alpha _8(t)+\alpha
   _5 x\nonumber\\&+&\alpha _6 y+\alpha _7 z]]{}^2]]/[[\alpha _9+[\alpha _4(t)+\alpha _1 x+\alpha _2 y+\alpha _3 z]{}^2\nonumber\\&+&[\alpha
   _8(t)+\alpha _5 x+\alpha _6 y+\alpha _7 z]{}^2]{}^2],
\end{eqnarray}}where $\alpha _4(t)$, $\alpha _4(t)$ and $\varrho(t)$ satisfy constraint (10).

\quad The physical structures for  $(u^{(III)})$ are shown in Figs.
6-8 with some 3d graphs and contour plots.  When $\gamma (t)=-1,
\delta(t)=\beta(t)=1$, the spatial structure called the bright lump
wave is seen in Fig. 6 at $x=-10; 0; 10$. When $\gamma (t)=-t,
\delta(t)=\beta(t)=t$, interaction behaviors of two bright lump
waves are presented in Fig. 7 at $y=-3; 0; 3$. As the value of $y$
changes, the two bright lump waves move towards each other, and
finally merge together. When $\gamma (t)=\delta(t)=\beta(t)=\cos t$,
a periodic-shape bright lump wave is found in Fig. 8 at $z=-10; 0;
10$ and $x=0$.

\includegraphics[scale=0.4,bb=20 270 10 10]{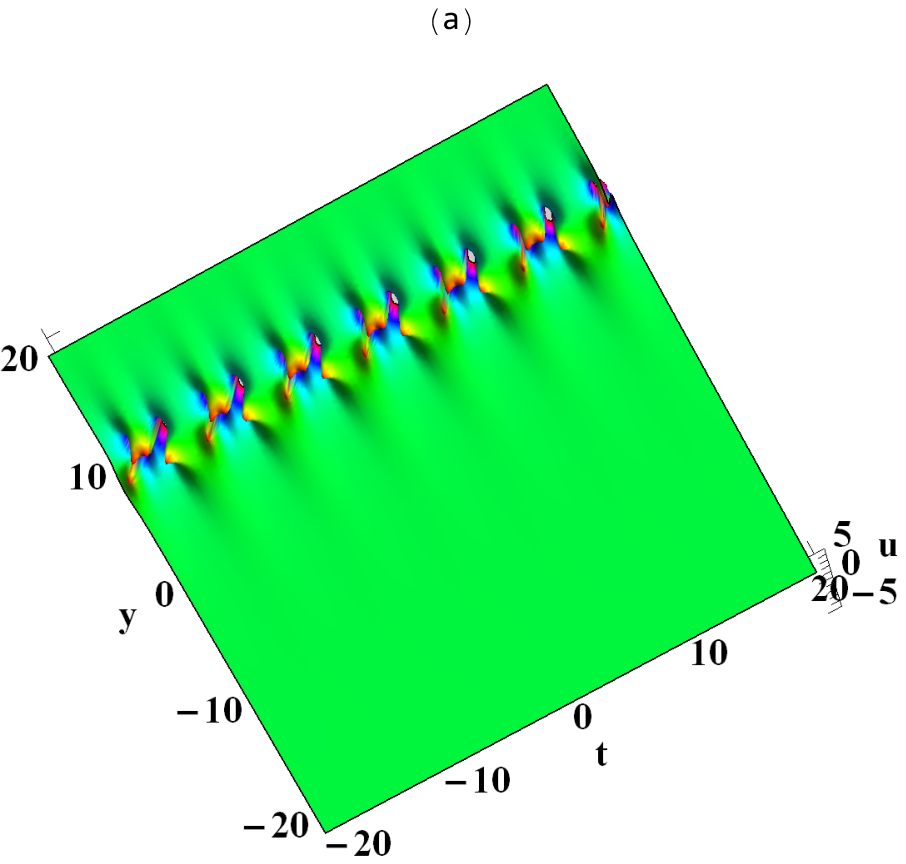}
\includegraphics[scale=0.4,bb=-255 270 10 10]{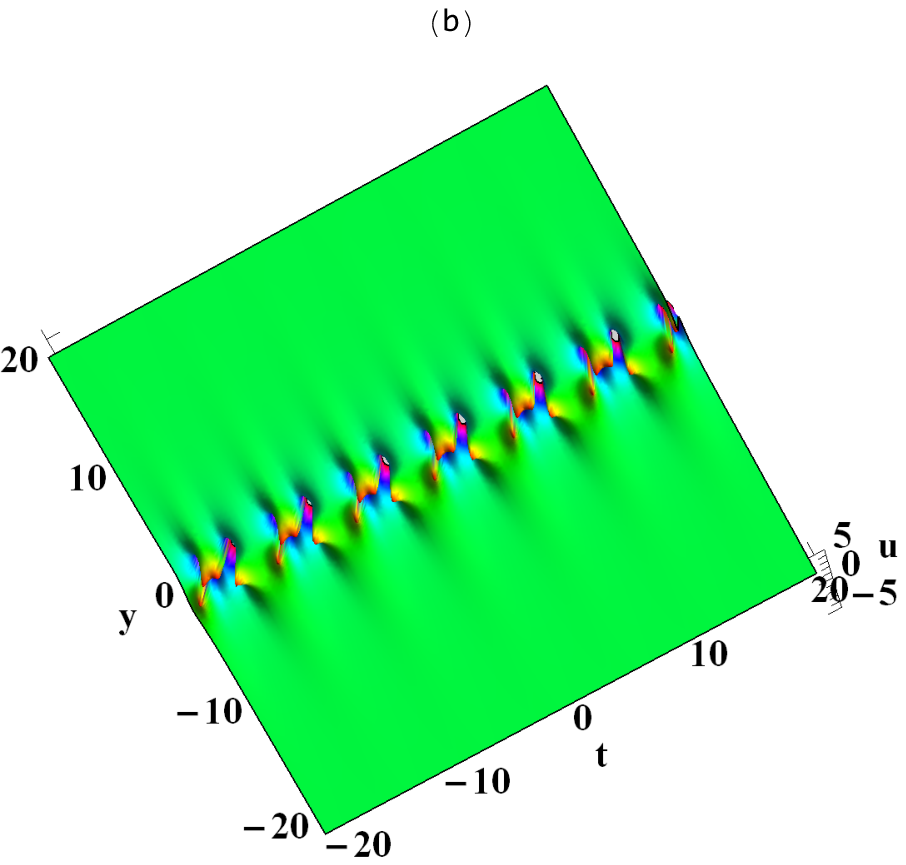}
\includegraphics[scale=0.4,bb=-260 270 10 10]{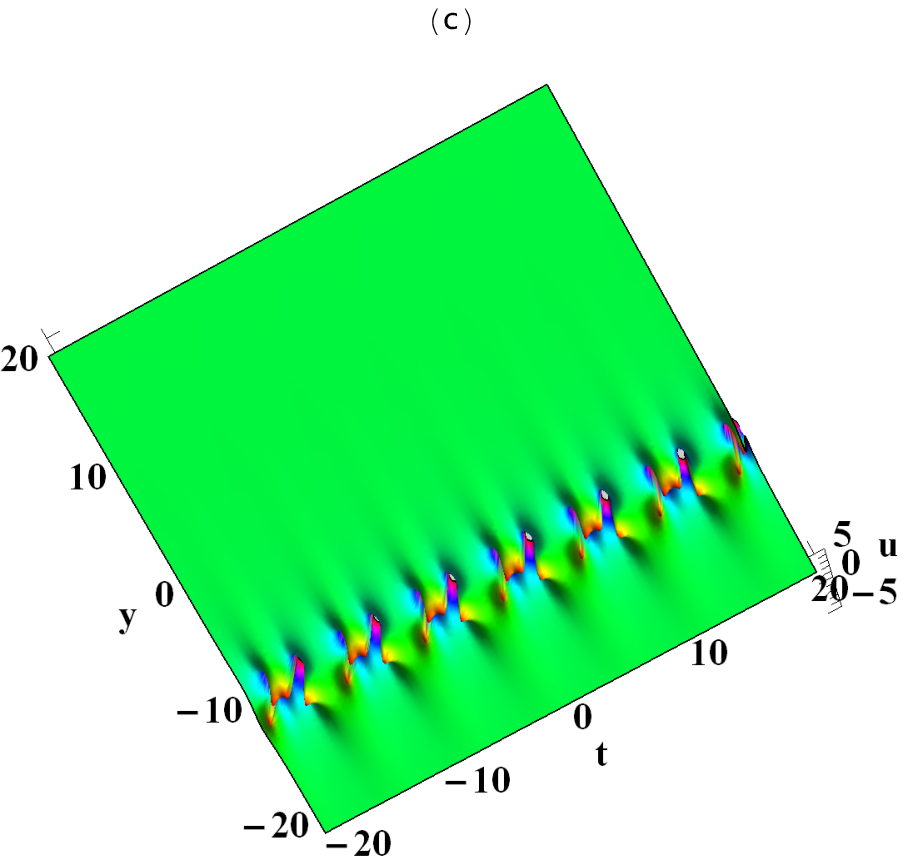}
\includegraphics[scale=0.35,bb=750 620 10 10]{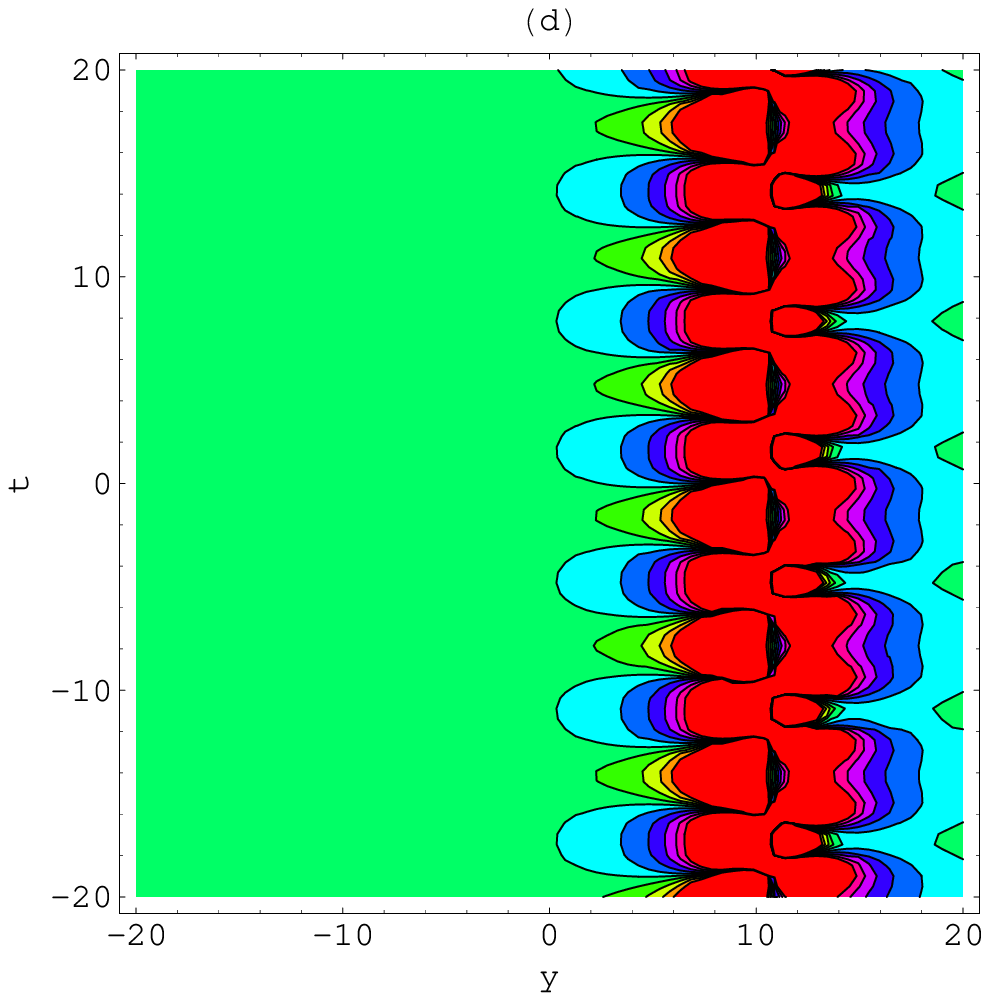}
\includegraphics[scale=0.35,bb=-290 620 10 10]{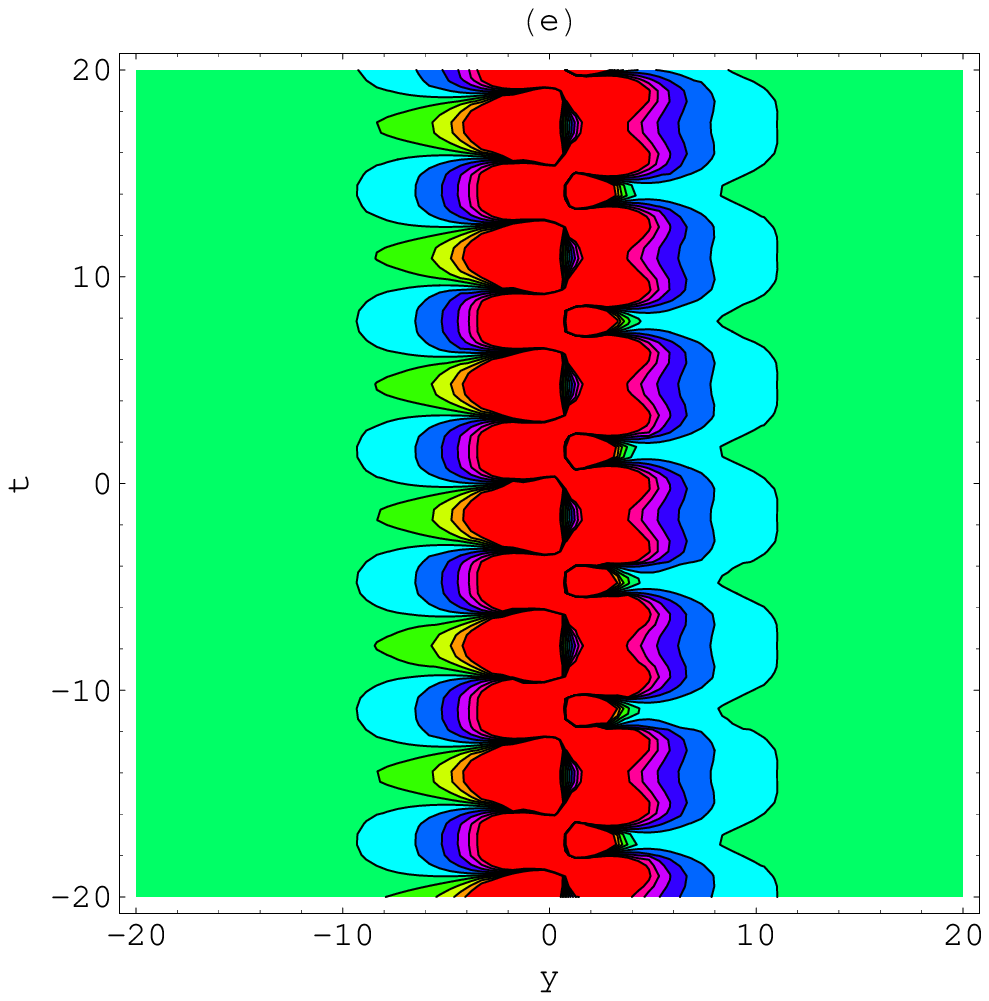}
\includegraphics[scale=0.35,bb=-290 620 10 10]{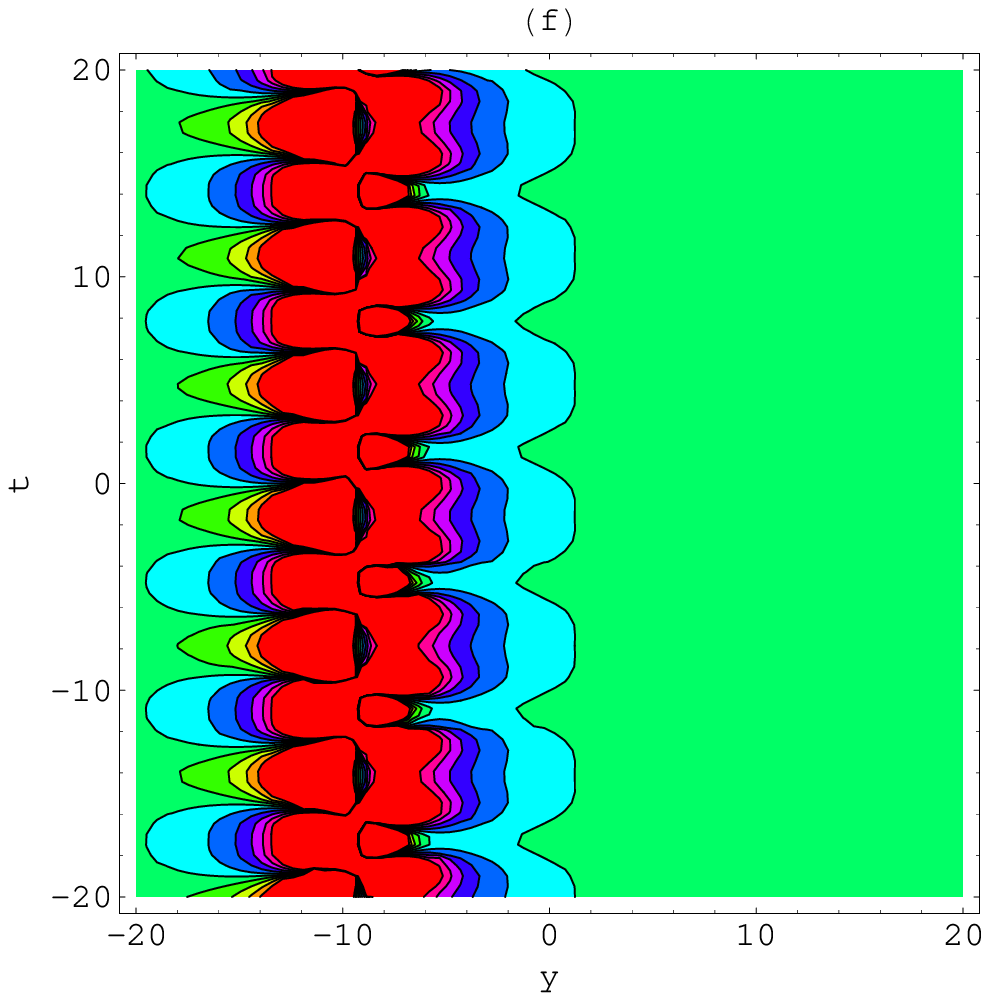}
\vspace{7.5cm}
\begin{tabbing}
\textbf{Fig. 8}. Lump solution (11) with $\alpha_1=
\alpha_2=\alpha_3=\alpha _5=-1$, $\alpha_6=\alpha_7= \eta_5=3$,\\
$\eta _6=-2$, $\alpha_9=2$, $x=0$, when $z= -10$  in (a) (d), $z= 0$
in (b) (e) and\\ $z = 10$ in (c) (f).
\end{tabbing}

\section{Interaction solutions between lump
 and one solitary wave} \label{sec:2}
\quad In order to find the interaction solutions between lump
 and one solitary wave, we add an exponential function in Eq. (5) as
 follows
{\begin{eqnarray} \zeta&=&x \alpha _1+y \alpha _2+z
\alpha _3+\alpha _4(t),\nonumber\\
\varsigma&=&x \alpha _5+y \alpha _6+z \alpha _7+\alpha _8(t),\nonumber\\
\xi&=& \zeta^2+\varsigma^2+\alpha _9(t)+\alpha _{14}(t) e^{\alpha
_{13}(t)+\alpha _{10} x+\alpha _{11} y+\alpha _{12} z},
\end{eqnarray}}where $\alpha_{10}$, $\alpha_{11}$ and $\alpha_{12}$ are
unknown constants. $\alpha_{13}(t)$ and $\alpha_{14}(t)$ are unknown
real functions. Substituting Eq. (12) into Eq. (4) through
Mathematical software, we have {\begin{eqnarray}
\alpha_6&=&-\frac{\alpha _1 \alpha _2}{\alpha _5},
\alpha_7=-\frac{\alpha _1 \alpha _3}{\alpha _5},
\alpha_{11}=\alpha_{12}=0, \alpha_9(t)=\frac{\alpha _1^2+\alpha _5^2}{\alpha _{10}^2},\nonumber\\
\delta(t)&=&-\frac{3 \alpha _5^2 \alpha _{10}^2 \beta (t)+\alpha
_3^2 \varrho (t)}{\alpha _2^2}, \alpha_8(t)=\eta _8-\alpha _5 \int
[3 \alpha _{10}^2 \beta (t)+\gamma (t)] \,
dt,\nonumber\\
\alpha_4(t)&=&\eta _9-\alpha _1 \int [3 \alpha _{10}^2 \beta
(t)+\gamma (t)] \, dt,\nonumber\\ \alpha_{13}(t)&=& \eta
_{10}-\alpha _{10} \int [\alpha _{10}^2 \beta (t)+\gamma (t)] \,
dt-\ln \alpha _{14}(t),
\end{eqnarray}}with  $\alpha_2 \neq 0$, $\alpha_5 \neq 0$ and  $\alpha_{10} \neq 0$.
Substituting Eq. (5) and Eq. (13) into the transformation
$u=12\,[ln\xi(x,y,z,t)]_{xx}$, we get

\includegraphics[scale=0.4,bb=20 270 10 10]{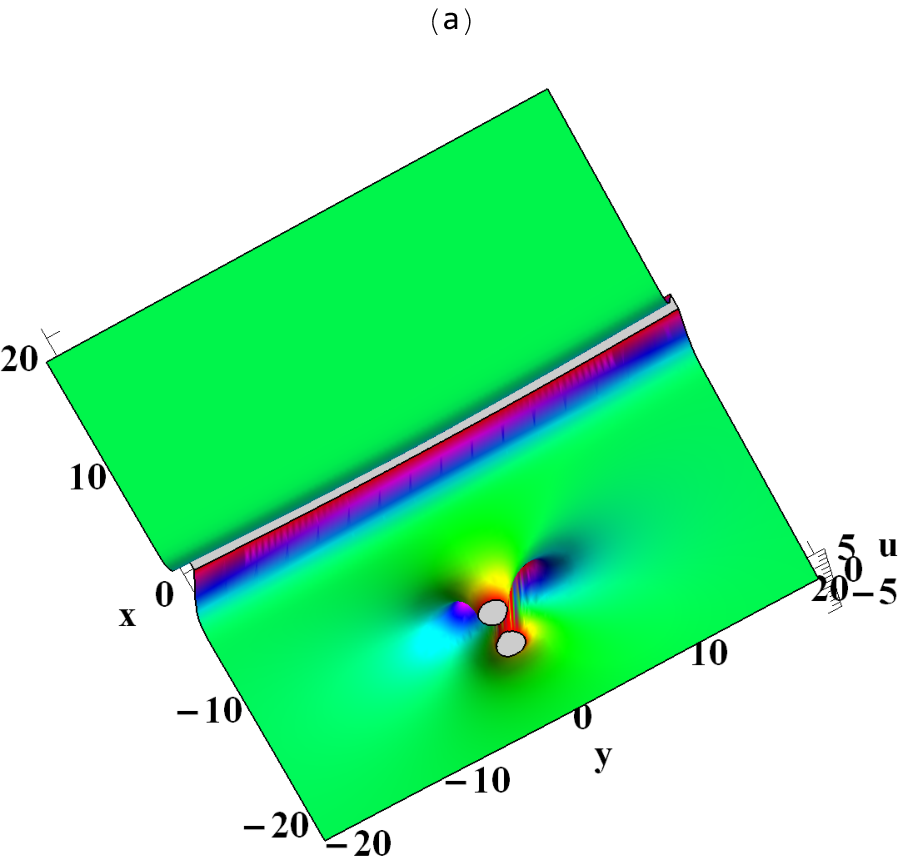}
\includegraphics[scale=0.4,bb=-255 270 10 10]{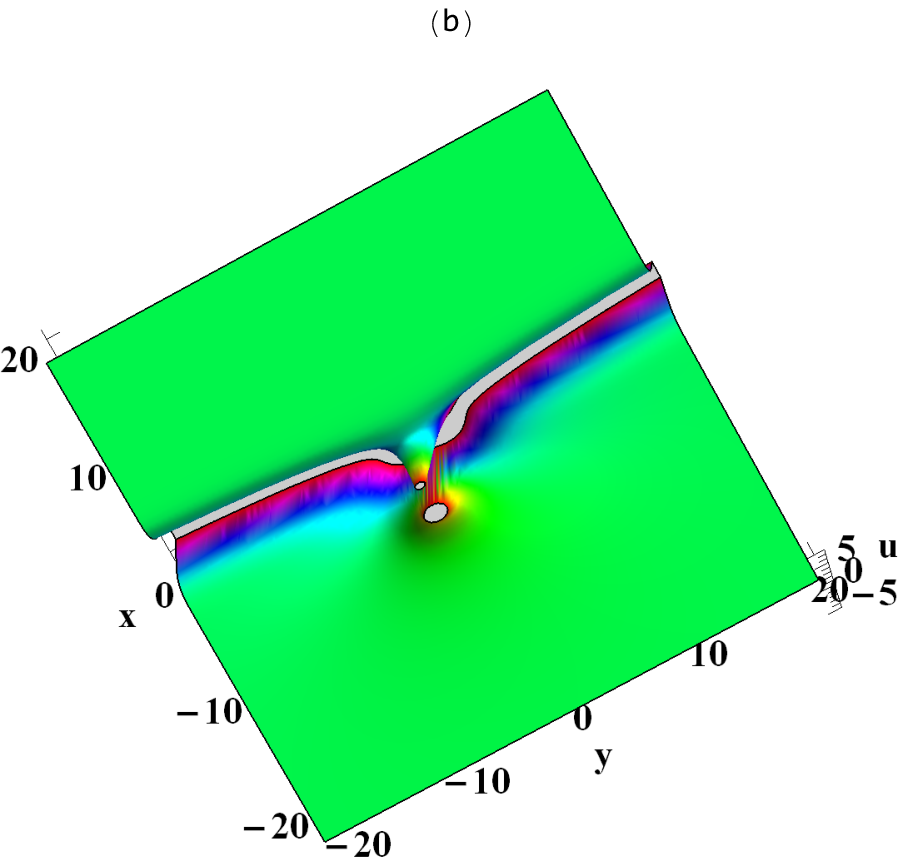}
\includegraphics[scale=0.4,bb=-260 270 10 10]{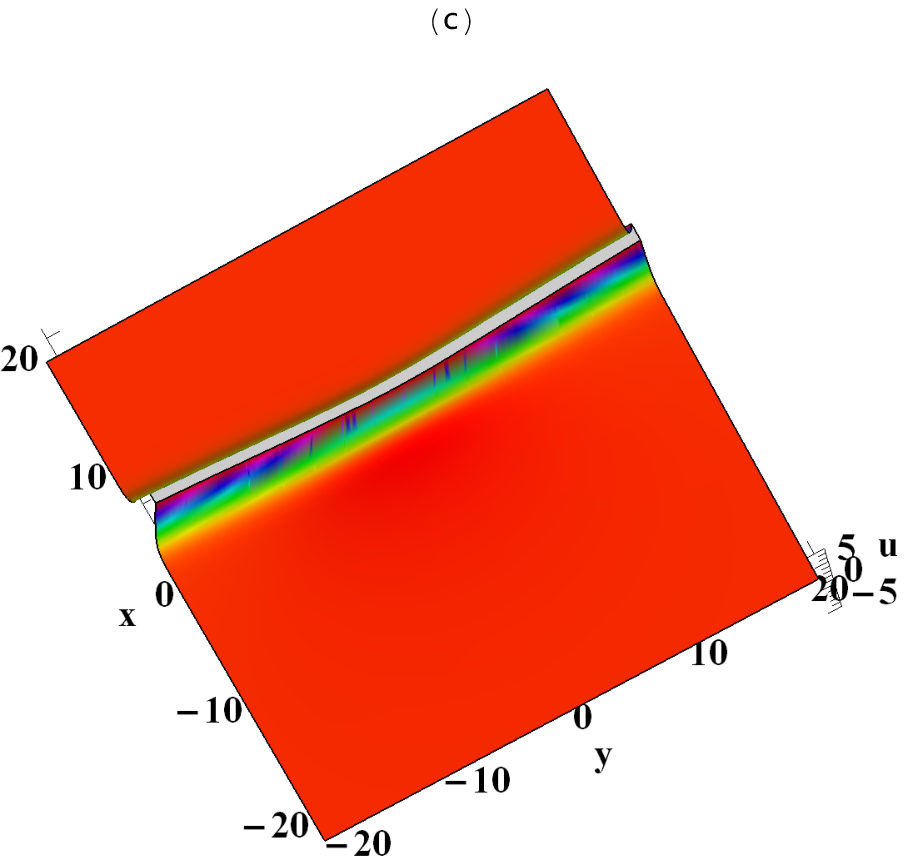}
\includegraphics[scale=0.35,bb=750 620 10 10]{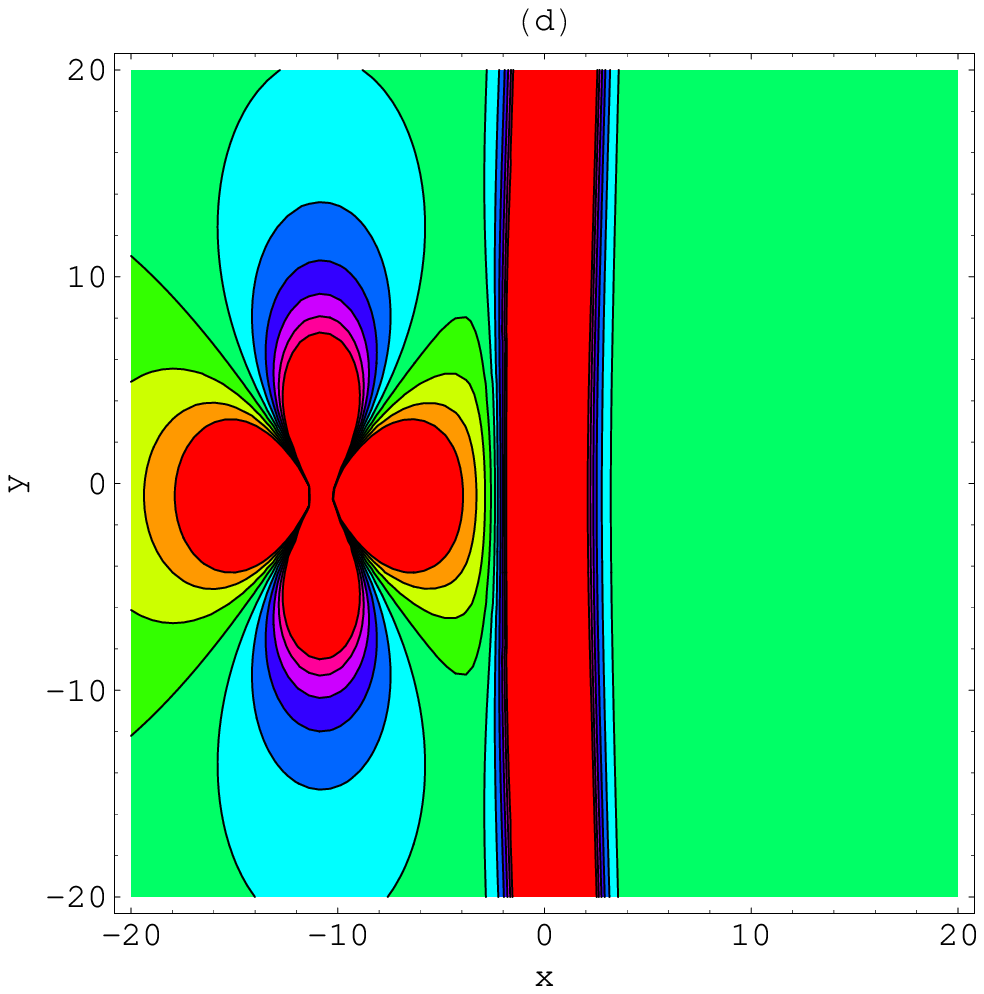}
\includegraphics[scale=0.35,bb=-290 620 10 10]{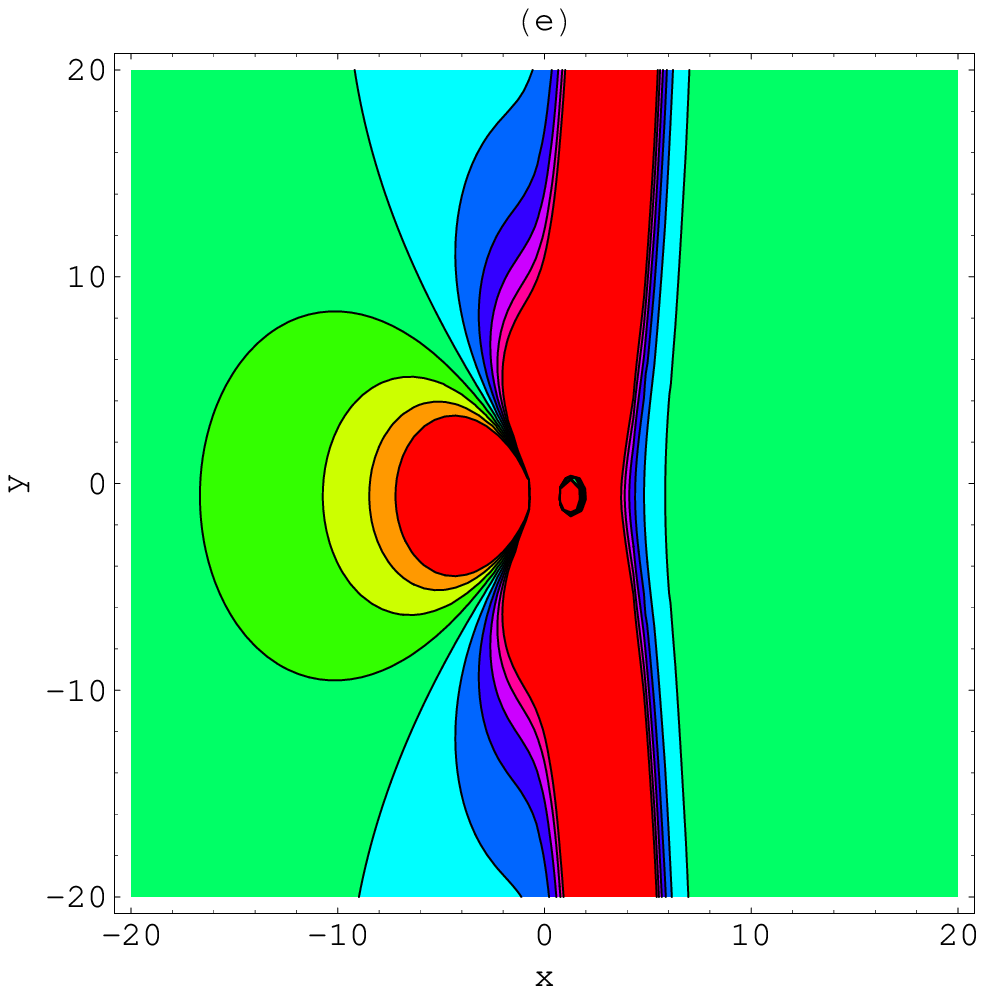}
\includegraphics[scale=0.35,bb=-290 620 10 10]{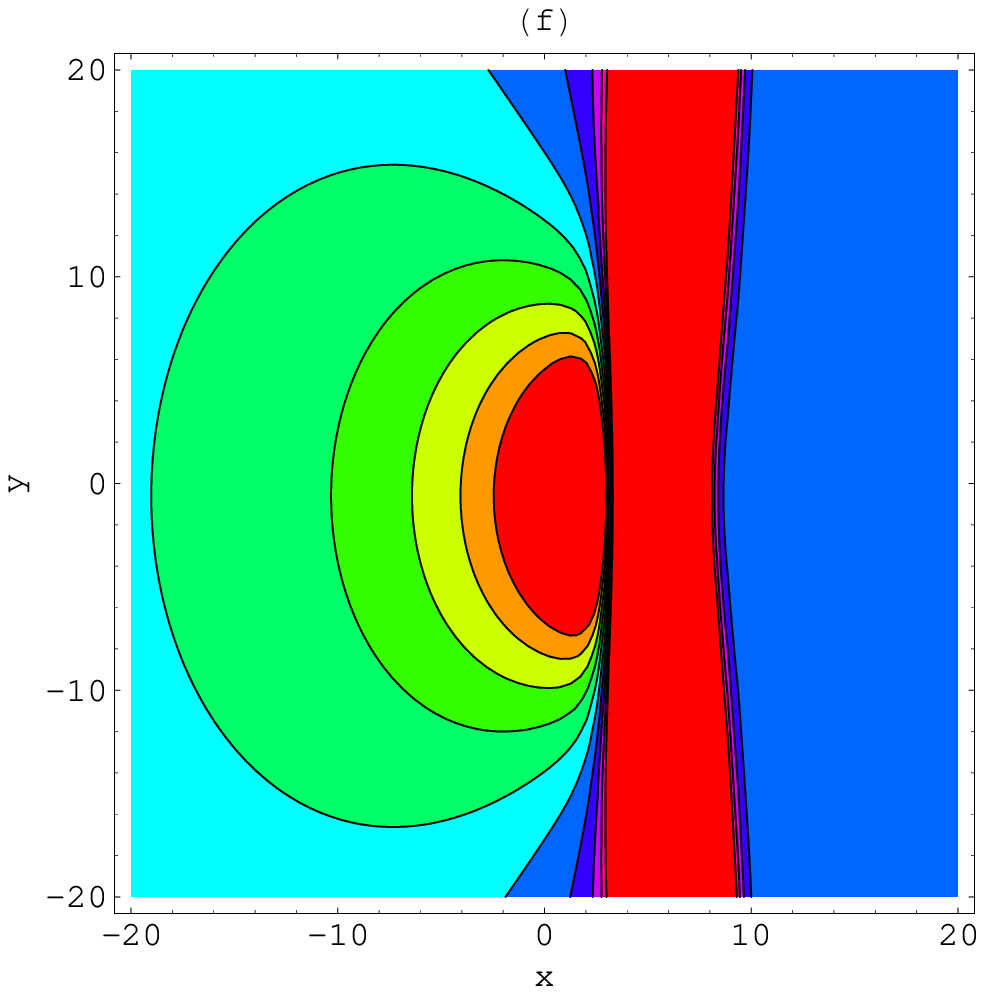}
\vspace{7.5cm}
\begin{tabbing}
\textbf{Fig. 9}. Interaction solution (14) with
$\alpha_1=\eta_8=\eta_9=\eta_{10}=\beta(t)=1$,\\ $\alpha
_3=\gamma(t)=-1$, $\alpha_2=\alpha_{10}= 2$,
 $\alpha_5=-3$, $z=0$, when $t= -1$  in (a) (d),\\ $t=
0$ in (b) (e) and $t = 1$ in (c) (f).
\end{tabbing}

  {\begin{eqnarray}
u^{(IV)}&=&[12 [[2 \left(\alpha _1^2+\alpha _5^2\right)+\alpha
_{10}^2 \exp [\eta _{10}+\alpha _{10} [x-\int \left(\alpha _{10}^2
\beta (t)+\gamma (t)\right) \,
   dt]]]\nonumber\\&*& [\frac{\alpha _1^2+\alpha _5^2}{\alpha _{10}^2}+\exp [\eta _{10}+\alpha _{10} \left(x-\int \left(\alpha _{10}^2 \beta (t)+\gamma (t)\right) \,
   dt\right)]\nonumber\\&+&\frac{[\alpha _1 \left(\alpha _2 y+\alpha _3 z\right)-\alpha _5 [\eta _8+\alpha _5 \left(x-\int \left(3 \alpha _{10}^2 \beta (t)+\gamma (t)\right) \,
   dt\right)]]{}^2}{\alpha _5^2}\nonumber\\&+&\left(\eta _9+\alpha _1 \left(x-\int \left(3 \alpha _{10}^2 \beta (t)+\gamma (t)\right) \, dt\right)+\alpha _2 y+\alpha _3
   z\right){}^2]\nonumber\\&-&[2 \alpha _1 \eta _9+\alpha _{10} \exp \left(\eta _{10}+\alpha _{10} \left(x-\int \left(\alpha _{10}^2 \beta (t)+\gamma (t)\right) \, dt\right)\right)\nonumber\\&+&2
   \alpha _5 [\eta _8+\alpha _5 \left(x-\int \left(3 \alpha _{10}^2 \beta (t)+\gamma (t)\right) \, dt\right)]\nonumber\\&+&2 \alpha _1^2 [x-\int \left(3 \alpha _{10}^2 \beta
   (t)+\gamma (t)\right) \, dt]]{}^2]]/[[\frac{\alpha _1^2+\alpha _5^2}{\alpha _{10}^2}\nonumber\\&+&
   \exp \left(\eta _{10}+\alpha _{10} \left(x-\int \left(\alpha _{10}^2 \beta
   (t)+\gamma (t)\right) \, dt\right)\right)+[[\alpha _1 \left(\alpha _2 y+\alpha _3 z\right)\nonumber\\&-&
   \alpha _5 \left(\eta _8+\alpha _5 \left(x-\int \left(3 \alpha _{10}^2 \beta
   (t)+\gamma (t)\right) \, dt\right)\right)]{}^2]/\alpha _5^2\nonumber\\&+&[\eta _9+\alpha _1
   [x-\int \left(3 \alpha _{10}^2 \beta (t)+\gamma (t)\right) \, dt]+\alpha _2
   y+\alpha _3 z]{}^2]{}^2],
\end{eqnarray}}where $\eta _8$, $\eta _9$  and $\eta _{10}$ are integral
constants. Interaction phenomena between lump
 and one solitary wave in Eq. (14) is shown in Fig. 9 and Fig. 10. Obviously, it can be seen
  a solitary wave and a lump wave in Fig. 9(a) and  Fig. 9(d). In Fig. 9(b) and  Fig. 9(e), the solitary  and lump wave are slowly
  approaching at $t=0$. In Fig. 9(c) and  Fig. 9(f), the solitary  and lump wave merge together to propagate
  forward at $t=1$. Fig. 10 displays the effect of  variable coefficient $\gamma(t)$ on the interaction phenomena between  lump
 and one solitary wave.

\includegraphics[scale=0.4,bb=20 270 10 10]{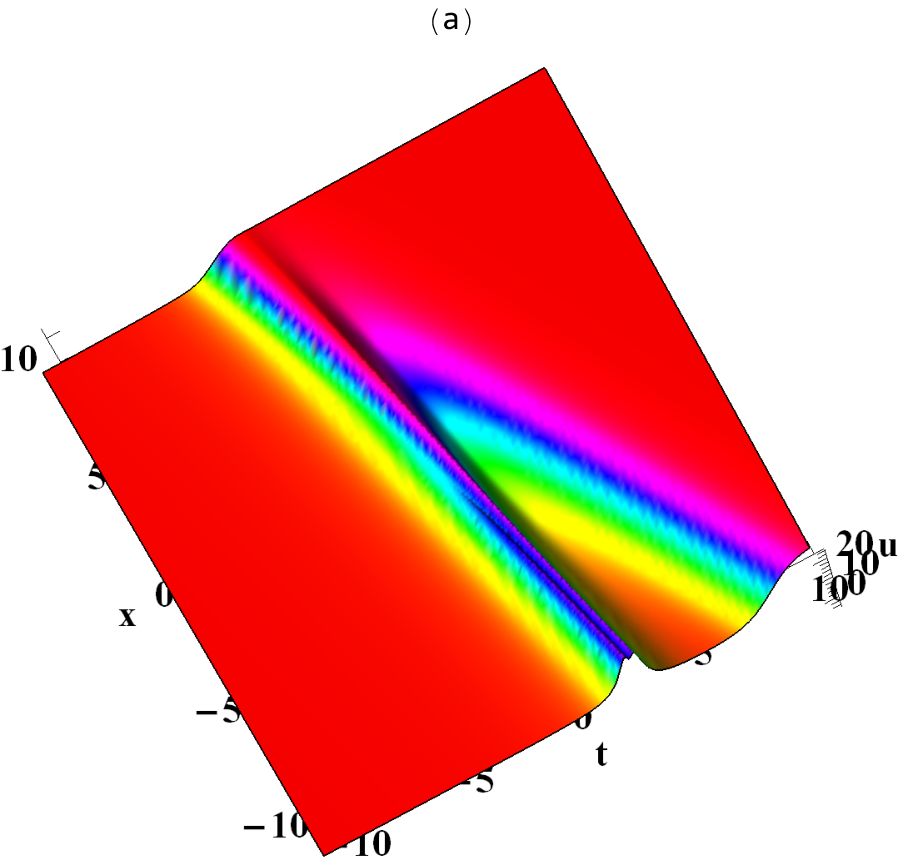}
\includegraphics[scale=0.4,bb=-255 270 10 10]{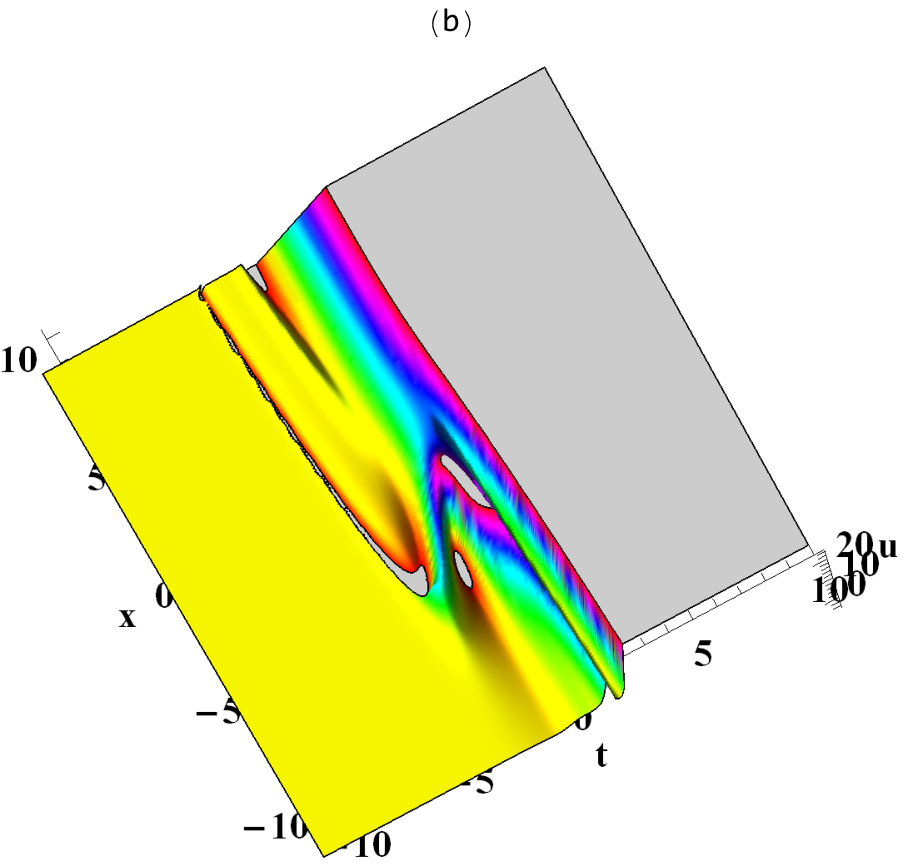}
\includegraphics[scale=0.4,bb=-260 270 10 10]{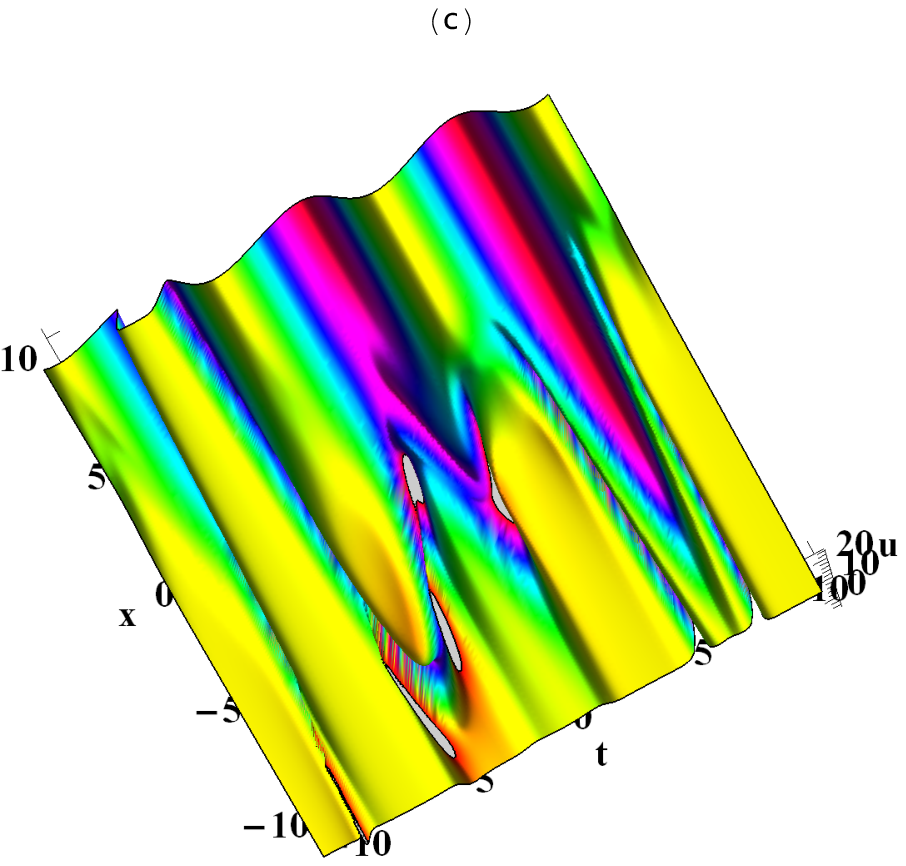}
\includegraphics[scale=0.35,bb=750 620 10 10]{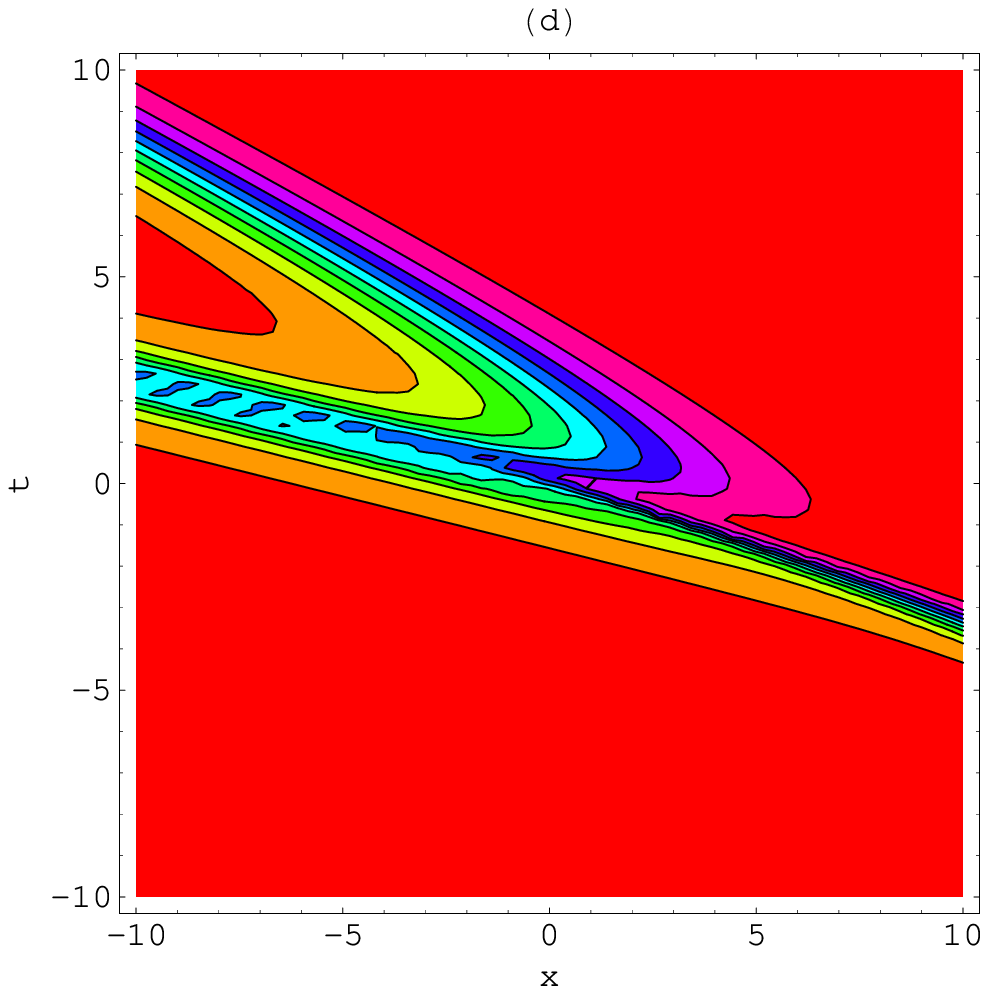}
\includegraphics[scale=0.35,bb=-290 620 10 10]{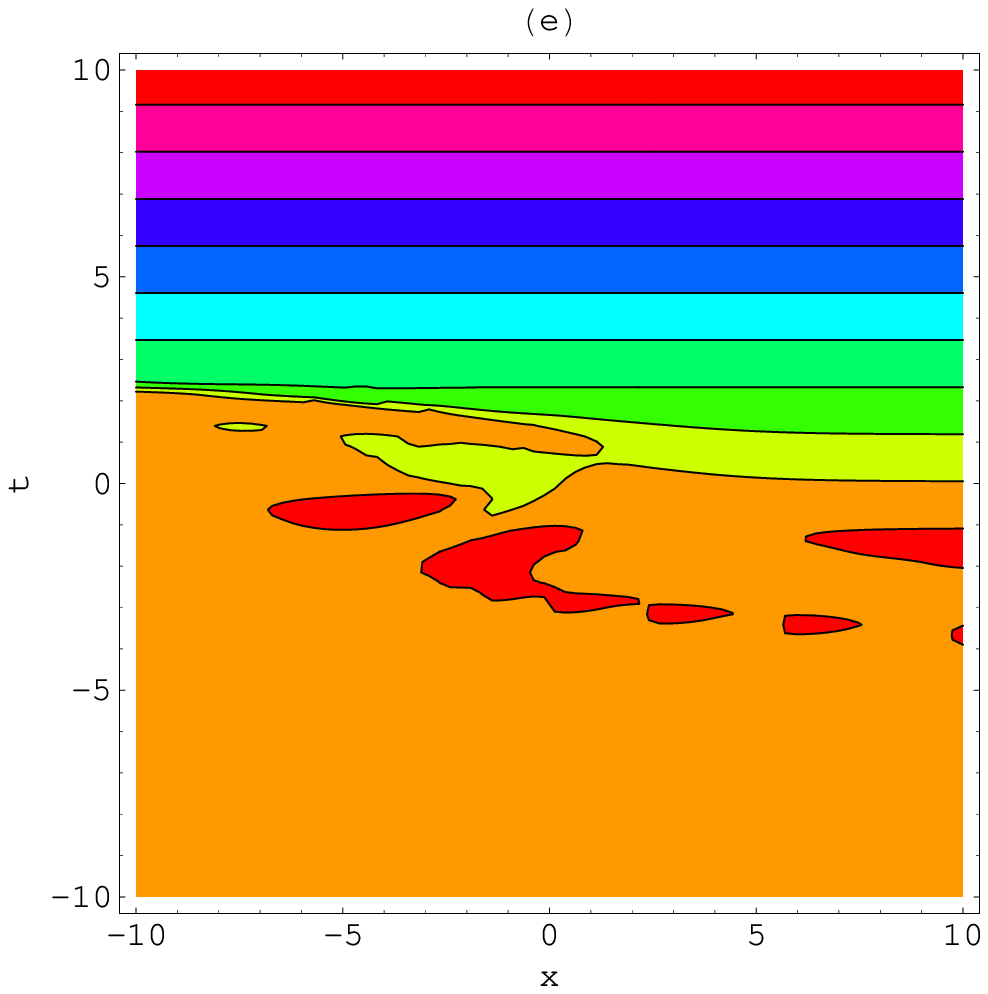}
\includegraphics[scale=0.35,bb=-290 620 10 10]{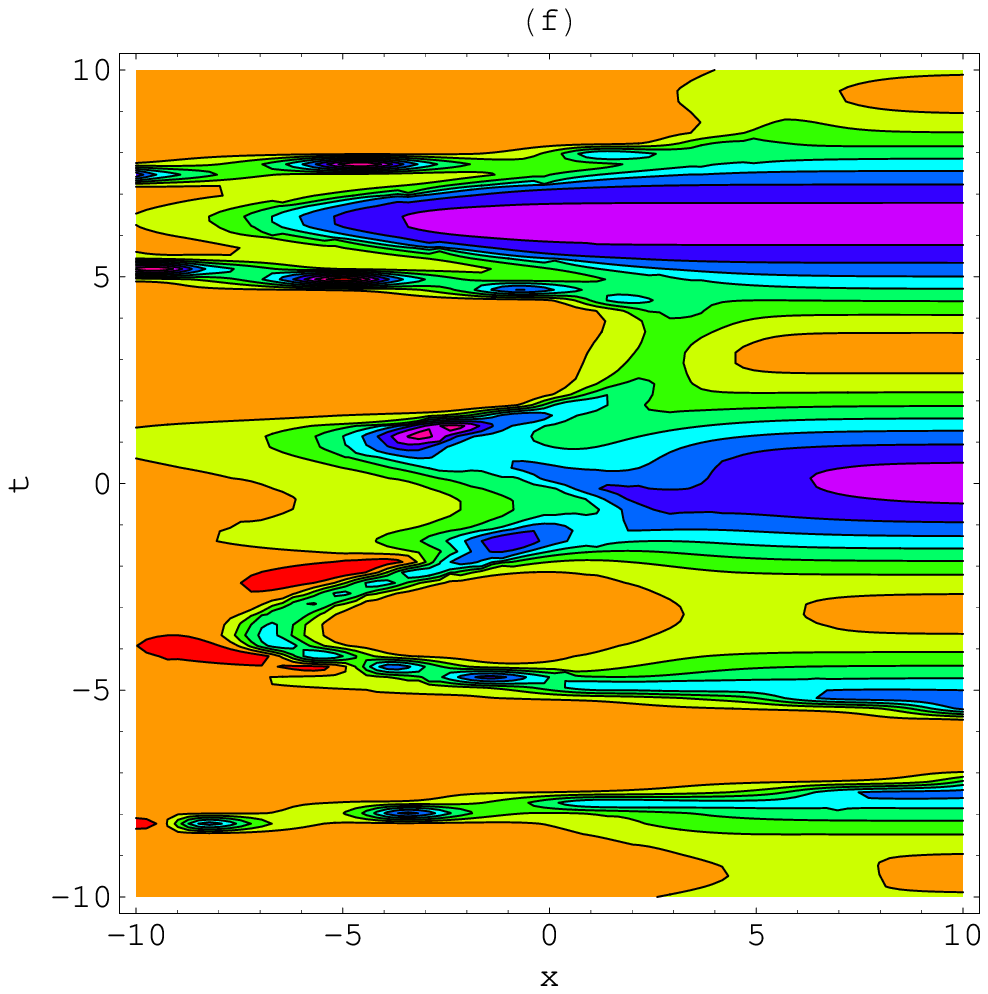}
\vspace{7.5cm}
\begin{tabbing}
\textbf{Fig. 10}. Interaction solution (14) with
$\alpha_1=\eta_8=\eta_9=\eta_{10}=\alpha_{10}=\beta(t)= 1$,\\
$\alpha _3=\alpha_5=-1$, $\alpha_2=2$, $y=z=0$, when $\gamma(t)= 1$
in (a) (d), $\gamma(t)= t$ in (b) (e)\\ and $\gamma(t) = \cos t$ in
(c) (f).
\end{tabbing}

\section{Interaction solutions between lump
 and two solitary waves} \label{sec:2}
\quad In order to derive the interaction solutions between lump
 and two solitary waves, we add two exponential functions in Eq. (5) as
 follows
{\begin{eqnarray} \zeta&=&x \alpha _1+y \alpha _2+z
\alpha _3+\alpha _4(t),\nonumber\\
\varsigma&=&x \alpha _5+y \alpha _6+z \alpha _7+\alpha _8(t),\nonumber\\
\xi&=& \zeta^2+\varsigma^2+\alpha _9(t)+\alpha _{14}(t) e^{\alpha
_{13}(t)+\alpha _{10} x+\alpha _{11} y+\alpha _{12}
z}\nonumber\\&+&\alpha _{15}(t) e^{-\alpha _{13}(t)-\alpha _{10}
x-\alpha _{11} y-\alpha _{12} z},
\end{eqnarray}}where $\alpha_{15}(t)$ are unknown
real functions. Substituting Eq. (15) into Eq. (4) through
Mathematical software, we obtain {\begin{eqnarray}
\alpha_6&=&-\frac{\alpha _1 \alpha _2}{\alpha _5},
\alpha_7=-\frac{\alpha _1 \alpha _3}{\alpha _5},
\alpha_9(t)=\frac{\alpha _{10}^4 \eta _{12}+\alpha _1^4+2 \alpha
_5^2
\alpha _1^2+\alpha _5^4}{\left(\alpha _1^2+\alpha _5^2\right) \alpha _{10}^2},\nonumber\\
\varrho(t)&=& \frac{\alpha _2^2 (-\delta (t))-3 \alpha _5^2 \alpha
_{10}^2 \beta (t)}{\alpha _3^2}, \alpha_8(t)=\eta _{13}-\alpha _5
\int [3 \alpha _{10}^2 \beta (t)+\gamma (t)] \,
dt,\nonumber\\
\alpha_4(t)&=&\eta _{14}-\alpha _1 \int [3 \alpha _{10}^2 \beta
(t)+\gamma (t)] \, dt, \alpha_{11}=\alpha_{12}=0,\nonumber\\
\alpha_{13}(t)&=& \eta _{15}-\alpha _{10} \int [\alpha _{10}^2 \beta
(t)+\gamma (t)] \, dt-\ln \alpha _{14}(t), \alpha
_{15}(t)=\frac{\eta _{12}}{\alpha _{14}(t)},
\end{eqnarray}}with  $\alpha_3 \neq 0$, $\alpha_5 \neq 0$, $\alpha _{14}(t) \neq 0$,  $\alpha _1^2+\alpha _5^2 \neq 0$  and  $\alpha_{10} \neq 0$.
Substituting Eq. (5) and Eq. (16) into the transformation
$u=12\,[ln\xi(x,y,z,t)]_{xx}$, we get

\includegraphics[scale=0.4,bb=20 260 10 10]{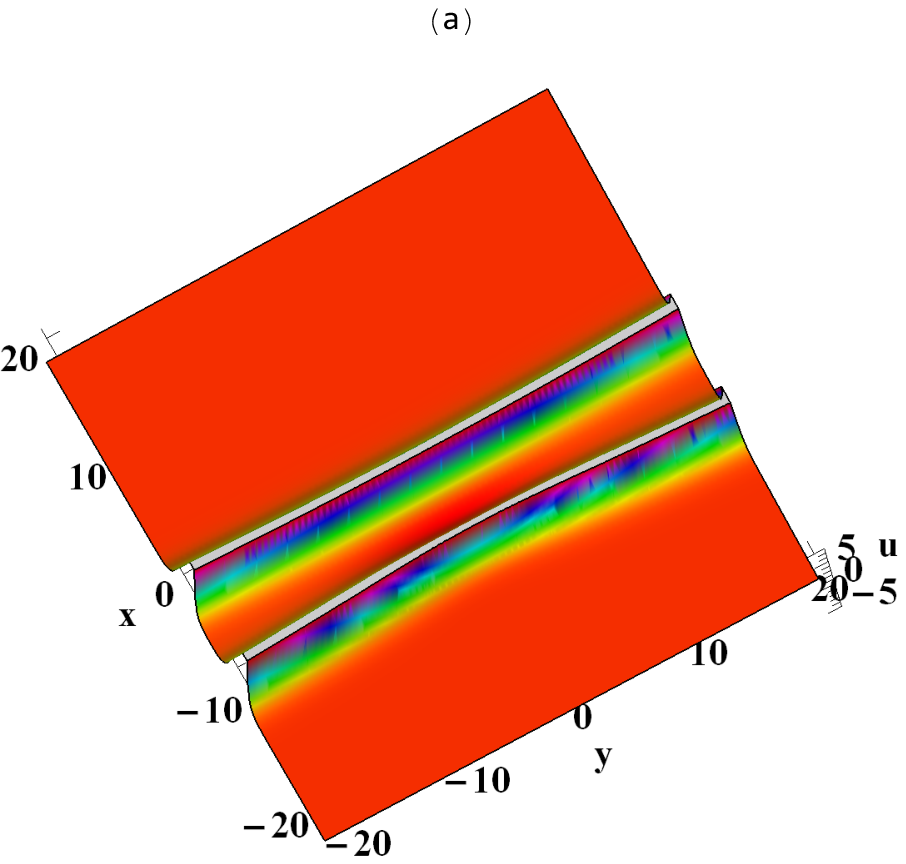}
\includegraphics[scale=0.4,bb=-255 260 10 10]{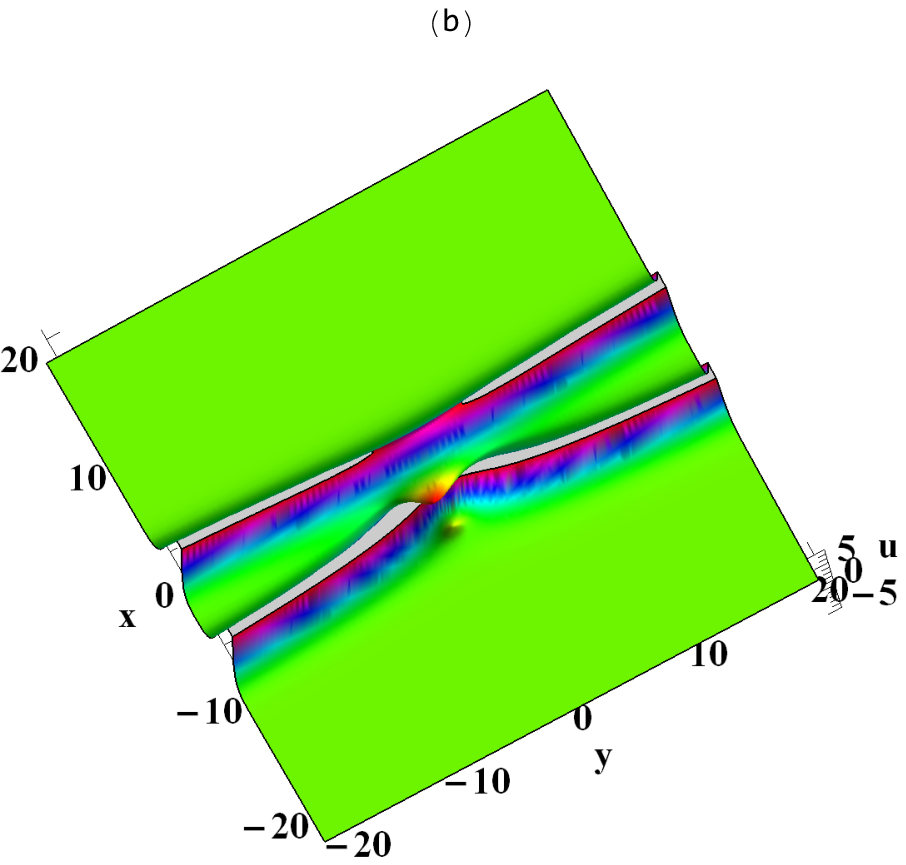}
\includegraphics[scale=0.4,bb=-260 260 10 10]{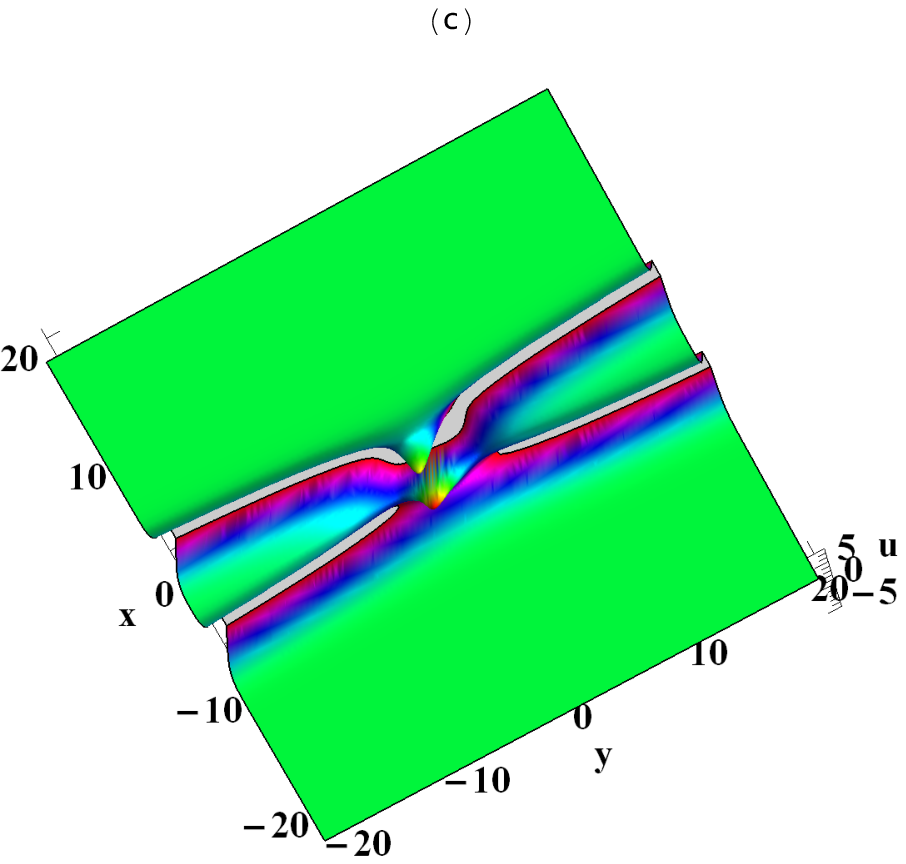}
\includegraphics[scale=0.4,bb=450 525 10 10]{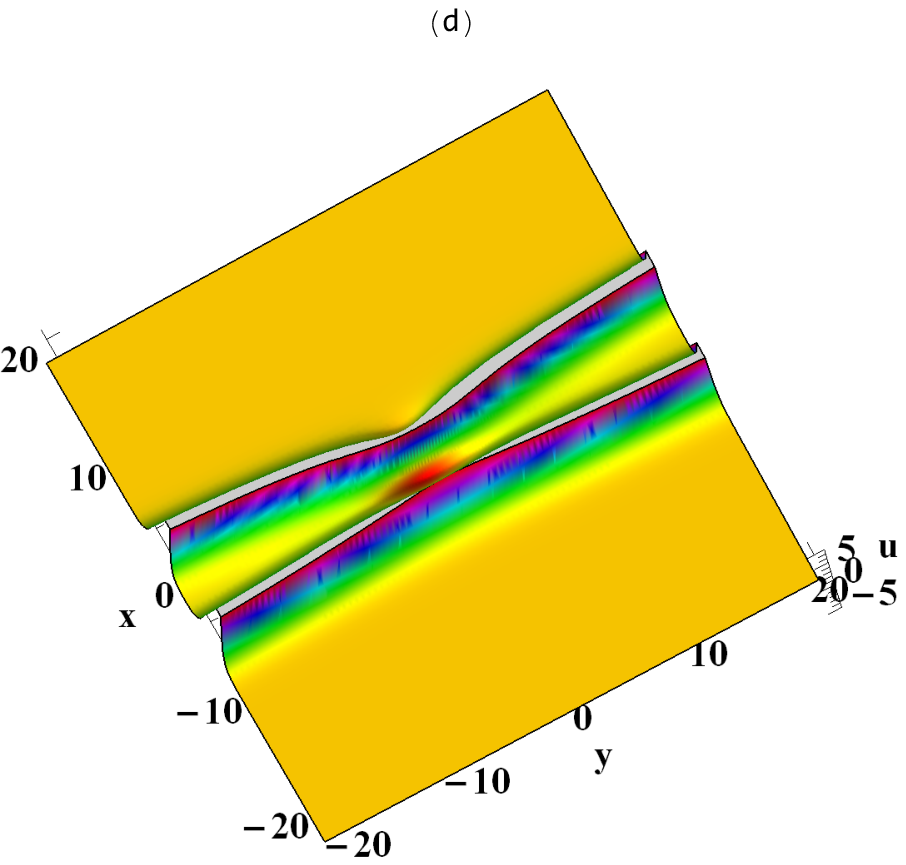}
\includegraphics[scale=0.4,bb=-300 525 10 10]{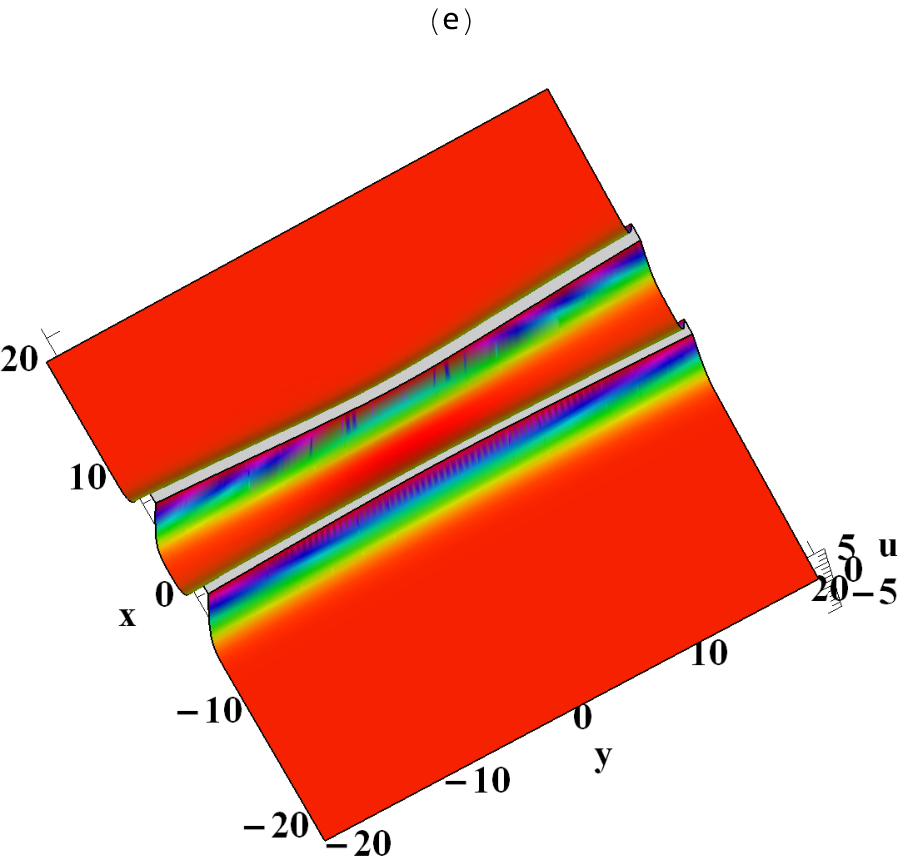}
\vspace{7.5cm}
\begin{tabbing}
\textbf{Fig. 11}. Interaction solution (16) with $\alpha
_3=\gamma(t)=-1$, $\alpha_2=\alpha_{10}= 2$,\\
 $\alpha_5=-3$, $\alpha_1=\eta_{12}=\eta_{13}=\eta_{14}=\eta_{15}=\beta(t)=1$, $z=0$, when $t= -1$ \\ in (a), $t= -0.3$  in (b), $t=
0$ in (c), $t= 0.3$  in (d), $t = 1$ in (e).
\end{tabbing}

  {\begin{eqnarray}
u^{(V)}&=&12 [[2 \alpha _1^2+2 \alpha _5^2+\alpha _{10}^2 \alpha
_{14}(t) e^{\alpha _{13}(t)+\alpha _{10} x}+\alpha _{10}^2 \alpha
_{15}(t) e^{-\alpha _{10} x-\alpha
   _{13}(t)}]\nonumber\\&/&[\alpha _9(t)+\alpha _{14}(t) e^{\alpha _{13}(t)+\alpha _{10} x}+\alpha _{15}(t)
    e^{-\alpha _{10} x-\alpha _{13}(t)}+[\alpha _4(t)+\alpha _1 x\nonumber\\&+&\alpha _2 y+\alpha _3
   z]{}^2+\left(\alpha _8(t)+\alpha _5 x+\alpha _6 y+\alpha _7 z\right){}^2]\nonumber\\&-&[[\alpha _{10} \alpha _{14}(t)
    e^{\alpha _{13}(t)+\alpha _{10} x}-\alpha _{10} \alpha
   _{15}(t) e^{-\alpha _{10} x-\alpha _{13}(t)}+2 \alpha _1 [\alpha _4(t)\nonumber\\&+&\alpha _1 x+\alpha _2 y+\alpha _3
   z]
   +2 \alpha _5 \left(\alpha _8(t)+\alpha _5 x+\alpha _6
   y+\alpha _7 z\right)]{}^2]/[[\alpha _9(t)\nonumber\\&+&\alpha _{14}(t) e^{\alpha _{13}(t)+\alpha _{10} x}+\alpha _{15}(t)
    e^{-\alpha _{10} x-\alpha _{13}(t)}+[\alpha
   _4(t)+\alpha _1 x\nonumber\\&+&\alpha _2 y+\alpha _3 z]{}^2+[\alpha _8(t)+\alpha _5 x+\alpha _6 y+\alpha _7 z]{}^2]{}^2]],
\end{eqnarray}}where $\eta _{12}$, $\eta _{13}$,  $\eta _{14}$  and $\eta _{15}$ are integral
constants. Interaction phenomena between lump
 andtwo solitary waves in Eq. (16) is shown in Fig. 11 and Fig. 12. Two solitary waves can be found in Fig. 11(a).
 A lump wave appears in one of two solitary waves in
Fig. 11(b). In Fig. 11(c) and Fig. 11(d), the lump wave slowly
shifts to another solitary wave, until vanishes in Fig. 11(e). Fig.
12 displays the corresponding contour plots of Fig. 11.

\includegraphics[scale=0.35,bb=140 290 10 10]{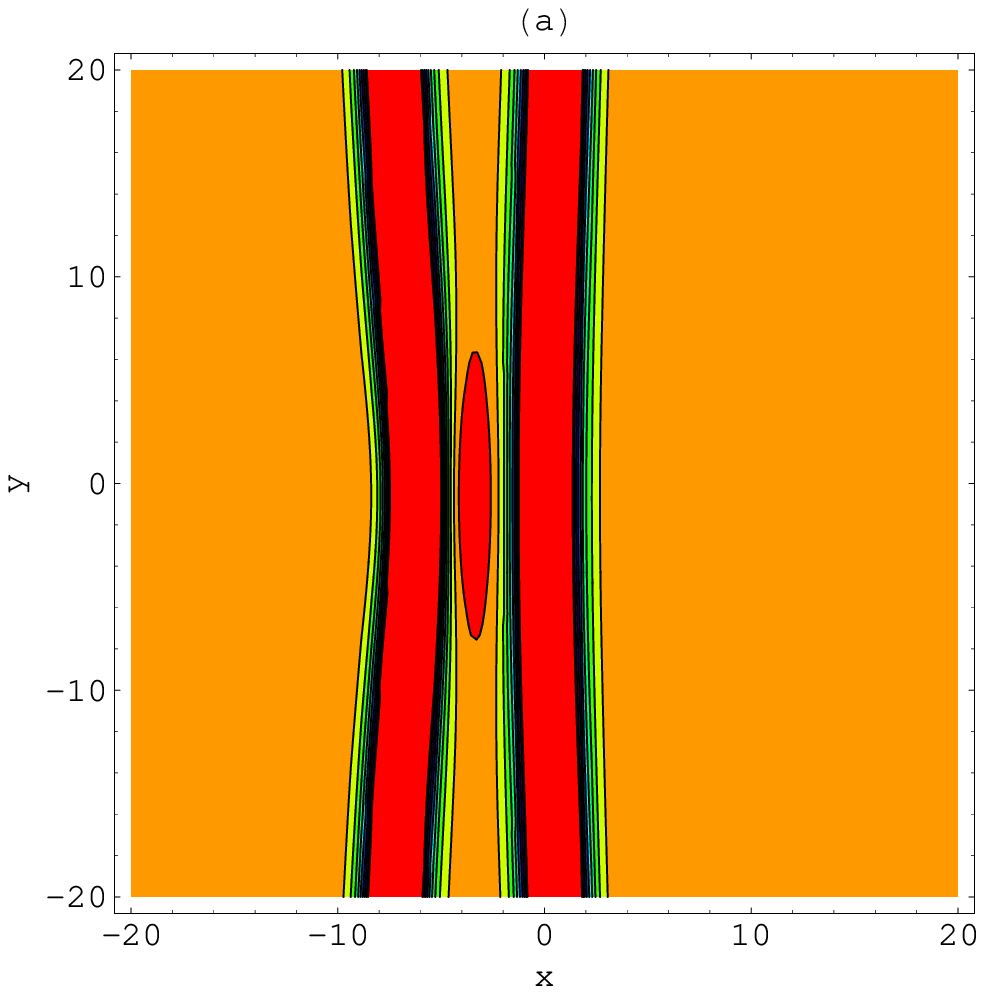}
\includegraphics[scale=0.35,bb=-275 290 10 10]{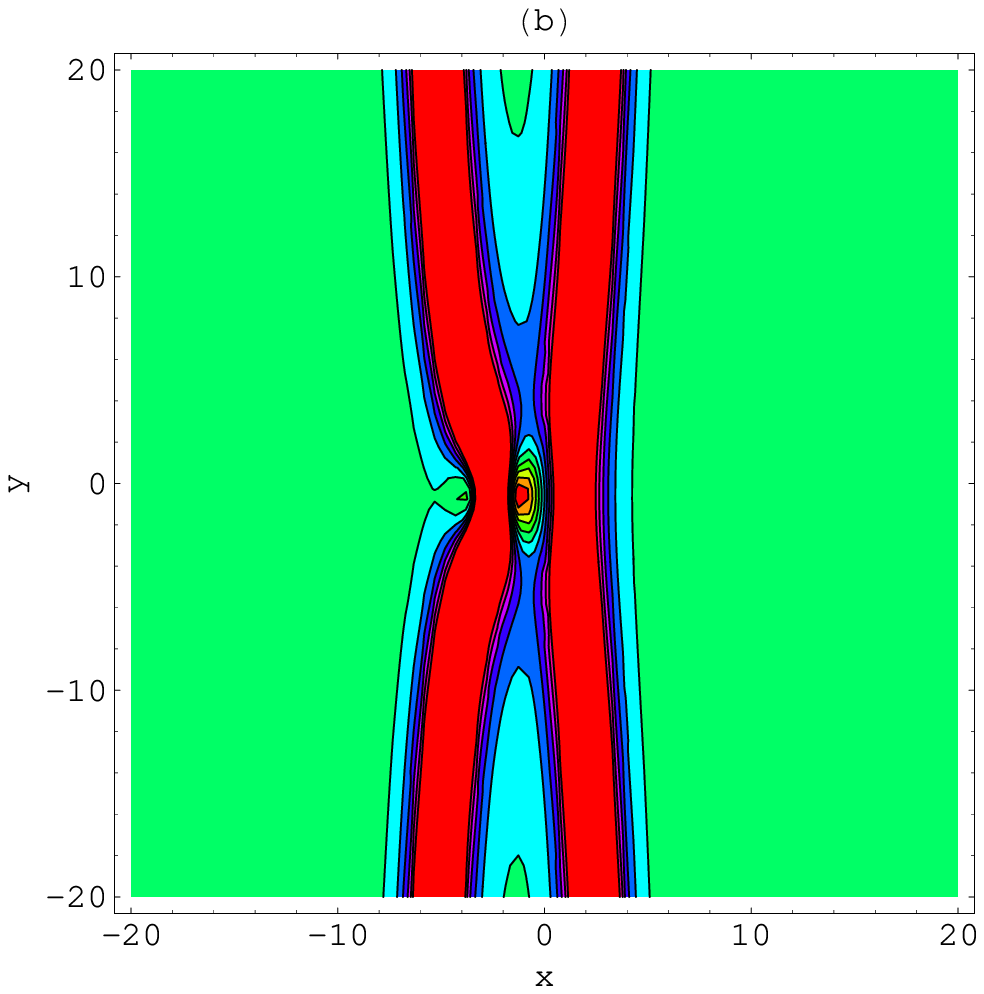}
\includegraphics[scale=0.35,bb=-280 290 10 10]{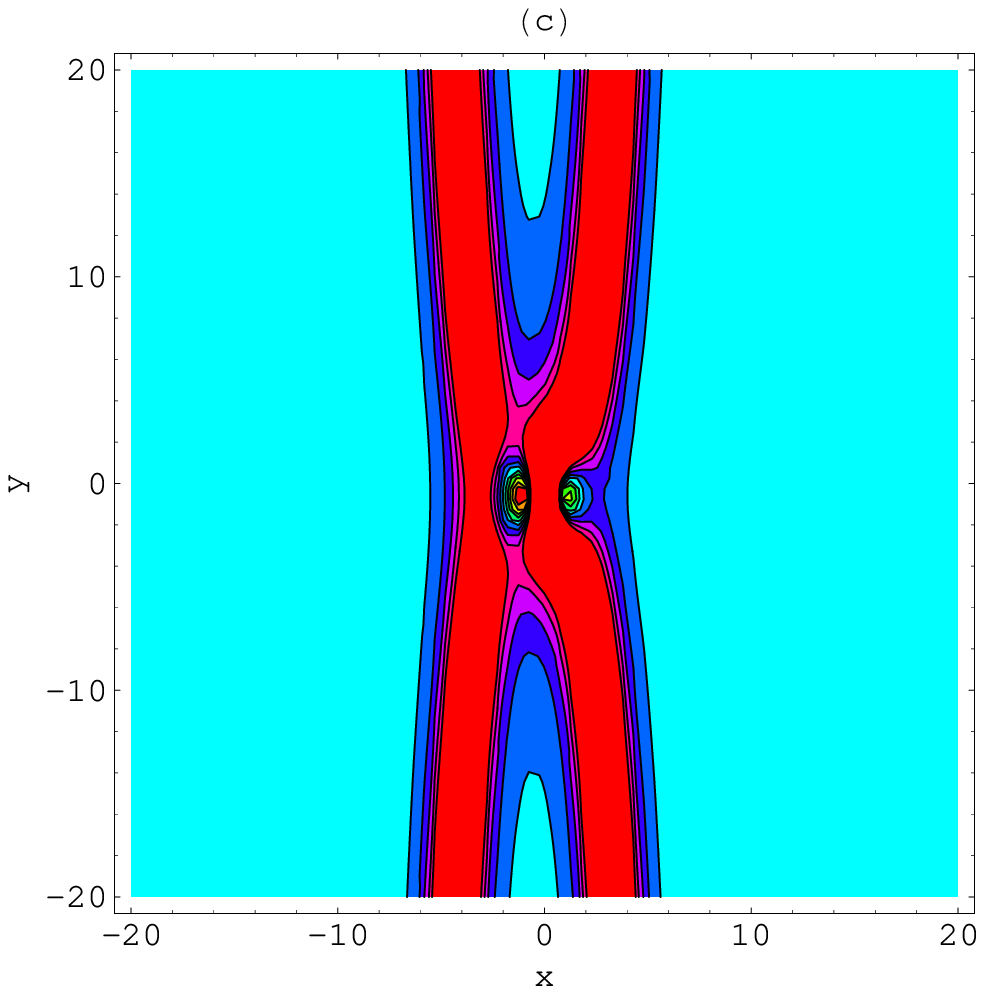}
\includegraphics[scale=0.35,bb=500 585 10 10]{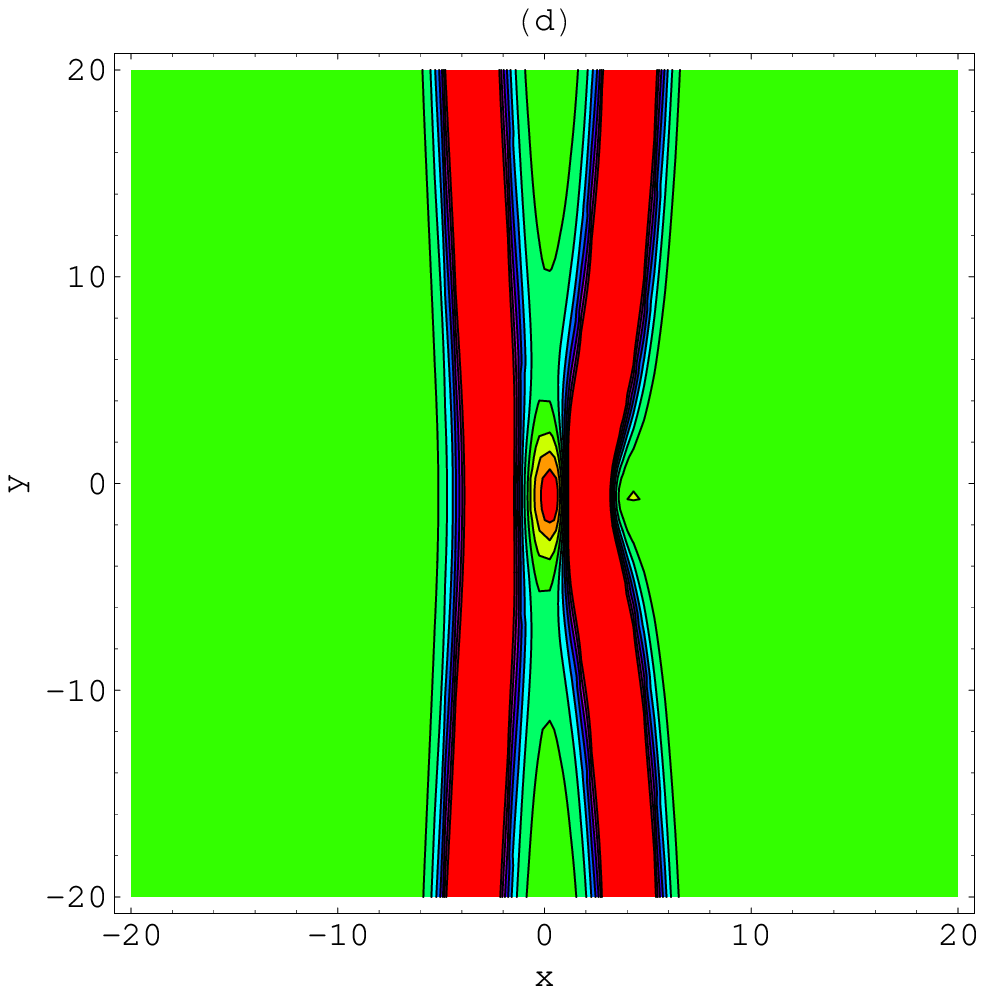}
\includegraphics[scale=0.35,bb=-360 585 10 10]{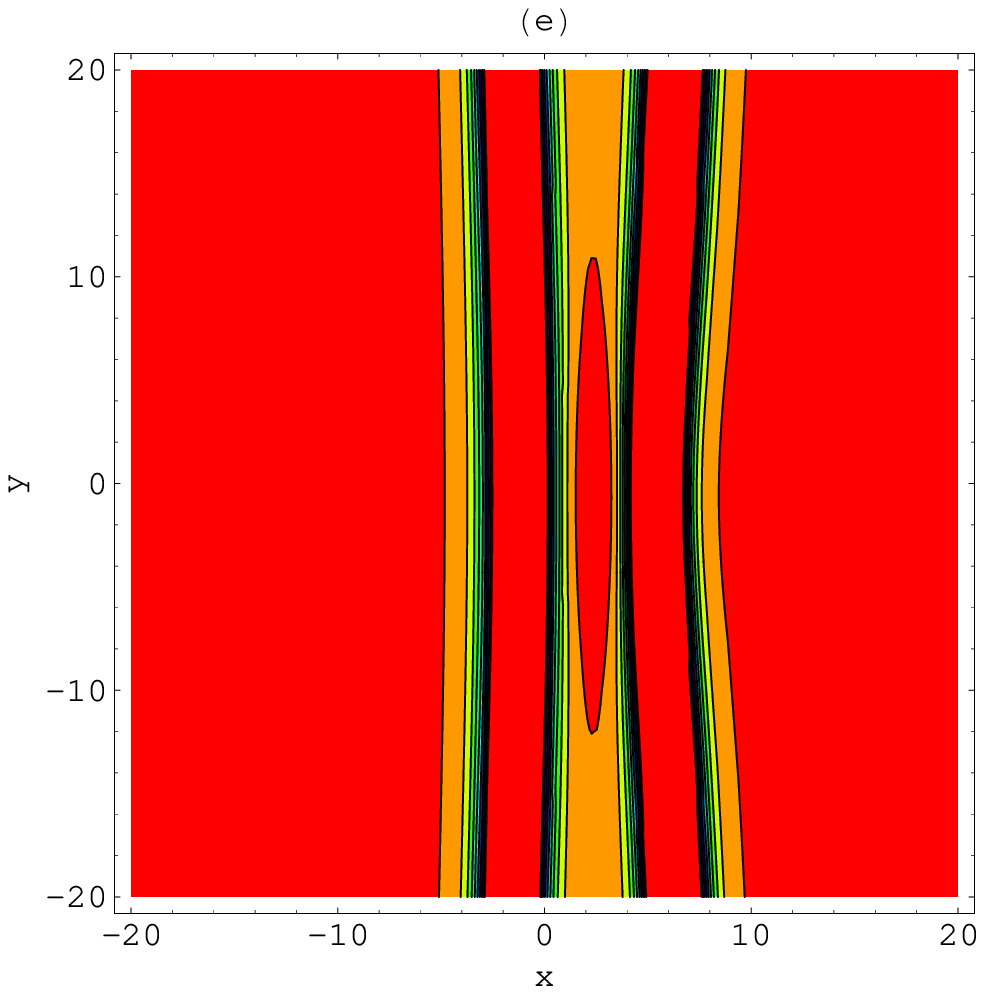}
\vspace{7.4cm}
\begin{tabbing}
\textbf{Fig. 12}. The corresponding contour plots of Fig. 11..
\end{tabbing}

\section{ Conclusion}
\label{sec:4} \quad In this paper, based on the Hirota's bilinear
form and Mathematical software, the lump and interaction solutions
between lump
 and solitary wave of a
generalized (3 + 1)-dimensional variable-coefficient nonlinear-wave
equation in liquid with gas bubbles are studied. Their physical
structures  are described  in  some 3d graphs and contour plots.
  A periodic-shape rational solution is listed in Fig. 1(a) and Fig. 1(d).
  A parabolic-shape rational solution is presented in Fig. 1(b) and Fig. 1(e).
  A cubic-shape rational solution is shown in Fig. 1(c) and Fig. 1(f).
  In lump solutions  $(u^{(II)})$, The spatial structure called the bright lump wave is seen in Fig. 2,
   the spatial structure called the bright-dark lump wave is shown in Fig. 3.
   Interaction behaviors of two bright-dark lump waves are presented in Fig. 4.
   A periodic-shape bright lump wave is found in Fig. 5. In lump solutions  $(u^{(III)})$,
   the spatial structure called the bright lump wave is seen in Fig. 6.
   Interaction behaviors of two bright lump waves are presented in Fig. 7.
   A periodic-shape bright lump wave is found in Fig. 8. Fig. 9 and Fig. 10 display the interaction phenomena between lump
 and one solitary wave. Fig. 11 and Fig. 12 discuss the interaction phenomena between lump
 and two solitary waves.\\

\noindent {\bf Compliance with ethical standards}\\

\quad {\bf Conflict of interests} The authors declare that there is
no conflict of interests regarding the publication of this article.

\quad {\bf Ethical standard} The authors state that this research
complies with ethical standards. This research does not involve
either human participants or animals.



\end{document}